\documentclass[a4paper,11pt]{article}
\pdfoutput=1

\usepackage{jheppub} 

\usepackage[T1]{fontenc}
\usepackage[latin9]{inputenc}
\usepackage[latin9]{inputenc}
\usepackage{amsmath}
\usepackage{amssymb}
\usepackage{esint}
\usepackage{lmodern}
\usepackage{dsdshorthand}
\usepackage{graphicx}
\usepackage{colonequals}

\usepackage[most]{tcolorbox}
  \newcommand{\LAdS}{\ensuremath{L_{\text{AdS}}}}



\title{Towards Bootstrapping RG flows: Sine-Gordon in AdS}

\author[a]{Ant\'{o}nio Antunes,}
\author[a]{Miguel S. Costa,}
\author[b]{Jo\~{a}o Penedones,}
\author[a]{Aaditya Salgarkar,}
\author[c]{Balt C. van Rees}

\affiliation[a]{Centro de F\'\i sica do Porto, Departamento de F\'\i sica e Astronomia, \\
Faculdade de Ci\^encias da Universidade do Porto\\
Rua do Campo Alegre 687, 4169-007 Porto, Portugal}
\affiliation[b]{Fields and Strings Laboratory, Institute of Physics,\\
\'Ecole Polytechnique F\'ed\'erale de Lausanne, Switzerland}
\affiliation[c]{CPHT, CNRS, \'Ecole Polytechnique, Institut Polytechnique de Paris, \\
Route de Saclay, 91128 Palaiseau, France}

\emailAdd{alantunes@fc.up.pt}
\emailAdd{miguelc@fc.up.pt}
\emailAdd{joao.penedones@epfl.ch}
\emailAdd{salgarkaraaditya@fc.up.pt}
\emailAdd{balt.van-rees@polytechnique.edu}

\abstract{The boundary correlation functions for a Quantum Field Theory (QFT) in an Anti-de Sitter (AdS) background can stay conformally covariant even if the bulk theory undergoes a renormalization group (RG) flow. Studying such correlation functions with the numerical conformal bootstrap leads to non-perturbative constraints that must hold along the entire flow. In this paper we carry out this analysis for the sine-Gordon 
RG flows in AdS$_2$, which start with a free (compact) scalar in the UV and end with  well-known massive integrable theories that saturate many S-matrix bootstrap bounds. We numerically analyze the correlation functions of both breathers and kinks and provide a detailed comparison with perturbation theory near the UV fixed point. Our bounds are often saturated to one or two orders in perturbation theory, as well as in the flat-space limit, but not necessarily in between.
}

\begin{document} 
\setlength{\abovedisplayskip}{5pt}
\setlength{\belowdisplayskip}{5pt}
\maketitle
\flushbottom

%!TEX root = draft.tex
\section{Introduction}
In this work we study quantum field theories in a fixed AdS background. Such a setup was first discussed long ago in \cite{Callan:1989em}, but it has gained more attention in recent years because of the applicability of novel conformal bootstrap methods \cite{Rattazzi:2008pe}. Indeed, as is well-known from the AdS/CFT correspondence, if the AdS isometries are respected then the correlation functions of boundary operators obey almost all the axioms of conformal field theory (CFT) and in particular can be studied with all the usual conformal bootstrap tools. Not only does this allow one to investigate non-perturbative properties of theories in AdS, but by taking a \emph{flat-space limit} one can even obtain quantitative results for the S-matrix of flat-space non-conformal QFTs, as was demonstrated in \cite{Paulos:2016fap,Paulos:2016but,Paulos:2017fhb,Homrich:2019cbt}. In this latter limit the boundary correlation functions in particular are expected to transform into S-matrix elements, as can be seen in several ways \cite{Penedones:2010ue,Paulos:2016fap,Dubovsky:2017cnj,Hijano:2019qmi,Komatsu:2020sag}.

From this prehistory let us highlight the recovery of a maximal coupling for a bound state in two-dimensional S-matrices with a $\Z_2$ symmetry discussed in \cite{Paulos:2016fap}. To obtain this result from a QFT in AdS approach one proceeds as follows. Assuming a one-dimensional boundary operator product expansion of the form
\begin{equation}
\label{generalZ2OPE}
     \cO_1 \times \cO_1 =  1 +c_{112} \cO_2 + \ldots \text{(operators with $\De > 2 \De_1$)} \ldots\,,
 \end{equation} 
one can numerically bound the coupling $c_{112}$ as a function of $\De_1$ and $\De_2$. In the flat-space limit $\De_1 \approx m_1\LAdS$ and $\De_2 \approx m_2 \LAdS$ become both large, but an extrapolation of the numerical bootstrap methods yields an upper bound on the three-point coupling that is in excellent agreement with a bound obtained from the analytic S-matrix bootstrap \cite{Paulos:2016but}. Moreover, for $\sqrt{2} < m_2/m_1 < 2$ the flat-space scattering amplitude that extremizes this coupling is physical: it corresponds to the elastic amplitude of two `breathers' in the integrable sine-Gordon theory.

This particular result invites the question of the physical relevance of the numerical bootstrap results at finite $\De$. We recall that $\LAdS$ can play the role of a renormalization group scale, and the spectrum $\De(\LAdS)$ and OPE coefficients $c(\LAdS)$ can generally be expected to vary smoothly between the BCFT in the UV as $\LAdS \to 0$ and the flat-space gapped theory as $\LAdS \to \infty$. Therefore, it is natural to ask whether the numerical upper bound on $c_{112}$ at finite $\De$ is perhaps also saturated by sine-Gordon theory, now in an AdS space with a finite curvature radius. And if this is not the case, are there perhaps other numerical bootstrap bounds that are saturated by quantum field theories in AdS? If so then this would be a compelling example of our ability to \emph{bootstrap an entire RG flow} using only conformal methods.

One of the aims of this paper is to explore this line of thought for the $\Z_2$ preserving RG flows emanating from the free boson $\phi$ in AdS$_2$. A general such flow will begin at the conformal point where the AdS curvature is unimportant and we simply have a BCFT setup with  well understood dynamics. For example, with the choice of Dirichlet boundary conditions there is always the simple operator $\partial_{\perp} \phi$ with $\Delta = 1$ and with generalized free boson correlation functions. We can then switch on a potential, which in the most general $\Z_2$ preserving case would take the form
\begin{equation}
\label{generalZ2deformation}
    \int_{\text{AdS}} d^2 x \, \sqrt{g} \, \sum_{n \geq 0} \lambda_n \phi^{2 n}\,.
\end{equation}
Without further tuning, the deformed theory will flow to a gapped phase and in particular all the boundary scaling dimensions will become parametrically large as $\LAdS \to \infty$. The objective of this paper is to investigate to which extent such RG flows can be constrained or bootstrapped.

For the sine-Gordon theory the deformation has the form
\begin{equation}
    \lambda \int_{\text{AdS}} d^2 x \, \sqrt{g} \cos(\beta \phi)\,,
    \label{SGdef} 
\end{equation}
with $\phi$ a compact boson, $\phi \sim \phi + 2 \pi/\beta$. The dimension of the deforming operator is $\Delta_\beta = \beta^2 / (4 \pi)$. It will be important to consider $\Delta_\beta\leq2$ for the perturbation to be relevant. The parameter $\beta$ also determines the flat space spectrum as we explain in the beginning of Appendix \ref{sec:appsG}. For example, 
for $\Delta_\beta < 2/3$, the infrared is gapped and there are at least two breathers. As already mentioned,  the scattering amplitude of the lightest breather saturates the S-matrix bootstrap bound on the cubic coupling $g_{112}\propto c_{112}$.
In the ultraviolet the picture is as follows. The boundary operator with the quantum numbers of the lightest breather is $\cO_1 = \partial_\perp \phi$ with $\Delta_1 = 1$. At the free point its self-OPE is indeed of the form \eqref{generalZ2OPE} with $\Delta_2 = 2$ just saturating the imposed gap, and fortuitously we find that $c_{112} = \sqrt{2}$ saturates its numerical upper bound for these values of $\De_1$ and $\De_2$.

In section \ref{sec:breathers} we discuss the saturation of this bound by perturbative results around the free points. We first show that the bound is saturated by the first-order perturbative result, which is encouraging. At the second order things are however more involved. The sine-Gordon theory at fixed $\beta$ is `lost' in the sense that it moves into the bulk of the numerically allowed region. On the other hand, one can also consider sending $\lambda \to \infty$ and $\beta \to 0$ so as to only retain the $\phi^4$ perturbation at the second order, and with this scaling the perturbative results \emph{do} appear to saturate the numerical bounds. (For a specific value of the external dimension the second-order equivalence between the numerical bounds and the $\phi^4$ theory was observed earlier in \cite{Paulos:2019fkw}.) This is however where we believe our luck will run out, and at higher orders we expect numerics and analytics to diverge for any scaling of $\lambda$ and $\beta$. Concretely this is because the extremal spectrum of the numerical bounds does not match the perturbative expectations; see subsection \ref{sec:sparse} for a detailed discussion. As far as any of these breather bootstrap bounds are concerned, then, we must conclude that the sine-Gordon theory in AdS can only be recovered in the deep UV and the deep IR. This does not suffice to achieve our stated goal of bootstrapping an RG flow. 

Starting at subsection \ref{sec:multicorr}, the remainder of section \ref{sec:breathers} is dedicated to a multi-correlator study of two operators that should become two different breathers in the infrared. We introduce a natural five-dimensional space of OPE data in which we carve out various allowed regions with a numerical bootstrap analysis. With the exception of the free point, we unfortunately find that our perturbative predictions always appear to lie strictly below the numerical bounds. Therefore, the conclusion that the `breather correlators' are not extremal holds also for this setup.

In the sine-Gordon theories there are more elementary objects than breathers: the \emph{kinks} which correspond to field configurations that interpolate between different minima of the cosine potential.
These are the subject of section \ref{sec:chargedcorr}. They correspond to winding modes in the free compact boson theory, and a first-order perturbative analysis is provided in subsection \ref{subsec:cptkinks}. We also perform a first-order analysis around the free Dirac fermion in subsection \ref{subsec:cptdiracfermion}, which describes essentially the same theory because of the bosonization duality between the sine-Gordon and the Thirring model \cite{Coleman:1974bu}.

In the remainder of section \ref{sec:chargedcorr} we turn to the numerical analysis. An \emph{a priori} reason for optimism is that kink states do not exist for non-compact bosons and so general interactions of the form \eqref{generalZ2deformation} no longer provide viable deformation of the UV correlators. At a practical level, the main difference with the breather setup is that the kinks are charged under a global $O(2)$ symmetry. We have chosen to numerically bound the \emph{value} of the correlators at the crossing symmetric point. This analysis yields a three-dimensional `menhir' shape displayed in figure \ref{fig:menhirs}. Just as for the breathers, we once more find that the free and first-order perturbative theories lie on the boundary of the allowed (menhiresque) space, and so does the flat-space S-matrix if we extrapolate the bounds to large scaling dimensions $\Delta$. The sine-Gordon flows must lie within this menhir all the way from the UV to the IR, offering a definite bootstrap constraint on an RG flow.

Further conclusions and an outlook are provided in section \ref{sec:conclusions}. We in particular point out that, beyond low orders in perturbation theory, physical theories are not expected to exactly saturate bounds with a finite number of correlators. Instead we expect that bounds are saturated by extremal correlators with a very sparse and unphysical spectrum. Some technical results are collected in the appendices: in appendix \ref{sec:appsG} we give details of the perturbative calculations for sine-Gordon breathers; in appendix \ref{app:multicorrbounds} we describe how multi-correlator bounds can be limited by the existence of unphysical solutions to crossing; in appendix \ref{sec:fermionsAppendix} we explain the computation of the correlation functions of charged fermions in the AdS$_2$ Thirring model; and appendix \ref{sec:appopemax} provides some further numerical data for the kink correlation functions. 

%!TEX root = draft.tex

\section{Breather scattering}
\label{sec:breathers}
In this section we focus on breather states in sine-Gordon theory. These can be viewed as bound states of kinks and anti-kinks that are neutral under the continuous $O(2)$ symmetry, but can still be charged under the $\Z_2$ symmetry that sends $\phi \to - \phi$. In the UV theory with Dirichlet boundary conditions in AdS, the first boundary operator with the corresponding quantum numbers is $\cO_1=\partial_{\perp} \phi$ and so we will assume that it generates the lightest $\Z_2$ odd breather state. 
We will denote the lightest $\Z_2$ even operator by $\cO_2$, which in the UV theory is given by $(\partial_{\perp} \phi)^2$.
We will therefore be investigating the four-point functions of $\cO_1$ and $\cO_2$.

As explained in the introduction, our initial interest with these correlation functions is to see if we can track the sine-Gordon RG flow from highly curved AdS in the UV all the way to the flat-space limit. Unfortunately the operators in questions are not sensitive to the compactification radius $r$ of the boson $\phi$, and the physically allowed deformations of the free correlator therefore involve all the possible $\phi^{2n}$ couplings mentioned above. From the viewpoint of the numerical bootstrap it will turn out that the sine-Gordon theory at fixed $\beta$ does not occupy a distinguished place in the space of all these flows.
 
The organization of this section is as follows. We begin by analyzing the four-point function of $\cO_1$ analytically and numerically near the fixed point, to first and to second order in perturbation theory. We will provide evidence that the sine-Gordon theory in AdS saturates the (extrapolated) numerical bounds to the first order but not to the second order. In subsection \ref{sec:multicorr} we do a multiple correlator analysis involving also the operator $\cO_2$. In this case the parameter space is five-dimensional and we provide numerical bounds along various cross-sections, which we can match to first-order perturbation theory.  We in particular show that the sine-Gordon theory does not seem to saturate the bounds away from the free point.

\subsection{The free boson and its perturbations}
Our background is Euclidean AdS$_2$, with the metric
\begin{equation}
ds^2= \frac{L_{AdS}^2}{y^2} \left(dy^2 +dx^2\right),
\end{equation}
with $y>0$ and with $x \in \mathbb{R}$ the boundary coordinate. In this background we consider a free massless boson with the action
\begin{equation}
\label{eq:freeactionagain}
S=  \frac{1}{2} \int_{AdS_2} d^2x \sqrt{g} \, (\partial \phi)^2 \,,
\end{equation}
and with Dirichlet boundary condition, so $\phi \to 0$ as $y \to 0$. The simplest non-trivial boundary operator is then $\cO_1 = \partial_\perp \phi(x)$ whose correlation functions are just those of a generalized free boson with $\Delta_1 = 1$. For example, if we write its four-point function as
\begin{equation}
  \big\langle \cO_1(x_1)\cO_1(x_2)\cO_1(x_3)\cO_1(x_4) \big\rangle  = \frac{1}{x_{12}^2 x_{34}^2}\,  f(z)\,,
\label{eq:4ptfunction}
\end{equation}
with
\begin{equation}
z= \frac{x_{12}x_{34}}{x_{13}x_{24}}\,,
\label{eq:cross_ratio}
\end{equation}
where $x_{ij} = x_i-x_j$, then in the free theory
\begin{equation}
f^{(0)}(z) =1 + z^2 + \frac{z^2}{(1-z)^2} \,,
\label{eq:f(0)}
\end{equation}
and all higher-point functions of $\cO_1$ are equally easily obtained by Wick contractions.

In this section we will be interested in small perturbations away from the free conformal point that preserve the $\Z_2$ reflection symmetry. As we stated in the introduction, at first sight one may want to consider an interaction Lagrangian of the form $\lambda_n \phi^{2n}$ which contains all the relevant operators in the theory. However, in principle we can also consider \emph{irrelevant} interactions, like $(\partial \phi)^4$ and more complicated operators. Irrelevant deformations certainly make sense to any finite order in perturbation theory, where only finitely many counterterms are needed to cancel all divergences. They can however also correspond to a non-perturbatively well-defined setup: any RG flow that ends on the free massless boson would locally be parametrized by such irrelevant deformations. This means that there is no reason to exclude them from our bootstrap studies. 

%%%%%%%%%%%%%%%%%%%%%%%%%%%%%%%%%%%%%%%%%%%
\subsection{Single correlator}
\label{sec:singlecorr}
%%%%%%%%%%%%%%%%%%%%%%%%%%%%%%%%%%%%%%%%%%%%

%%%%%%%%%%%%%%%%%%%%%%%%%%%%%%%%%%%%%%%%%%%%
 \subsubsection{\texorpdfstring{First-order $\phi^4$ perturbation theory}{First-order phi4 perturbation theory}}
 \label{sec:1stOrder}
%%%%%%%%%%%%%%%%%%%%%%%%%%%%%%%%%%%%%%%%%%%%
As discussed in the introduction, we are  interested in $\mathbb{Z}_2$ symmetric deformations of the massless boson and therefore we can add any $\phi^{2n}$ operator to the Lagrangian. At first order, however, only the $\phi^2$ and $\phi^4$ operators change the four-point function of $\partial_{\perp}\phi$, and so (for now) we will consider only the action 
\begin{equation}
S= \int_{AdS_2} d^2x \sqrt{g} \left[ \frac{1}{2}(\partial \phi)^2 + \lambda\left( \frac{g_2}{2!} \, \phi^2 +\frac{g_4}{4!} \, \phi^4\right)  \right] .
\end{equation}
Using the Feynman-Witten rules, the first-order correction to the correlator is then given by
\begin{align}
&\langle \cO_1(x_1)\cO_1(x_2)\cO_1(x_3)\cO_1(x_4) \rangle^{(1)} =
\nonumber \\
&=\int_{\text{AdS}_2} d^2x \sqrt{g} \left[ -\frac{\lambda g_2}{\pi} \left( \frac{1}{x_{12}^2}\Pi_3\Pi_4 + \text{5 permutations}\right)-\frac{\lambda g_4}{\pi^2}\Pi_1\Pi_2\Pi_3\Pi_4\right]  ,	
\end{align}
with
\begin{equation}
\Pi_i \equiv \frac{y}{y^2 + (x-x_i)^2}\,,
\label{eq:BBpropagator}
\end{equation}
the bulk-to-boundary propagator for $\Delta=1$. The integrals can be evaluated straightforwardly as they correspond to a mass shift and a basic D-function. The complete correlator, obtained after integration, is given below in section \ref{sec:multicorr}.

Using the results given in appendix \ref{sec:appfirstord}, we can extract until first order the relevant CFT data for our two-parameter family of CTs. The result is
 \begin{equation}
 (\Delta_1,\Delta_2,c^2_{112})
 =   \left( 1+\lambda g_2 , 2+2\lambda g_2+ \lambda \frac{g_4}{4\pi},2 - \lambda \frac{g_4}{2 \pi}\right) .
 \label{eq:1stOrderData}
 \end{equation}
We can understand the $g_2$-dependent contributions as coming from disconnected diagrams with a mass shift. The $g_4$ correction is derived from the connected quartic Witten diagram. 
It will be convenient for comparison with the numerics to work in terms of physical quantities only. Therefore we restate the previous result as relations between conformal data. To first order in perturbation theory we can write
\begin{equation}
\label{firstorderplane}
c_{112}^2 =2 - 2\Delta_2 + 4\Delta_1\,.
\end{equation}
This defines a plane in the 3-d space $(\Delta_1,\Delta_2,c_{112}^2)$.
 
%%%%%%%%%%%%%%%%%%%%%%%%%%%%%%%%%%%%%%%%%%%%
\subsubsection{Comparison with numerics}
%%%%%%%%%%%%%%%%%%%%%%%%%%%%%%%%%%%%%%%%%%%%
It is well-known that the generalized free boson saturates the upper bound $c_{112}^2 \leq 2$ for $\De_1 = 1$ and $\De_2 = 2$. This alone indicates that the result of first-order perturbation theory should be tangential to the bound. Indeed, to first order we can always switch on both $g_2$ and $g_4$ with arbitrary signs because we can stabilize the potential with higher-order terms. But if every direction is physical then no direction can exit the allowed region, which geometrically is only possible if the bound is tangential to the plane defined by \eqref{firstorderplane} at $\De_1 = 1$ and $\De_2 = 2$ \cite{Paulos:2019fkw}.

\begin{figure}
	\centering
		\centering
		\includegraphics[width=.45\linewidth]{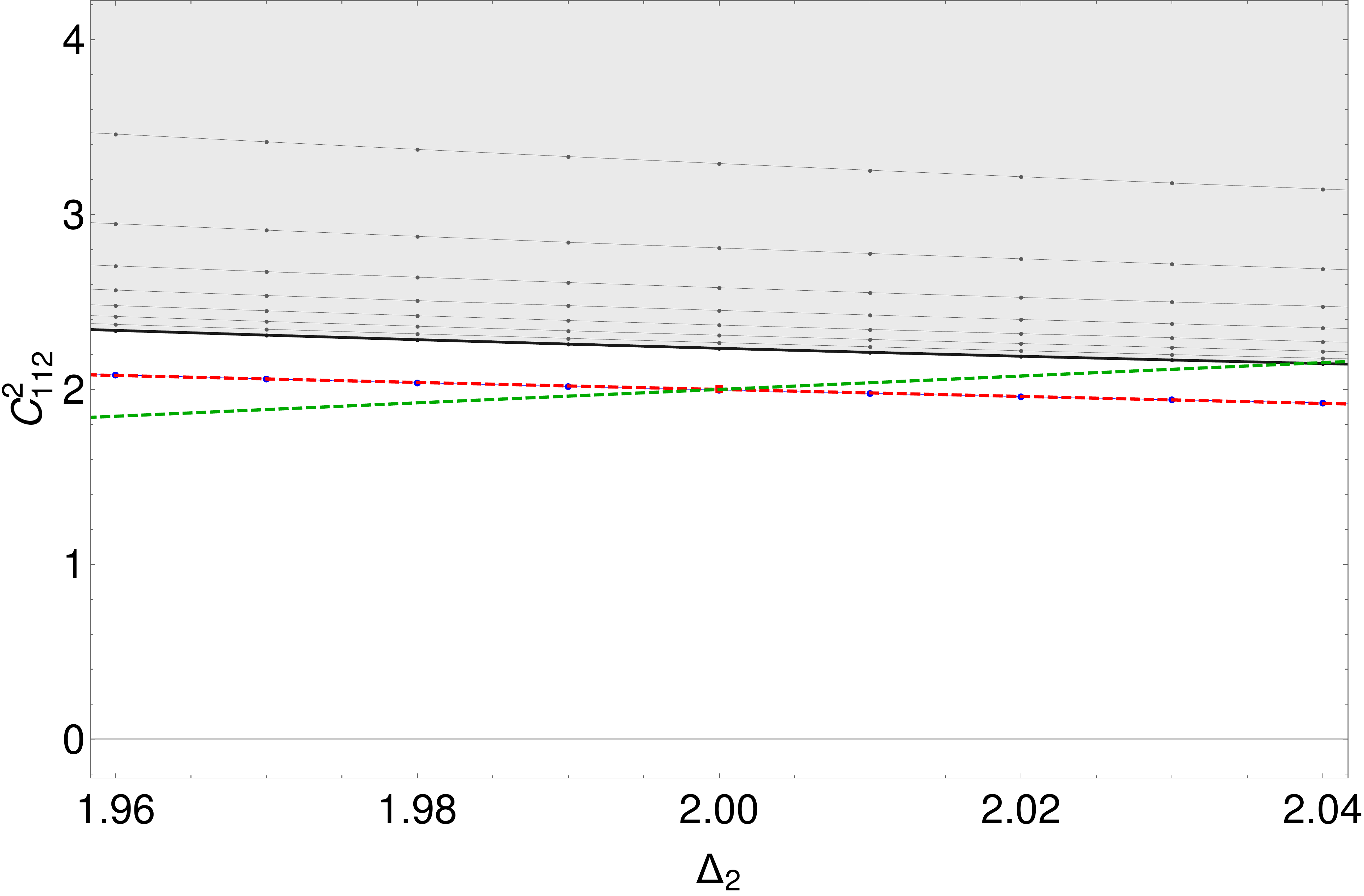}
		\includegraphics[width=.45\linewidth]{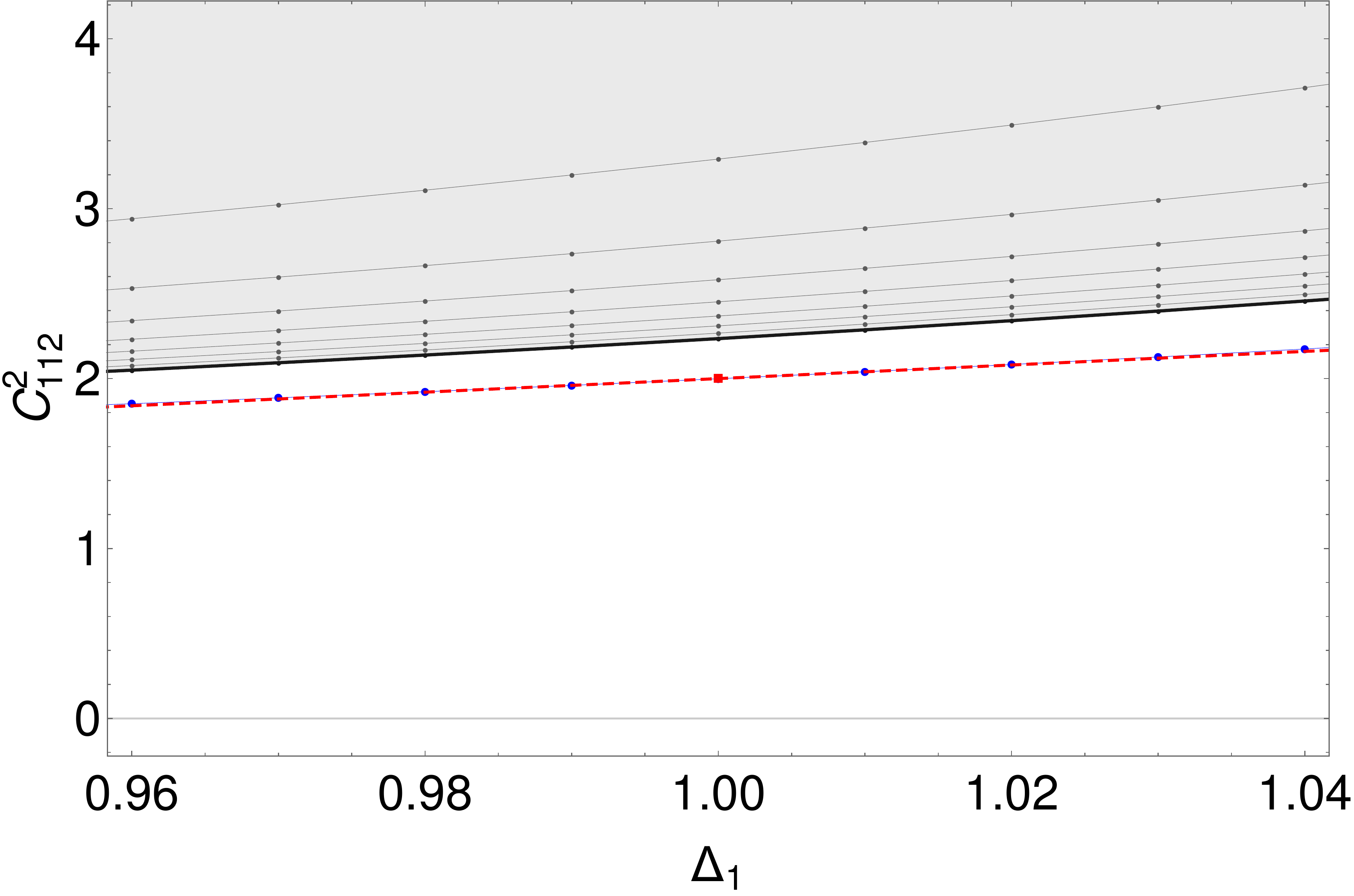}
	\caption{Bounds on the OPE coefficient $c_{112}^2$ in the vicinity of the free point. In the first plot we keep $\Delta_1=1$ fixed and in the second $\Delta_2=2$. The raw data points range from $\Lambda = 5$ (upper gray line) to $\Lambda = 29 $ (lower black line) in steps of 4, where $\Lambda$ is the number of derivatives of the crossing equation that we used. (We show the same values of $\Lambda$ in figures \ref{fig:secderc112} and \ref{fig:firstseconddergap}.) The blue points are an extrapolation to $\Lambda = \infty$ which fit well the first-order perturbative result (red line) around the free theory (red point). The green line corresponds to the irrelevant deformation discussed below.}
	\label{fig:openearfree}
\end{figure}

We have verified that this is indeed what happens in the entire plane.\footnote{The numerical bootstrap analyses in this paper were all done using SDPB \cite{Simmons-Duffin:2015qma,Landry:2019qug}. The numerical setup is entirely analogous to \cite{Paulos:2016fap}.} To illustrate this we show in figure \ref{fig:openearfree} the two slices given by the lines with fixed $\De_1$ and fixed $\De_2$. The dark areas are the rigorously ruled out region and we observe that the slope already matches first-order perturbation theory quite well. Furthermore, if we extrapolate the numerical results to infinite numerical precision we obtain an excellent match for all the shown data points. This confirms our expectation that the numerical bound matches first-order perturbation theory.

\subsubsection{Other deformations}
Now let us consider other deformations of the free massless bosons. First of all, we could have set $g_2 = g_4 = 0$. Then the first-order deviations given above would vanish trivially, and instead the leading deviation from the free theory would be given (at some loop order) by the first non-zero coupling like $g_6$ or $g_8$. The same argument as above would show that these deviations are necessarily \emph{also} tangential to the numerical bound. In this way the entire infinite space of RG flows emanating from the free boson appears to collapse to the lines in figure \ref{fig:openearfree}.

As mentioned at the beginning of this section, to first order it is also completely acceptable to study irrelevant deformations. Out of all of those we will consider only the $(\partial \phi)^4$ interaction. Physically one may think of this interaction as the least irrelevant operator in a theory that preserves both the reflection and the shift symmetry of $\phi$, and whose RG flow ends in the free massless boson. In higher dimensions this situation would for example arise whenever $\phi$ is a Goldstone boson, and then it is well-known that the coefficient of $(\partial \phi)^4$ must be positive in flat space \cite{Adams:2006sv}. For the two-dimensional theory in Euclidean AdS the action
\begin{equation}
S= \int_{AdS_2} d^2x   \sqrt{g} \left[ \frac{1}{2}(\partial \phi)^2 - \tilde \lambda (\partial \phi)^4 \right] ,
\end{equation}
yields the first-order correction to the OPE data
\begin{equation}
  (\De_1, \De_2, c_{112}^2) = \left(1, 2 - \frac{\tilde{\lambda}}{6\pi}, 2 - \tilde \lambda \frac{23}{36 \pi}\right) + O(\tilde{\lambda}^2)\,.
\end{equation}
This perturbative result corresponds to the green line in the left plot in figure \ref{fig:openearfree}. However, the upper half of this line is excluded by the (extrapolated) numerical bootstrap bound. We therefore conclude that this leading-order perturbation cannot exponentiate to a valid solution to the crossing symmetry equations, and therefore
\begin{equation}
  \tilde \lambda \geq 0\,,
\end{equation}
just as in higher dimensions.

It is interesting that we could so easily bound the coefficient of the leading irrelevant operator. In future work it might be worthwhile to see if this idea can be used to derive similar bounds in higher-dimensional theories and for the subleading irrelevant terms. In this way the numerical bootstrap can perhaps re-derive or improve the analytic results of \cite{Caron-Huot:2020cmc,Caron-Huot:2021rmr,Caron-Huot:2021enk} and \cite{Kundu:2021qpi} for effective field theories in AdS.

%%%%%%%%%%%%%%%%%%%%%%
\subsubsection{Second-order}
%%%%%%%%%%%%%%%%%%%%%%
Starting with the second order in perturbation theory we have a choice to make. Suppose the $\phi^4$ interaction strength is proportional to a parametrically small coupling $\lambda$. Then how should we scale the $\phi^6$ and higher interactions? Our first natural option is to consider the sine-Gordon interaction at fixed $\beta$ as discussed in the introduction. Then we can heuristically write
\begin{equation}
\lambda (\cos(\beta \phi) - 1) = \lambda \sum_{n > 0} \frac{(-1)^n}{(2n)!}\beta^{2n}\phi^{2n}\,,
\end{equation}
and deduce that the $\phi^6$ coupling should simply scale as $\lambda$. (In practice we should work directly with the compact boson and regard the cosine term as a real vertex operator, as explained in detail in appendix \ref{sec:appsecondorder}.)

The other choice is obtained by replacing $\beta \to \lambda \xi$ and $\lambda \to \lambda^{-1}$ so the interaction becomes
\begin{equation}
  \sum_{n > 0}\frac{(-1)^n}{(2n)!} \lambda^{2 n - 1} \xi^{2n} \phi^{2n}\,.
\end{equation}
In this case the $\phi^6$ interaction scales as $\lambda^{2}$. The advantage of the second scaling is that $\lambda$ is now a true loop counting parameter, as is easily verified by drawing a few Feynman diagrams. It is also the scaling that was used in \cite{Dorey:1996gd}  to give an elegant intuitive argument for the integrability of the classical theory in flat space.\footnote{If we introduce the $\phi^{2k}$ interactions order by order then we necessarily have to consider the boson to be non-compact and then the spectrum of bulk operators is continuous. Fortunately, this does not pose any problem for the correlation functions of boundary operators because with our choice of Dirichlet boundary conditions the boundary spectrum remains discrete.}

The different choices of expanding the interaction potential lead to different ways of perturbing the fixed point and a priori we can consider all of them in connection with the numerical results. In both cases we will get an expansion of the form
\begin{equation}
  \begin{split}
  \De_1 &= 1 + \gamma_1^{(1)} \lambda + \gamma_1^{(2)} \lambda^2 + \ldots \,,\\
  \De_2 &= 2 + \gamma_2^{(1)} \lambda + \gamma_2^{(2)} \lambda^2 + \ldots\,,\\
  c_{112}^2 &= 2 + c^{(1)} \lambda + c^{(2)} \lambda^2 + \ldots\,,
  \end{split}
\end{equation}
where the coefficients are functions of the single remaining parameter $\beta$ or $\xi$. The computation of these coefficients can be found in appendix \ref{sec:appsecondorder}; for the sine-Gordon theory at fixed $\beta$ the computations are far from trivial and $c^{(2)}(\beta)$ and $\gamma_2^{(2)}(\beta)$ can only be obtained numerically, with a computational cost that increases quickly with $\beta$. If we keep $\xi$ fixed  then the computation is significantly easier, and only the $\phi^2$ and $\phi^4$ interaction vertices contribute. Either way, in both cases the equations are seen to lead to a one-parameter family of RG flows that emanate from the free point. For comparison with the numerics it is useful to eliminate $\lambda$ and the parameter in favor of $(\De_1 -1, \De_2 -2)$, obtaining a quadratic equation for $c_{112}^2$ in terms of $\Delta_2 - 2$ and $\Delta_1 - 1$. Doing so for the second scaling, which is the same as the $\phi^4$ perturbation, yields
\begin{equation}
\label{c112squaredphi4}
  c_{112}^2 = 2-2(\Delta_2-2\Delta_1) +\left( \frac{\pi^4}{15}-4\zeta(3)+\frac{5}{2}\right) (\Delta_2-2\Delta_1)^2 + 4(\Delta_2-2\Delta_1)(\Delta_1-1)\,,
\end{equation}
and one may envisage a similar equation for the sine-Gordon perturbation at fixed $\beta$, which is however much more difficult to write down. Notice that we can no longer deduce the individual RG flows from the parametrization given in equation \eqref{c112squaredphi4} --- instead we only see the surface that is foliated by all the flows together. This is however also all we are able to see numerically.

\begin{figure}[h]
\centering
\includegraphics[width=0.65\linewidth]{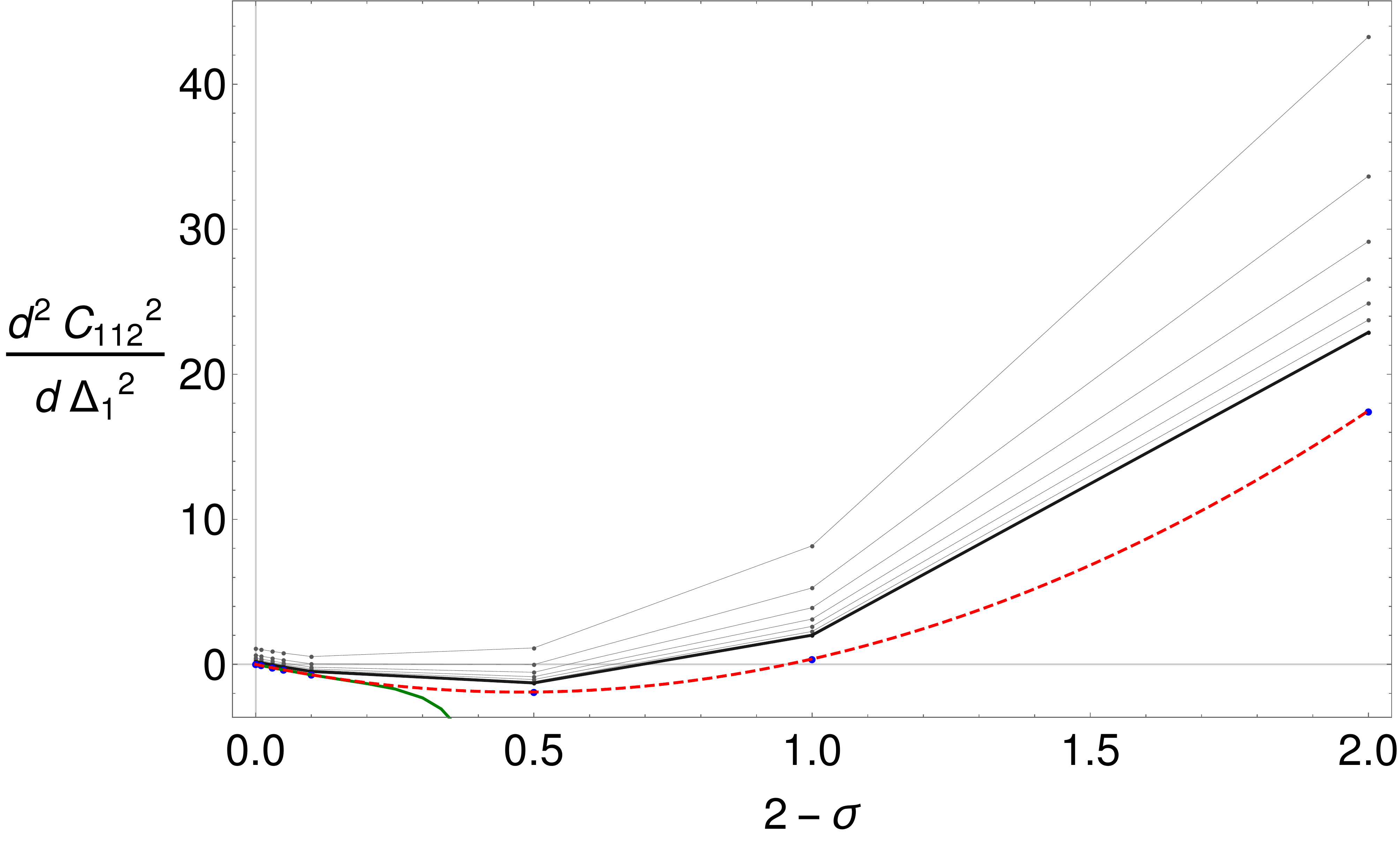}
\caption{The second derivative of $c_{112}$ with respect to $\Delta_1$,
at the free theory point, as a function of $\sigma = (\Delta_2 - 2) / (\Delta_1 - 1)$. The dashed red $\phi^4$ curve coincides to high precision with the extrapolation of the numerical results to infinite $\Lambda$ which are represented by the blue points. On the other hand, the sine-Gordon curve (in green) is subleading.}
\label{fig:secderc112}
\end{figure}

For the numerical experiment we have chosen to compute the second derivative of  the maximal value of $c_{112}^2$ along the straight lines given by
\begin{equation}
  \Delta_2 - 2 = \sigma (\Delta_1 - 1)\,.
\end{equation}
The results are shown in figure \ref{fig:secderc112} for $0 \leq \sigma \leq 2$ where the sine-Gordon deformation is relevant. 
For this figure we estimated the second derivative of the numerical bound using finite differences, and then extrapolated to infinite $\Lambda$. Our first observation is that the $\phi^4$ theory, and therefore also the sine-Gordon theory at fixed $\xi$, provides an excellent match with the numerical data.\footnote{For $\Delta_1 = 1$ this was also observed in \cite{Paulos:2019fkw}.}
At this order sine-Gordon is a maximal theory. We shall argue below that we do not expect such property to hold at higher orders.
At this order the sine-Gordon deformation at fixed $\beta$ is however no longer maximal, as we anticipated in the introduction.

\subsubsection{Relation to gap maximization}
\begin{figure}
  \centering
  \includegraphics[width=0.49\linewidth]{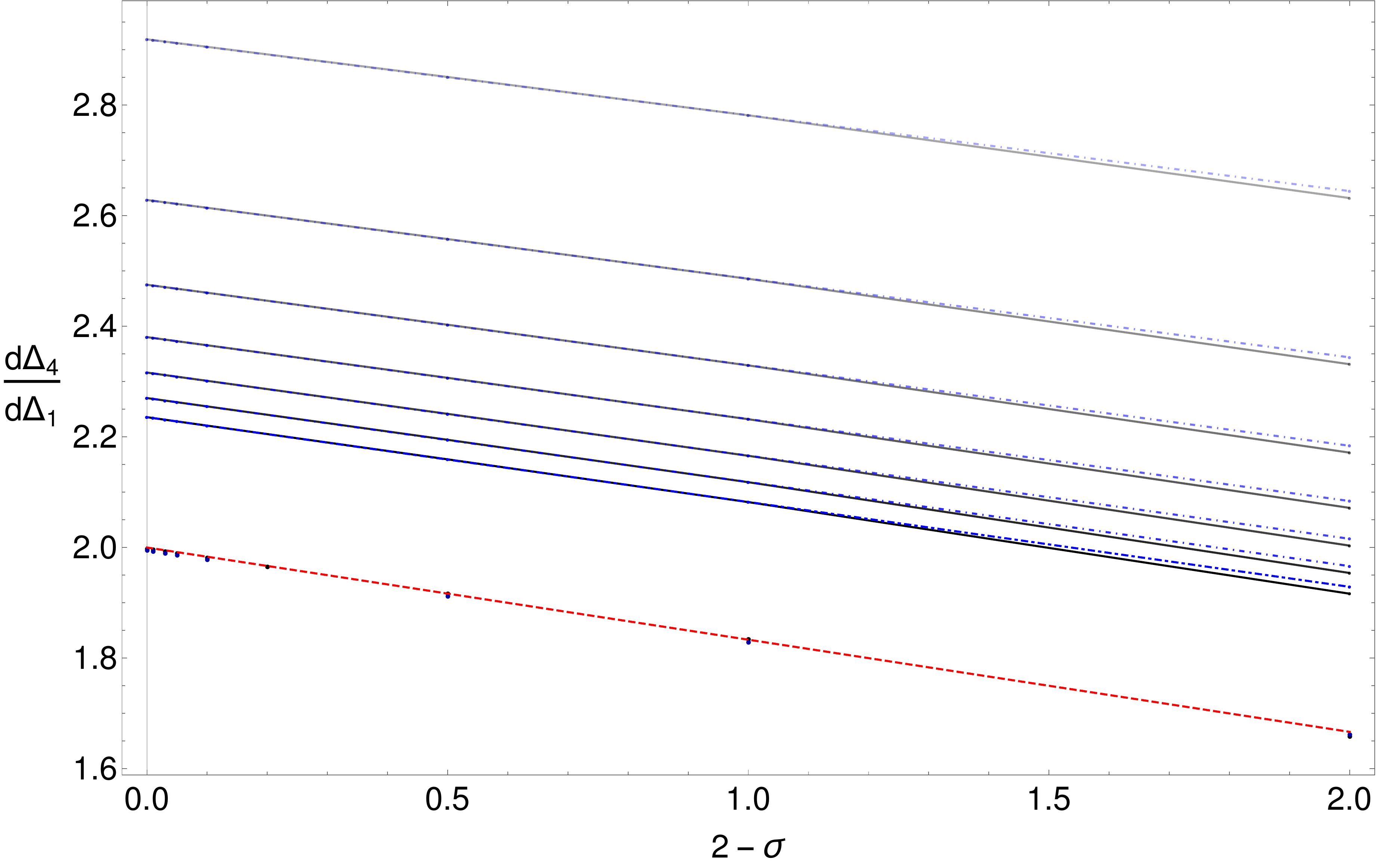}
  \includegraphics[width=0.49\linewidth]{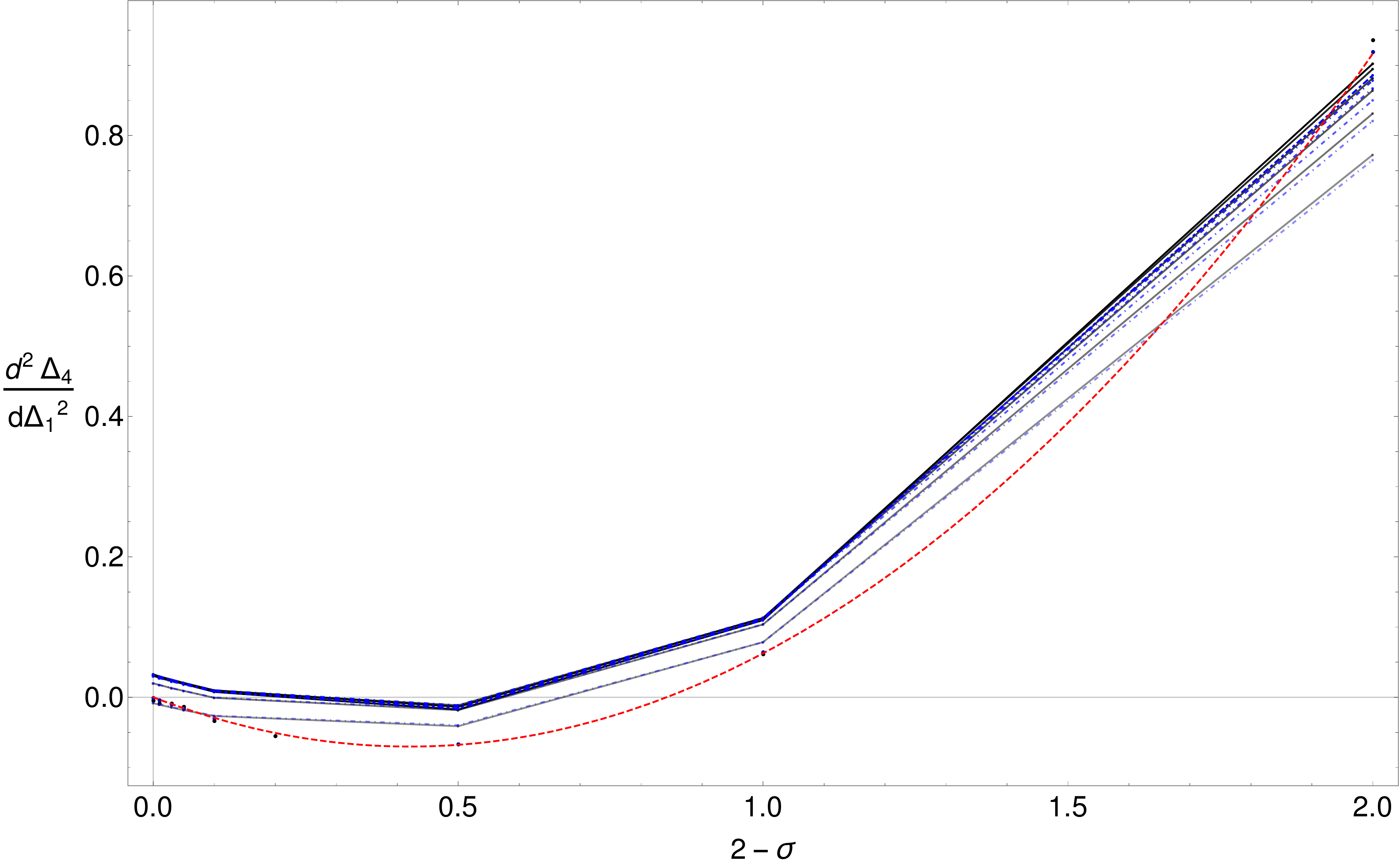}
  \caption{First and second derivative of $\Delta_4$ at the free theory point, as a function of $2-\sigma$. The $\phi^4$ curve, or the SG theory at fixed $\xi$, in red coincide to high precision with the large $\Lambda$ extrapolation of the spectrum extraction from OPE maximization points in blue and the black gap maximization points. Both numerical approaches give very similar results. Note that the second derivatives, as estimated from finite differences, are not monotonic in the cutoff $\Lambda$.}
  \label{fig:firstseconddergap}
\end{figure}

It turns out that we can trace the $\phi^4$ theory to second order also in a different manner: we can try to maximize the gap to the operator after $\cO_2 = (\partial_\perp \phi)^2$ rather than maximizing the OPE coefficient $c_{112}^2$. At the free conformal point there is a degeneracy since these next operators are given by
\begin{equation}
  O_{4}= (\partial_\perp \phi)\Box(\partial_\perp \phi)\,, \qquad
O_{4'}= (\partial_\perp \phi)^4 \,,
\end{equation} 
which both have dimension $4$. Of course, this degeneracy generically gets lifted as we switch on the $\phi^4$ or even the $\phi^2$ terms in the Lagrangian. But to second order only the first of these operators makes an appearance in the four-point function of $\cO_1 = \partial_\perp \phi$ because $c^2_{114'} = O(\lambda^4)$. For this operator ${\cal O}_4$ we find, in a manner analogous to before, that 
\begin{align}
  \Delta_4 &= 4 +2(\Delta_1-1) + \frac{1}{6}(\Delta_2-2\Delta_1)
  \\&+ \frac{1}{6}(\Delta_2-2\Delta_1)(\Delta_1-1)
  +\left(\frac{317}{144}-\frac{5}{3}\zeta(3)\right) (\Delta_2-2\Delta_1)^2 \,.
  \nonumber
\end{align}
This quadratic curve once more precisely traces the numerical bounds as can be seen in figure \ref{fig:firstseconddergap}. Using the uniqueness of the extremal solution it is then clear that
\begin{align}
& \text{Boundary dual of} \,\phi^4\, \text{theory in AdS}   \nonumber\\
=\ &\text{Extremal theory that maximizes the OPE coefficient}\;  c_{112}=c_{112}(\Delta_1,\Delta_2) \\
=\ &\text{Extremal theory that maximizes the gap} \; \Delta_4=\Delta_4(\Delta_1,\Delta_2)\,, \nonumber
\end{align}
to second order around $\De_1 - 1$ and $\De_2 - 2 \De_1$. We also note that we empirically found that the OPE and gap maximization problems have the same solution at finite truncation order,  which was also observed in \cite{Paulos:2019fkw} before.

\subsubsection{Comments on higher orders}
\label{sec:sparse}
We have seen that the sine-Gordon theory can be extremal around the free point, albeit only with a specific scaling of the parameters, to second order in perturbation theory. Unfortunately this extremality property is unlikely to persist at higher orders, as we will now proceed to explain. The overall picture will therefore be that sine-Gordon theory in AdS saturates the bootstrap bound to zeroth, first and second order in the UV and also in the deep IR, but not in between.

Rather than working out the details of the third-order perturbative result we will provide an indirect argument for non-extremality. First we recall that the numerical bootstrap procedure allows us to extract an approximate solution to the crossing symmetry equations precisely at the extremal value of the OPE or gap bound. Now, for any $\De_1$ and $\De_2$ in the vicinity of the free point this so-called extremal spectrum appears to be quite special in the sense that it is relatively sparse: as we explain in more detail below, it contains at most a single operator per `bin' of width $2$ in $\Delta$ space. 

For reference we first discuss this sparseness property in physical theories. It is clearly obeyed at the free point: the spectrum in the generalized free four-point function of $\cO_1 = \partial_\perp \phi$ contains operators of dimensions $2\Delta_1 + 2n$ with $\Delta_1 = 1$. In reality, however, there are multiple such operators for each $n \geq 1$ and the free spectrum is highly degenerate. Perhaps surprisingly these degeneracies remain hidden to first and second order in perturbation theory. For example, the operator $\cO_{4'} = (\partial_\perp \phi)^4$ only appears at fourth order in $\lambda$ in the four-point function of $O_1$ and the same is true for other operators at higher $n$. Therefore, whereas the spectrum up to third order is sparse enough to be extremal, at fourth and higher orders this is generally no longer the case. 

On the numerical side we simply observed a sparse extremal spectrum for all the values of $\Delta_1$ and $\Delta_2$ that we tried, with no hint of resolved degeneracies at any $\Lambda$. The sparseness was also already discussed in some detail in \cite{Paulos:2019fkw}. In that paper it is reflected not only in the choice of functional basis, but the numerical results (for $\Delta_1 = 1$ and varying $\Delta_2$) also provide substantial evidence that there is indeed a single operator per bin. Finally, the sparseness property also fits in nicely with the extremal functionals in one dimension that were found in \cite{Mazac:2018mdx,Mazac:2018ycv} which also always have a single operator per bin.

It remains an interesting open question whether \emph{every} extremal solution has at most a single operator per bin, and whether a similar sparseness can be true even for multi-correlator bootstrap bounds. This is however beyond the scope of the present work.\footnote{We can offer some comments nevertheless. Of course the mean-field spectrum of a multi-correlator bootstrap setup involving $\cO_1$ and $\cO_2$ would generally contain 3 operators per bin, corresponding to the different double-twist operators $\cO_1 \partial^{2n}\cO_1$, $\cO_2 \partial^{2n}\cO_2$ and $\cO_1 \partial^{n}\cO_2$. But this is not necessarily an \emph{extremal} spectrum. On the other hand, let us recall the dictionary and numerical results of \cite{Paulos:2016fap} which state that correlators (of identical operators) with a single operator per bin must converge to scattering amplitudes which saturate elastic unitarity in the flat-space limit. But in \cite{Homrich:2019cbt} it was shown that some multi-correlator systems (or actually the bounds obtained from them) converge to multi-amplitude systems (or actually the bounds obtained from them) whose individual amplitudes \emph{do not} all saturate elastic  unitarity. We therefore believe that these extremal correlators do not contain a single operator per bin. It would be nice to check this, but the authors of \cite{Homrich:2019cbt} did not analyze the extremal spectra for their bounds.
}

Are there mechanisms that could retain the sparsity of the spectrum and therefore extremality? We can for example imagine tuning the couplings such that the entire spectrum of the theory remains degenerate also at higher orders, or tuning the OPE coefficients such that the spectrum of operators appearing in $\langle\cO_1\cO_1\cO_1\cO_1\rangle$ remains sparse. (In the latter case we would still observe non-sparseness in other correlation functions, for example the ones studied in the next subsection.) Some counting arguments however show that either scenario is unlikely to be achievable with only $\phi^{2k}$ interactions: at every order there is simply too much OPE data to tune given the finite number of coefficients. A more promising avenue would be to also allow for irrelevant deformations. Indeed, every primary operator can also be used to deform the theory and one might therefore imagine tuning their coefficients precisely such that sparsity is retained.

It would be interesting to see whether there indeed exists a tuning of relevant and irrelevant interactions such that the spectrum remains sparse at finite coupling. Such a tuning bears some resemblance to the flat-space analysis of \cite{Dorey:1996gd}  where the flat-space sine-Gordon theory is recovered by dialing the interactions so as to eliminate particle production. Indeed, according to \cite{Paulos:2016fap,Komatsu:2020sag}, a correlator with a single operator per bin produces an elastic amplitude in the flat-space limit. It is therefore likely to be this fine-tuned and likely non-local theory that saturates the numerical upper bound on $c_{112}^2$ all the way from the free boson at $\Delta = 1$ until the flat-space sine-Gordon theory at $\Delta \to \infty$.
      
%%%%%%%%%%%%%%%%%%%%%%%%%%%%%%%%%%%%%%%%%%%%
\subsection{Multiple correlators}
\label{sec:multicorr}
%%%%%%%%%%%%%%%%%%%%%%%%%%%%%%%%%%%%%%%%%%%%

We will now analyze the following system of correlators
\begin{equation}
\label{multiplecorrs}
\begin{split}
\langle\cO_1(x_1) \cO_1(x_2) \cO_1(x_3) \cO_1(x_4) \rangle\,,\\
\langle\cO_2 (x_1)\cO_2(x_2) \cO_1(x_3) \cO_1(x_4)\rangle \,,\\
 \langle \cO_2(x_1) \cO_2(x_2) \cO_2(x_3) \cO_2(x_4)\rangle \,.
\end{split}
\end{equation}
We will again probe this system in the vicinity of the generalized free boson point with $\De_1 = 1$ and $\De_2 = 2$, where we can identify $\cO_1=\partial_{\perp}\phi$ and $\cO_2=(\partial_{\perp}\phi)^2$.

The operators appearing in this mixed one-dimensional correlator system are labeled by their quantum numbers under the $\Z_2$ reflection symmetry sending $\phi \mapsto - \phi$, as well as under boundary parity $x \mapsto - x$. 
The latter symmetry is what remains of a rotational symmetry in one space dimension. Parity odd operators cannot appear in the OPE of two identical operators, which exemplifies that it can be useful to think of the parity odd operators as spin 1 and the parity even operators as spin 0. The operators $\cO_1$ and $\cO_2$ are parity even. The operator spectra will be assumed to have the form
\begin{center}
\begin{tabular}{l|l|l}
$\Z_2$ & P & assumed spectrum\\
\hline
$+$ & $+$ & $\mathbf 1$, $\cO_2$, % $(\partial_{\perp}\phi)^2$, 
and operators with $\De > \De_{\text{gap}}$\\
$-$ & $+$ & $\cO_1$ %$\partial_{\perp}\phi$ 
and operators with $\De > \De_1$\\
$-$ & $-$ & operators with $\De > \De_1$\\
$+$ & $-$ & no assumptions, as these do not feature in \eqref{multiplecorrs}
\end{tabular}
\end{center}
With these assumptions we are left with the following natural five-dimensional space of parameters
\begin{equation}
  \cP : \{\De_1, \De_2, \De_\text{gap}, c_{112}, c_{222} \} \,.
\end{equation}
As an example, it is easily verified that the generalized massless free boson point corresponds to $\{1,2,4,\sqrt{2},2\sqrt{2}\}$.

Below, we will study some first-order deformations away from the generalized free boson point, both numerically and perturbatively. For the perturbative computations we will assume that the $\Z_2$ symmetry remains preserved. If we furthermore only consider relevant perturbations then the most general first-order deformation is captured by the action
\begin{equation}
\label{phi8action}
	S= \int_{AdS_2} d^2x  \sqrt{g}\left[ \frac{1}{2}(\partial \phi)^2 + \lambda\left( \frac{g_2}{2!}  \,\phi^2 +\frac{g_4}{4!} \, \phi^4+\frac{g_6}{6!}  \,\phi^6+\frac{g_8}{8!} \, \phi^8\right)  \right] \,,
\end{equation}
with $\lambda$ infinitesimal and $g_2$, $g_4$, $g_6$ and $g_8$ arbitrary. As before, couplings of the form $\phi^{2k}$ for sufficiently large $k$ do not lead to a first-order change of the correlators in \eqref{multiplecorrs}. The action \eqref{phi8action} leads to a four-dimensional space of deformations emanating from the generalized massless free boson point, and our first goal is to compute how the OPE data in $\cP$ is affected by these deformations.

%%%%%%%%%%%%%%%%%%%%%%%%%%%%%%%%%%%%%%%%%%%%
\subsubsection{First-order perturbation theory: correlators}
%%%%%%%%%%%%%%%%%%%%%%%%%%%%%%%%%%%%%%%%%%%%
We begin our perturbative analysis by computing the correlators in \eqref{multiplecorrs} to first order in $\lambda$ with the action \eqref{phi8action}. In this subsection, with a small abuse of notation, it is understood that, in the free theory $\cO_2=(\partial_{\perp}\phi)^2$ is normalized to have  unit norm.

As explained in section \ref{sec:1stOrder}, the four-point function of $\cO_1$ to first order reads:
\begin{align}
&\langle \cO_1(x_1) \cO_1(x_2)\cO_1(x_3)\cO_1(x_4)\rangle= \frac{1}{x_{12}^{2\Delta_1}x_{34}^{2\Delta_1}}\Bigg[1 + z^2 + \frac{z^2}{(1-z)^2} -\frac{\lambda g_4}{4\pi}z^2 \bar{D}_{1111}(z)\nonumber\\ 
&
\qquad\qquad\qquad+\frac{2 \lambda g_2z^2}{(1-z)^2}\left((1-z)^2 \log(z)+ \log\left(\frac{z}{1-z}\right)\right)\Bigg]  + O(\lambda^2)\,,
\end{align}
where $\Delta_1=1+\lambda g_2$ and the D-function $\bar{D}_{1111}(z)$ is defined in appendix \ref{sec:appfirstord}. Notice that all terms proportional to $g_2$ come from disconnected diagrams,\footnote{Notice that 
$z^{2\Delta_1} = z^2+ 2 \lambda g_2z^2  \log(z) + O(\lambda^2)$.}
 the only connected term comes from the $g_4$ coupling, and the $g_6$ and $g_8$ couplings do not contribute.

\begin{figure}
  \centering
  \includegraphics[width=0.8\linewidth]{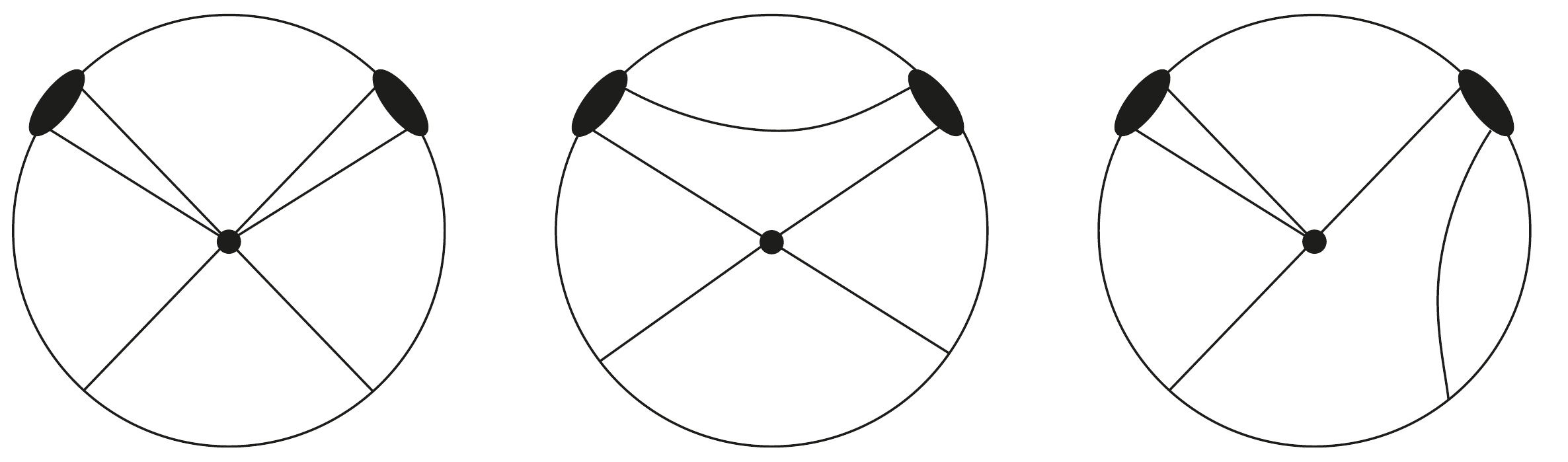}
  \caption{Connected diagrams contributing to the 4-pt function $\langle\cO_2 \cO_2 \cO_1 \cO_1 \rangle$.
   There are contributions from both the $g_6$ and $g_4$ couplings. Additional diagrams obtained by permuting the external operators must be added.}
  \label{fig:2211}
\end{figure}

For the other two correlation functions in \eqref{multiplecorrs}, let us first write their zeroth-order term. Simple Wick contractions yield
\begin{align}
  \langle\cO_2(x_1) \cO_2(x_2)\cO_1(x_3)\cO_1(x_4) \rangle^{(0)}& =\frac{1}{x_{12}^4 x_{34}^2} \left(1 +2z^2 +2 \frac{z^2}{(1-z)^2}\right)   ,
  \nonumber  
  \\
  \langle \cO_2(x_1) \cO_2(x_2)\cO_2(x_3)\cO_2(x_4)\rangle^{(0)} &=\frac{1}{x_{12}^4 x_{34}^4}\left[1 + z^4 + \frac{z^4}{(1-z)^4}\right.\label{eq:MixedTree}\\
   &\left.\quad+ 4\left(z^2 + \frac{z^2}{(1-z)^2} + \frac{z^4}{(1-z)^2}\right)  \right] . \nonumber
\end{align}

The first-order corrections to the first correlator come from the connected diagrams in figure \ref{fig:2211}, plus other 
disconnected diagrams. The first diagram in figure \ref{fig:2211} is proportional to the $g_6$ coupling and reads
\begin{equation}
-\lambda \frac{g_6}{2\pi^3}\int_{AdS_2} d^2x \sqrt{g}  \,\Pi_1^2 \Pi_2^2 \Pi_3 \Pi_4 \,,
\end{equation}
with $\Pi_i$ corresponding the bulk to boundary propagator for a field dual to an operator of dimension 1, defined previously in (\ref{eq:BBpropagator}). Notice that $\Pi_{i}^2$ is proportional to the bulk to boundary propagator for a field dual to an operator of dimension 2, and this means that this contribution to the correlator is simply the D-function $D_{2211}$. Taking into account all the other diagrams we obtain that
\begin{align}
  &\left\langle\cO_2(x_1) \cO_2(x_2)\cO_1(x_3)\cO_1(x_4) \right\rangle= \frac{1}{x_{12}^{2\Delta_2} x_{34}^{2\Delta_1}}\Bigg(h^{(0)}(z)-\frac{3\lambda g_6}{16 \pi^2} z^4 \bar{D}_{2211}(z)-\frac{\lambda g_4}{2\pi} z^2 \bar{D}_{1111}(z) 
  \nonumber\\
  & +\frac{ \lambda g_4  z^2 \left( z(z-2)(\log (1-z)-1)-2\right) }{2 \pi  (z-1)^2} +
  \frac{4 \lambda g_2 z^2 \left(\left(z^2-2 z+2\right) \log (z)-\log (1-z)\right)}{(z-1)^2} \Bigg)
    \nonumber\\
  & + O(\lambda^2) \,,
\end{align}
where $h^{(0)}(z)$ is defined by the tree level answer obtained from (\ref{eq:MixedTree}), $\Delta_1 = 1+\lambda g_2$ as before, and $\Delta_2 = 2 + 2 \lambda g_2 + \lambda g_4 / (4\pi)$. Notice that we previously also obtained these scaling dimensions from the four-point function of $\cO_1$ --- see equation (\ref{eq:1stOrderData}).

\begin{figure}[t]
	\centering
	\includegraphics[width=0.99\linewidth]{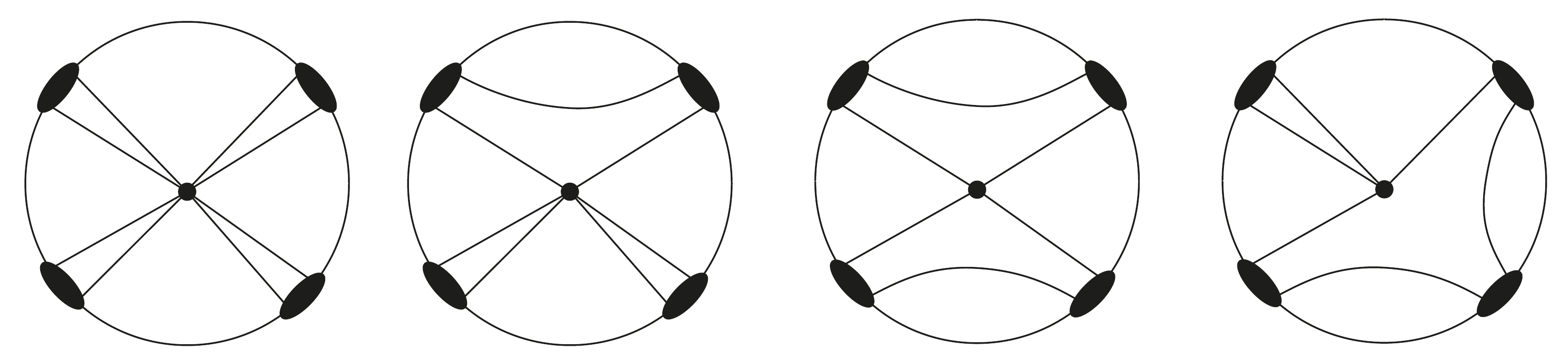}
	\caption{Tree level diagrams contributing to the 4-pt function $\langle\cO_2 \cO_2 \cO_2 \cO_2\rangle$
	 from $\phi^8,\phi^6$ and $\phi^4$ interactions. Additional diagrams obtained by permuting the external operators must be added.}
	\label{fig:2222}
\end{figure}

Finally, the four-point function of $(\partial_\perp \phi)^2$ is given by
\begin{align}
&\left\langle \cO_2(x_1) \cO_2(x_2)\cO_2(x_3)\cO_2(x_4)\right\rangle
= \frac{1}{x_{12}^{2\Delta_2} x_{34}^{2\Delta_2}} \left[ k^{(0)}(z) -\frac{15  \lambda g_8}{64 \pi ^3} \, z^4 \bar{D}_{2222}\right.\nonumber \\
&-\frac{3  \lambda g_6}{8 \pi ^2}\,  z^2 \bigg(\bar{D}_{1122}+z^2 \left(\bar{D}_{1212}+\bar{D}_{1221}+\bar{D}_{2121}+\bar{D}_{2211}+(z-1)^{-2}\bar{D}_{2112}\right)\bigg) \nonumber\\
&-\frac{\lambda g_4}{\pi} \,  \frac{z^2 \bar{D}_{1111} }{(z-1)^2} \, \Big((z-1) z+1\Big)^2
+ \frac{\lambda g_4}{2 \pi } \,  \frac{z^2}{ (z-1)^4} \left(z\big(-8 z^2+15 z-8\big) \log (1-z) \right.
\nonumber\\
&  \left.+z^2 \big(z^4-4 z^3+14 z^2-20 z+10\big) \log(z) -8(z-1)^2(z^2-z+1)\right) 
   \\
& -    \frac{4  \lambda g_2 z^2}{(z-1)^4} \left(\big(2 z^4-4 z^3+5 z^2-4 z+2\big) \log (1-z) \right.\nonumber\\
 & \left.\left. -\big(z^6-4 z^5+12 z^4-20 z^3+20 z^2-12z+4\big) \log (z) \right)\right] +  O(\lambda^2) \,,
 \nonumber
\end{align}
where $k^{(0)}(z)$ is again the tree level answer defined by (\ref{eq:MixedTree}) and $\De_1$ and $\De_2$ are as before. For this correlator the connected Witten diagrams are shown in \ref{fig:2222}. In particular, the first diagram introduces a contribution from the $g_8$ coupling
given by
\begin{equation}
-\lambda \frac{g_8}{4 \pi^4} \int_{AdS_2} d^2x  \sqrt{g} \, \Pi_1^2 \Pi_2^2 \Pi_3^2 \Pi_4^2 \,,
\end{equation}
which is just the D-function $D_{2222}$.

\subsubsection{First-order perturbation theory: OPE data}
To compare with the numerical bootstrap we will extract the OPE data in $\cP$ from the above correlators. The extraction of $\De_1$, $\De_2$ and $c_{112}$ is immediate and leads to the same answers given previously. 
We can then extract $c_{222}$ from either of the final two correlators in \eqref{multiplecorrs}, with the result 
\begin{equation}
  c_{222} = 2 \sqrt{2} -\frac{3 g_4 \lambda }{2 \sqrt{2} \pi }-\frac{3 g_6 \lambda }{16 \sqrt{2} \pi ^2}\,.
\end{equation}

We are left with the extraction of $\Delta_\text{gap}$. As explained above, at the massless free point the gap is set by two degenerate operators of dimension 4, namely $O_{4}= (\partial_\perp \phi)\Box(\partial_\perp \phi)$ and $O_{4'}=	(\partial_\perp \phi)^4$. At first order in $\lambda$ we need to resolve the mixing problem to derive the change in the gap. It is helpful to write $O_a$ and $O_b$ as the two orthonormal linear combinations of $O_4$ and $O_{4'}$. The variables to resolve are then the coefficients $p_{11a}^{(0)},p_{11b}^{(0)},p_{22a}^{(0)},p_{22b}^{(0)}$ of the conformal blocks corresponding to these operators as well as the two anomalous dimensions $\gamma_{a}^{(1)}, \gamma_{b}^{(1)}$. We can write
 \begin{align}
 	\langle \cO_1 \cO_1\cO_1\cO_1 \rangle^{(0)} &
		\sim  \left(p_{11a}^{(0)} +p_{11b}^{(0)}\right)G_4(z)+ \dots \,, 
		\nonumber\\
 		\left\langle\cO_2 \cO_2\cO_1\cO_1 \right\rangle^{(0)} &
		\sim \left(\sqrt{p_{22a}^{(0)}p_{11a}^{(0)}} +\sqrt{p_{22b}^{(0)}p_{11b}^{(0)}}\right)G_4(z)+ \dots\,,
		\\
 		\left\langle\cO_2 \cO_2\cO_2\cO_2 \right\rangle^{(0)} &
		\sim  \left(p_{22a}^{(0)} +p_{22b}^{(0)}\right)G_4(z)+ \dots 	\,,	
		\nonumber
 \end{align}
 and, similarly, at first order we should have:
  \begin{align}
  \langle\cO_1 \cO_1\cO_1\cO_1 \rangle^{(1)} &
  \sim  \left(p_{11a}^{(0)}\gamma_{a}^{(1)} +p_{11b}^{(0)}\gamma_{b}^{(1)}\right)G_4(z)\log(z)+ \dots\,,
  \nonumber\\
  \left\langle\cO_2 \cO_2\cO_1\cO_1 \right\rangle^{(1)} &
  \sim \left(\sqrt{p_{22a}^{(0)}p_{11a}^{(0)}}\gamma_{a}^{(1)} +\sqrt{p_{22b}^{(0)}p_{11b}^{(0)}}\gamma_{b}^{(1)}\right)G_4(z)\log(z)+ \dots\,,
  \nonumber\\
  \left\langle\cO_2 \cO_2\cO_2\cO_2 \right\rangle^{(1)} &
  \sim\left(p_{22a}^{(0)}\gamma_{a}^{(1)} +p_{22b}^{(0)}\gamma_{b}^{(1)}\right)G_4(z)\log(z)+ \dots 	\,.
  \end{align}
By matching these expressions to the conformal block decomposition of the correlators in the previous subsection we obtain
\begin{align}
	p_{11a}^{(0)} +p_{11b}^{(0)} = \frac{6}{5}\,,
	\qquad\qquad
	\sqrt{p_{22a}^{(0)}}\sqrt{p_{11a}^{(0)}} +\sqrt{p_{22b}^{(0)}}\sqrt{p_{11b}^{(0)}}= \frac{12}{5} \,,
	\nonumber
	\\
	p_{22a}^{(0)} +p_{22b}^{(0)} = \frac{54}{5}\,,
	\qquad\qquad
	p_{11a}^{(0)}\gamma_{a}^{(1)} +p_{11b}^{(0)}\gamma_{b}^{(1)} = \frac{g_4+48\pi g_2}{20 \pi}\,,
	\\
	\sqrt{p_{22a}^{(0)}}\sqrt{p_{11a}^{(0)}}\gamma_{a}^{(1)} +\sqrt{p_{22b}^{(0)}}\sqrt{p_{11b}^{(0)}}\gamma_{b}^{(1)}= \frac{-5g_6+8\pi g_4+384\pi^2g_2}{80 \pi^2}\,,
	\nonumber\\
	p_{22a}^{(0)}\gamma_{a}^{(1)} +p_{22b}^{(0)}\gamma_{b}^{(1)}=\frac{25 g_8+400\pi g_6+2944\pi^2 g_4+10752\pi^3 g_2}{320 \pi ^3}	\,.
	\nonumber
\end{align}
These equations admit the unique solution (up to permutation of $a$ and $b$)
\begin{align}
	p_{11a}^{(0)} &= \frac{3}{5}\left(1-\frac{u}{\sqrt{u^2+ 320\pi^2 g^2_6}} \right) ,
	\qquad\qquad
p_{11b}^{(0)}=  \frac{3}{5}\left(1+\frac{u}{\sqrt{u^2+ 320\pi^2 g^2_6}} \right) ,
	\nonumber\\
	p_{22a}^{(0)} &= \frac{27}{5}+ \frac{3(u-160\pi g_6)}{5\sqrt{u^2+ 320\pi^2 g^2_6}}\,,
	\qquad\qquad\qquad
	p_{22b}^{(0)} =\frac{27}{5}- \frac{3(u-160\pi g_6)}{5\sqrt{u^2+ 320\pi^2 g^2_6}}\,,
	\\
	\gamma_{a}^{(1)}&= 2g_2 + \frac{g_4}{24\pi} + \frac{u+\sqrt{u^2+ 320\pi^2 g^2_6}}{768\pi^3}\,,
	\qquad
	\gamma_{b}^{(1)}=  2g_2 + \frac{g_4}{24\pi} + \frac{u-\sqrt{u^2+ 320\pi^2 g^2_6}}{768\pi^3}\,,
	\nonumber
	\end{align}
where $u$ is the following linear combination of couplings
\begin{equation}
	u= 5g_8 +96\pi g_6 +560\pi^2g_4 +768\pi^3 g_2\,.
\end{equation}
Since the square root in the above expression is never negative, it follows that $\gamma_{b}^{(1)}$ is always the smallest of the two anomalous dimensions and therefore
\begin{equation}
\Delta_\text{gap} = 4 + \lambda \left( 2g_2 + \frac{g_4}{24\pi} + \frac{u-\sqrt{u^2+ 320\pi^2 g^2_6}}{768\pi^3}\right),
\end{equation}
where it is assumed that $\lambda > 0$ but the $g_{2k}$ couplings can have either sign.

\subsubsection{Numerical analysis}
The numerical analysis of the three correlators in  \eqref{multiplecorrs} proceeds exactly as in \cite{Homrich:2019cbt} and we refer to appendix K of that paper for the detailed conformal block decompositions and crossing symmetry equations. We recall in particular that the number of constraints is parametrized by an integer $\Lambda$; larger $\Lambda$ leads to better bounds but is computationally more demanding.

\begin{figure}[t]
  \centering
  \includegraphics[width=0.5\linewidth]{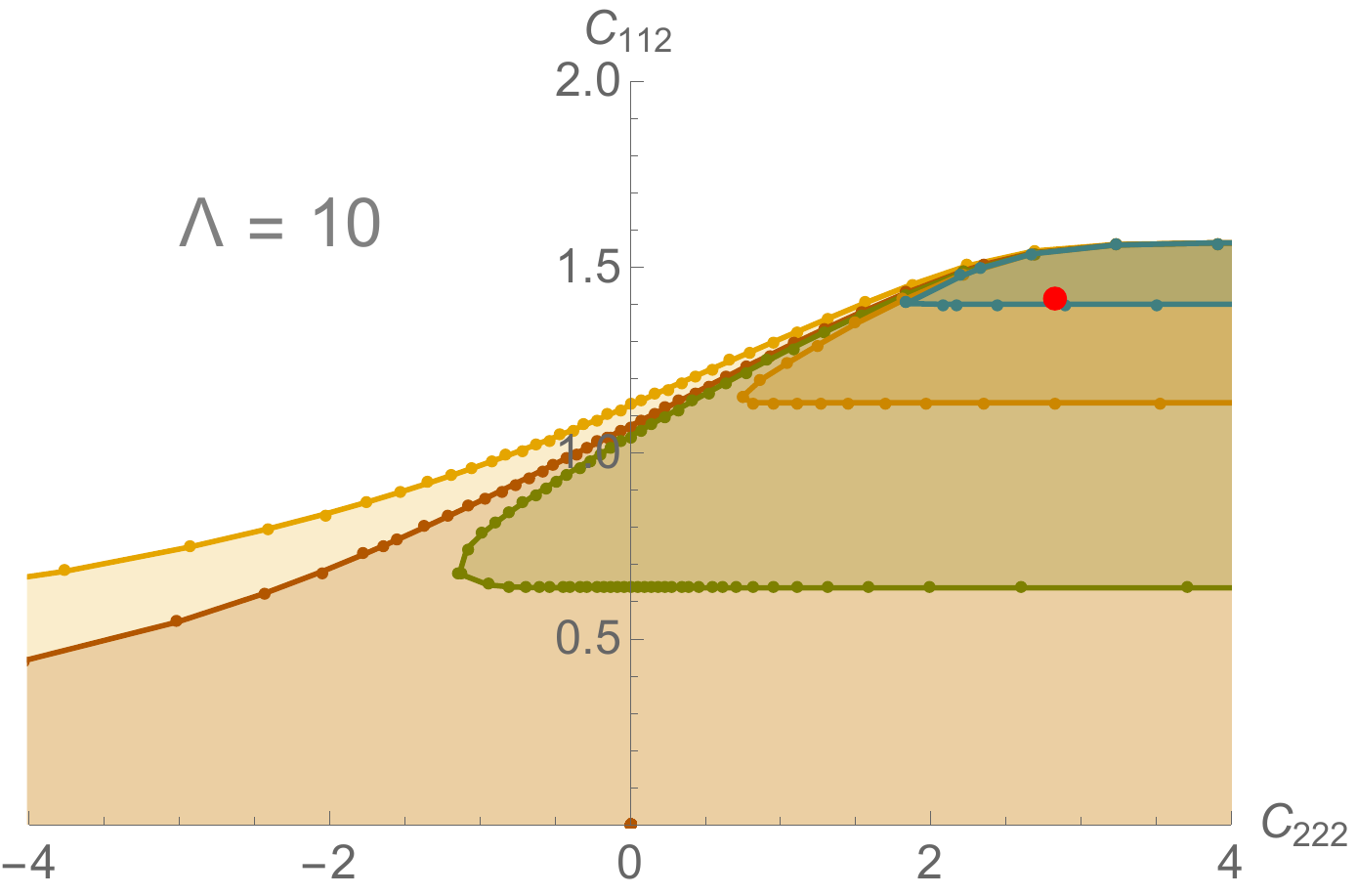}${}\qquad{}$\includegraphics[width=0.5\linewidth]{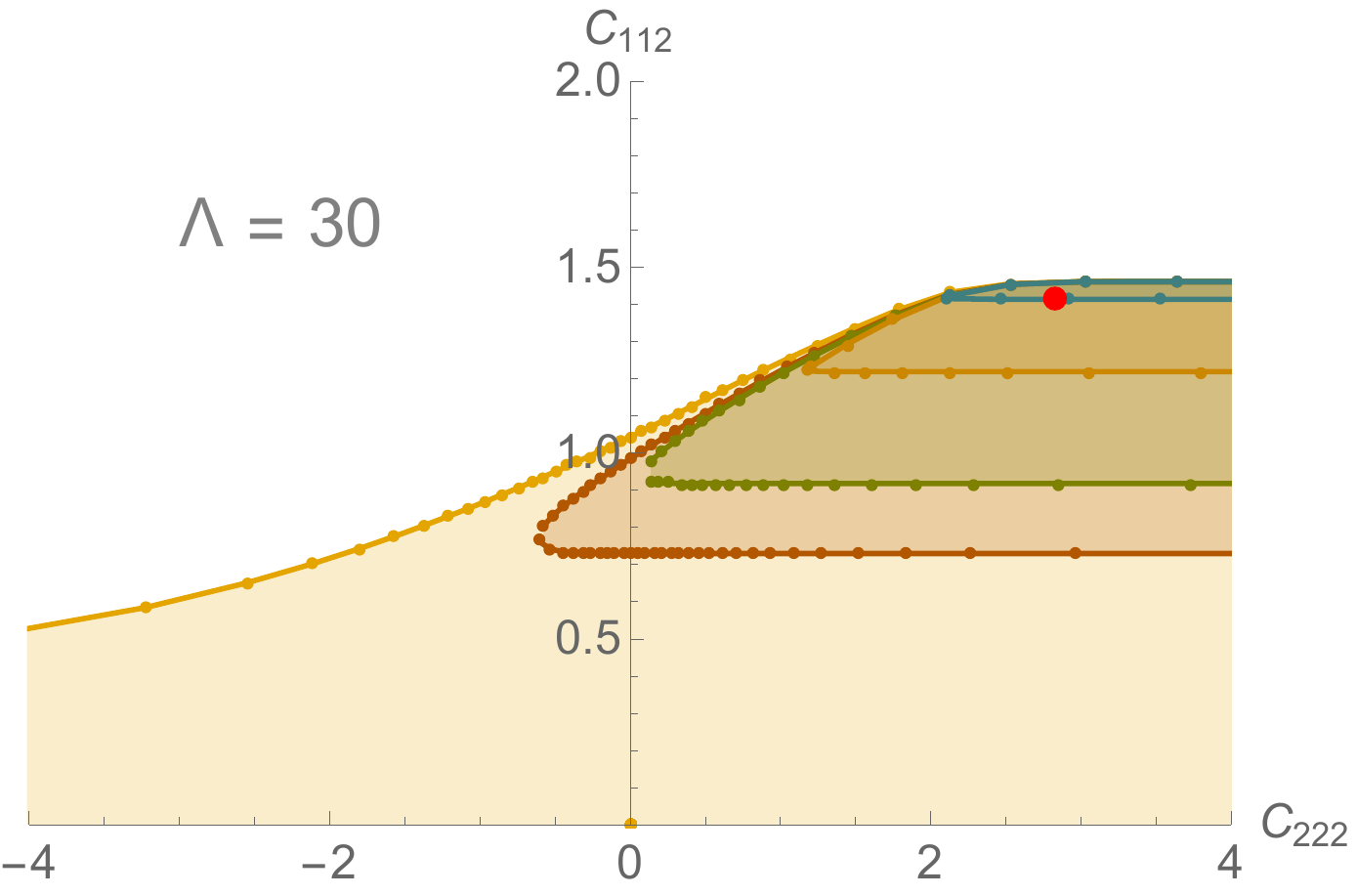}
  \caption{The space of allowed values for $(c_{112}, c_{222})$ for $\De_1=1$ and $\De_2=2$ and $\Delta_{\text{gap}}$ taking the values $3$ (outermost yellow), 3.2 (red), 3.3, 3.6 and ultimately 4 (innermost blue). Increasing the number of constraints from $\Lambda = 10$ to $\Lambda = 30$ shrinks all the regions. The red point corresponds to the generalized free field theory. }
  \label{fig:mixed1}
\end{figure}

Since our parameter space $\cP$ is five-dimensional we will have to restrict ourselves to various cross-sections around the massless free boson point. Our first attempt at visualizing the basic features of the allowed region inside $\cP$ is shown in figure \ref{fig:mixed1}. We fixed $\De_1 = 1$ and $\De_2 = 2$ and show an allowed region in the $(c_{112}, c_{222})$ space which clearly shrinks if we increase $\De_\text{gap}$ from $3$ to $4$.\footnote{When $\De_\text{gap} = 2$ the bound on $c_{222}$ disappears and the allowed region grows to a horizontal strip. Furthermore, the remaining bound on $c_{112}$ then equals the single-correlator bound. It is surprising that no extra information can be gleaned from a multi-correlator analysis in this case. In appendix \ref{app:multicorrbounds} we explain that this comes about because of a peculiar `identity-less' solution to the crossing equations.} We include plots for $\Lambda = 10$ and $\Lambda = 30$ to demonstrate that the numerical bounds have not quite converged yet, and especially for small $\De_\text{gap}$ further improvements can be expected by increasing $\Lambda$. We also assumed that $c_{112} \geq 0$; this can be done without loss of generality because CFT correlators are invariant under a simultaneous reflection of all operators $\cO_i \to -\cO_i$.

The red point in each panel of figure \ref{fig:mixed1} corresponds to the massless free boson. Interestingly, for $\De_\text{gap}$ very close to 4 the bounds appear to converge to a small sliver around this point. This would imply that it is impossible to change $c_{112}$ without lowering $\De_\text{gap}$ at the same time, but it does appear possible to change $c_{222}$ in both directions. We will explain this from the viewpoint of perturbation theory below.

To get an idea of the allowed region in the whole of $\cP$ we add that these plots do not qualitatively change if we vary $\De_1$ and $\De_2$ a little bit around the generalized free boson values.

%%%%%%%%%%%%%%%%%%%%%%%%%%%%%%%%%%
\subsubsection*{Comparison with first-order perturbation theory}
%%%%%%%%%%%%%%%%%%%%%%%%%%%%%%%%%%
Recall the first-order perturbative result of the previous subsection:
\begin{equation}\label{firstordermulticorrpert}
\begin{split}
  \De_1 &= 1 + \lambda g_2 + O(\lambda^2) \,,\\
  \De_2 &= 2 + 2 \lambda g_2 + \lambda \frac{g_4}{4\pi} + O(\lambda^2)\,,\\
  c_{112} &= \sqrt{2}- \lambda  \frac{g_4}{4 \sqrt{2} \pi} + O(\lambda^2)\,,\\
  c_{222} &= 2 \sqrt{2} -\lambda  \left( \frac{3 g_4 }{2 \sqrt{2} \pi }+\frac{3 g_6}{16 \sqrt{2} \pi ^2}\right) + O(\lambda^2)\,,\\
\Delta_\text{gap} &= 4 + \lambda \left( 2g_2 + \frac{g_4}{24\pi} + \frac{u-\sqrt{u^2+ 320\pi^2 g^2_6}}{768\pi^3}\right)+ O(\lambda^2)\,,
\end{split}
\end{equation}
with
\begin{equation}
  u= 5g_8 +96\pi g_6 +560\pi^2g_4 +768\pi^3 g_2\,,
\end{equation}
and where the four possible couplings $g_2$, $g_4$, $g_6$, $g_8$ can in principle take arbitrary real values.

We will now compare these results to the numerical bootstrap bounds along several different lines. For the `$g_2$ line' we set $g_4 = g_6 = 0$, for the `$g_4$ line' we set $g_2 = g_6 = 0$ and for the `$g_6$ line' we set $g_2 = g_4 = 0$. For each line we let $(\De_1, \De_2, c_{112}, c_{222})$ be parametrized as in \eqref{firstordermulticorrpert} and measure the tangent line at the free boson point for the bound on $\De_\text{gap}$. Notice that the $g_8$ dependence only enters in $\De_\text{gap}$ so we will not meaningfully be able to compare the numerical bootstrap bound to a `$g_8$ line' within $\cP$. Finally we will consider several `sine-Gordon' lines where the couplings are taken to be varied as dictated by the expansion of $\cos(\beta \phi)$.

\subsubsection*{The $g_2$ line}
If we set $g_4 = g_6 = 0$ then
\begin{equation}
  \Delta_\text{gap} = 4 + 2 (\De_1 - 1) + \lambda \,\frac{1}{768 \pi^3} 2 u \, \theta(- u) + O(\lambda^2)\,,
\end{equation}
with $\theta$ the Heaviside theta function and $u$ arbitrary since $g_8$ is arbitrary. The largest gap is therefore found by setting $u$ to any non-negative value. But since $u = 5 g_8 + 768 \pi^3 g_2$, this means we should take 
\begin{equation}
 g_8 \geq \max\left(- \frac{768 \pi^3}{5} g_2,0\right)
\end{equation}
to maximize the gap. Thus, for $g_2 > 0$, which means $\De_1 > 1$, the maximal gap is obtained by the non-interacting theory with $\phi^2$ deformation. On the other hand, for $g_2 < 0$, so for $\De_1 < 1$, we actually find that an \emph{interacting} theory is the one that maximizes the gap within our parameter space.

\begin{figure}
\begin{center}
  \includegraphics[width=10cm]{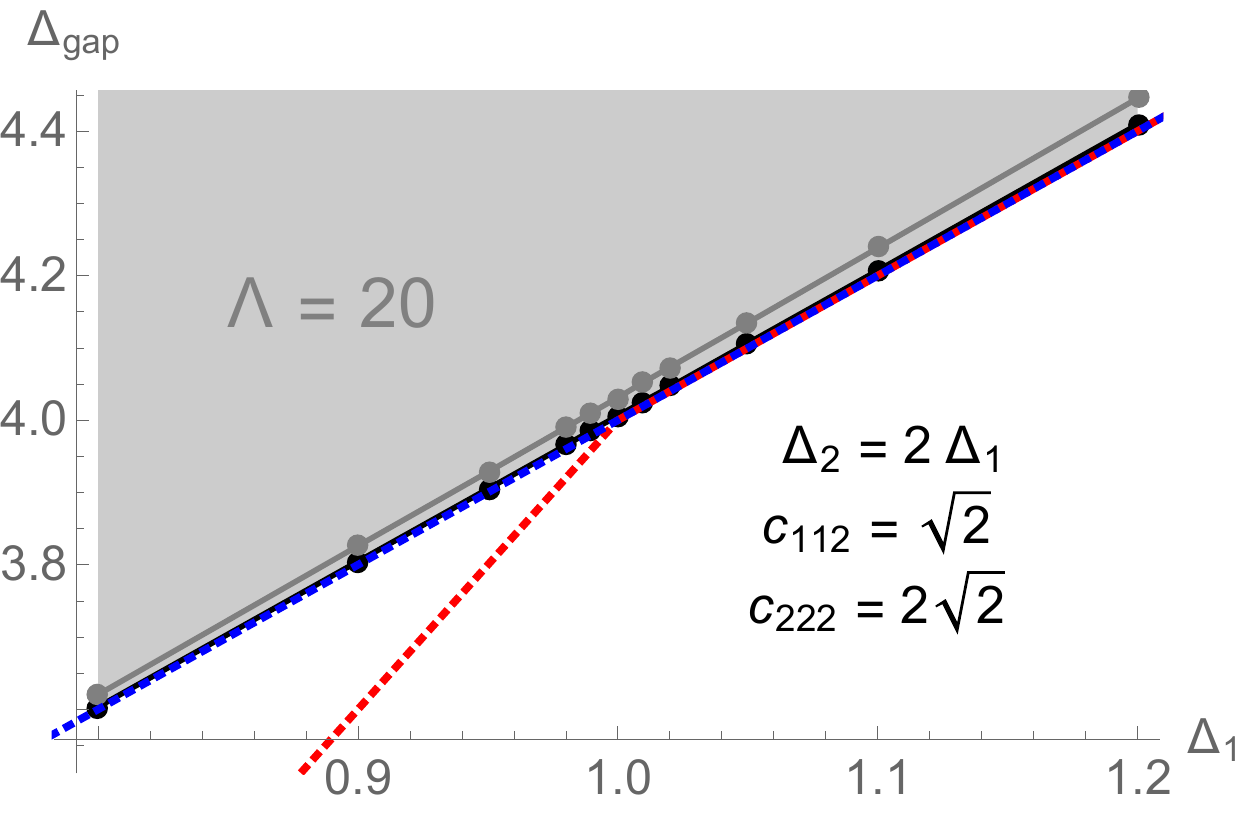}
  \caption{\label{fig:g2line}The maximal value of the gap as a function of $\Delta_1$ with $(\De_2, c_{112}, c_{222})$ as given, which to first order corresponds to switching on only the $g_2$ deformation. The numerical bounds, obtained with $\Lambda = 10$ in gray and $\Lambda = 20$ in black, appear to converge to the line $2 \Delta_1 + 2$. The free theory (red dashed line) can only explain this bound for $\Delta_1 > 1$. By selectively switching on a $\phi^8$ interaction (blue line) we can also saturate the bound to first order in perturbation theory for $\Delta_1 < 1$.}
\end{center}
\end{figure}

As we show in figure \ref{fig:g2line}, this observation is sufficient to explain the behavior of the numerical bootstrap bound near the generalized free point. Physically we observe that the gap at the free point is saturated by two operators $\cO_4 \sim (\partial_\perp \phi)\square (\partial_\perp \phi)$ and $\cO_{4'} \sim (\partial_\perp \phi)^4$, whose dimensions under the $g_2$ deformation change as
\begin{equation}
   \De_4 = 2 \De_1 + 2\,,\qquad\De_{4'} = 4 \De_1\,.
\end{equation}
Taking the minimum of these two values we obtain the red line in the figure, which only saturates the bound for $\De_1 > 1$. On the other hand, if we selectively switch on a $g_8$, so as to make $u \geq 0$ then we obtain the blue line which is nicely tangential to the bound on both sides of $\De_1 = 1$.

Notice that the multi-correlator bound appears to coincide with the single-correlator bound $2 \Delta_1 + 2$ for a large range of $\Delta_1$, and not just in a small neighbourhood of the free point. This indicates that there might be a not necessarily physical solution of the multi-correlator crossing equations whose gap equals the single-correlator bound, perhaps in the same style as the identity-less solution discussed in appendix \ref{app:multicorrbounds} for $\Delta_\text{gap} \leq 8 \De_1/3$. However we have shown that there also exists a \emph{physical} setup that saturates the bound in the vicinity of $\De_1 = 1$.

\subsubsection*{The $g_4$ line}
Along the $g_4$ line we set $g_2 = g_6 = 0$ and find that
\begin{equation}
\begin{split}
  \De_1 &= 1 + O(\lambda^2)\, ,\\
  \De_2 &= 2 + \lambda \frac{g_4}{4\pi} + O(\lambda^2)\,,\\
\Delta_\text{gap} &= 4 +  \frac{1}{6}(\De_2 - 2) + \frac{\lambda}{768\pi^3} 2 u \theta(-u) + O(\lambda^2)\,,
\end{split}
\end{equation}
and the smallest gap is obtained by setting
\begin{equation}
  g_8 \geq \max \left( - \frac{560 \pi^2}{5} g_4, 0\right),
\end{equation}
such that $u \geq 0$ always. This once more means that the gap along the $g_4 \phi^4$ deformation line has a kink at the free point, but by switching on $g_8$ for $\De_2 < 2$ so as to retain $u \geq 0$ we can avoid the kink and obtain a smooth tangent line in perturbation theory.

\begin{figure}
\begin{center}
  \includegraphics[width=10cm]{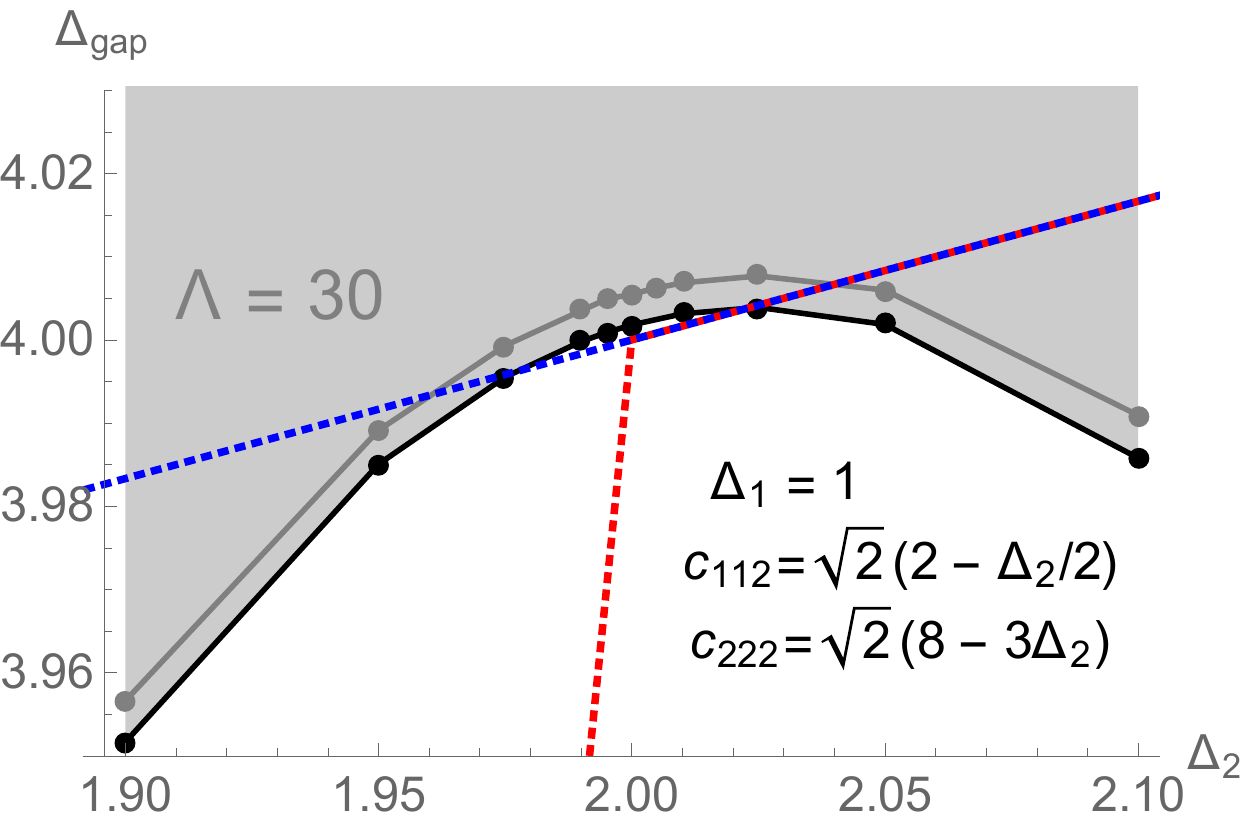}
  \caption{\label{fig:g4line}The maximal value of the gap as a function of $\Delta_2$ with $(\De_1, c_{112}, c_{222})$ as given, which to first order corresponds to switching on only the $g_4$ deformation. The best bound was obtained with $\Lambda = 30$; the slightly weaker bound with $\Lambda = 20$. At the free point the numerical bound appears to become tangent to the line $4 + \frac{1}{6}(\De_2 - 2)$. The free theory (red dashed line) can only explain this bound for $\Delta_2 > 2$. By selectively switching on a $\phi^8$ interaction (blue line) we can also saturate the bound to first order in perturbation theory for $\Delta_2 < 2$.}
\end{center}
\end{figure}

Upon comparison with the numerical results shown in figure \ref{fig:g4line} we once more see that the perturbative tangent line lies parallel to the numerical bootstrap curve around the free point, provided we switch on the $g_8$ interactions for $\De_2 < 2$. The full numerical result however deviates rather quickly from the straight line. It would be interesting to match this to second-order perturbation theory \cite{Paulos:2019fkw} for the multi-correlator system in the future.

Notice that both for the $g_2$ line and for the $g_4$ line there is always an extremal tangent direction with $u = 0$, implying that $\gamma_{a}^{(1)}$ and $\gamma_{b}^{(1)}$ actually become \emph{equal} to each other at first order. The extremal theory therefore maintains the degeneracy of the two operators, which is consistent with the `single operator per bin' observation for the extremal spectrum that we discussed above in the context of the single correlator analysis. We would like to stress again that it would be worth investigating the existence of any `single operator per bin' extremal theory beyond first-order perturbation theory.

%%%%%%%%%%%%%%%%%%%%%%%%%%%%
\subsubsection*{The $g_6$ line}
\begin{figure}
\begin{center}
  \includegraphics[width=10cm]{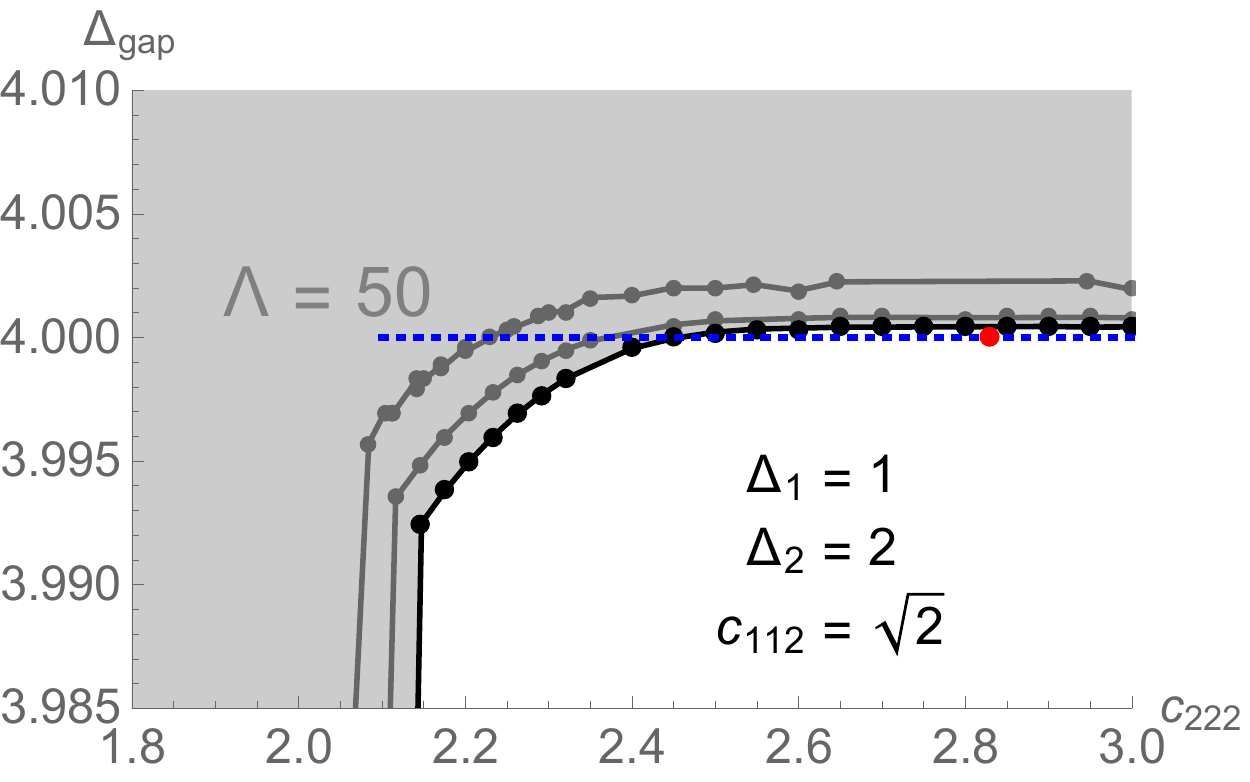}
  \caption{\label{fig:g6line}The maximal value of the gap as a function of $c_{222}$ with $(\De_1, \De_2, c_{112})$ as given, which to first order corresponds to switching on only the $g_6$ deformation. The numerical bound appears to converge to the horizontal line $ \Delta_{\rm gap}=4$. The free theory can only explain this bound at the single red point. We can venture away from this point by switching on $g_6$, but to obtain the blue line we need to simultaneously turn on a much larger $\phi^8$ interaction. Notice that the best bound (in black) corresponds to $\Lambda = 50$, whereas the gray bounds correspond to $\Lambda = 30$ and $\Lambda = 40$.}
\end{center}
\end{figure}

%%%%%%%%%%%%%%%%%%%%%%%%%%%%
Along the $g_6$ line we have
\begin{equation}
  \De_1 = 1, \qquad \De_2 = 2, \qquad c_{112} = \sqrt{2}\,,
\end{equation}
and only $c_{222}$ and $\Delta_\text{gap}$ can change, with a relation that we can write as:
\begin{equation}
 \Delta_\text{gap} = 4 + \frac{u -\sqrt{\frac{163840}{9} \pi ^6 \left(c_{222} - 2 \sqrt{2}\right)^2+u^2}}{768 \pi ^3} + O(\lambda^2)\,.
\end{equation}
Interestingly, to maximize the gap away from the free point we need to take $u \to \infty$. In other words, we can take $g_8 / g_6 \to \infty$ and then we would expect $\Delta_\text{gap}$ to remain approximately flat around the free point. Of course this limit is a bit singular but, as we show in figure \ref{fig:g6line}, it appears to accurately saturate the bound to the first order in perturbation theory.

The plot in figure \ref{fig:g6line} is more zoomed in than the previous plots and also evaluated at significantly higher $\Lambda$. This allowed us to clearly exhibit the sharp and somewhat intriguing kink in the maximal gap when we decrease $c_{222}$ below the free value. Since $\Delta_\text{gap}$ is below $4$ already at the shown value $\Lambda = 50$, it is unlikely that this kink merges with the free point as $\Lambda \to \infty$. (Notice that this means that the leftmost point of the blue `sliver' in figure \ref{fig:mixed1} will not merge with the free point as $\Lambda \to \infty$.) We do not have a good candidate theory that can explain this kink, but we may speculate that it corresponds to an extremal point in the space of all RG flows starting from the free massless boson. In more detail, we envisage that the (infinite-dimensional) space of all possible relevant deformations as in \eqref{generalZ2deformation} (which in turn is foliated by RG flows) must somehow map into the (infinite-dimensional) space of OPE data. It is natural to expect that extremal points in the image of this map are also physically interesting. For example, they may be points where the potential becomes unstable or a phase transition takes place. It would be very interesting to see if the image of such points in the space of OPE data can be reliably identified.

\subsubsection*{The Sine-Gordon lines}
\begin{figure}
\begin{center}
  \includegraphics[width=7.5cm]{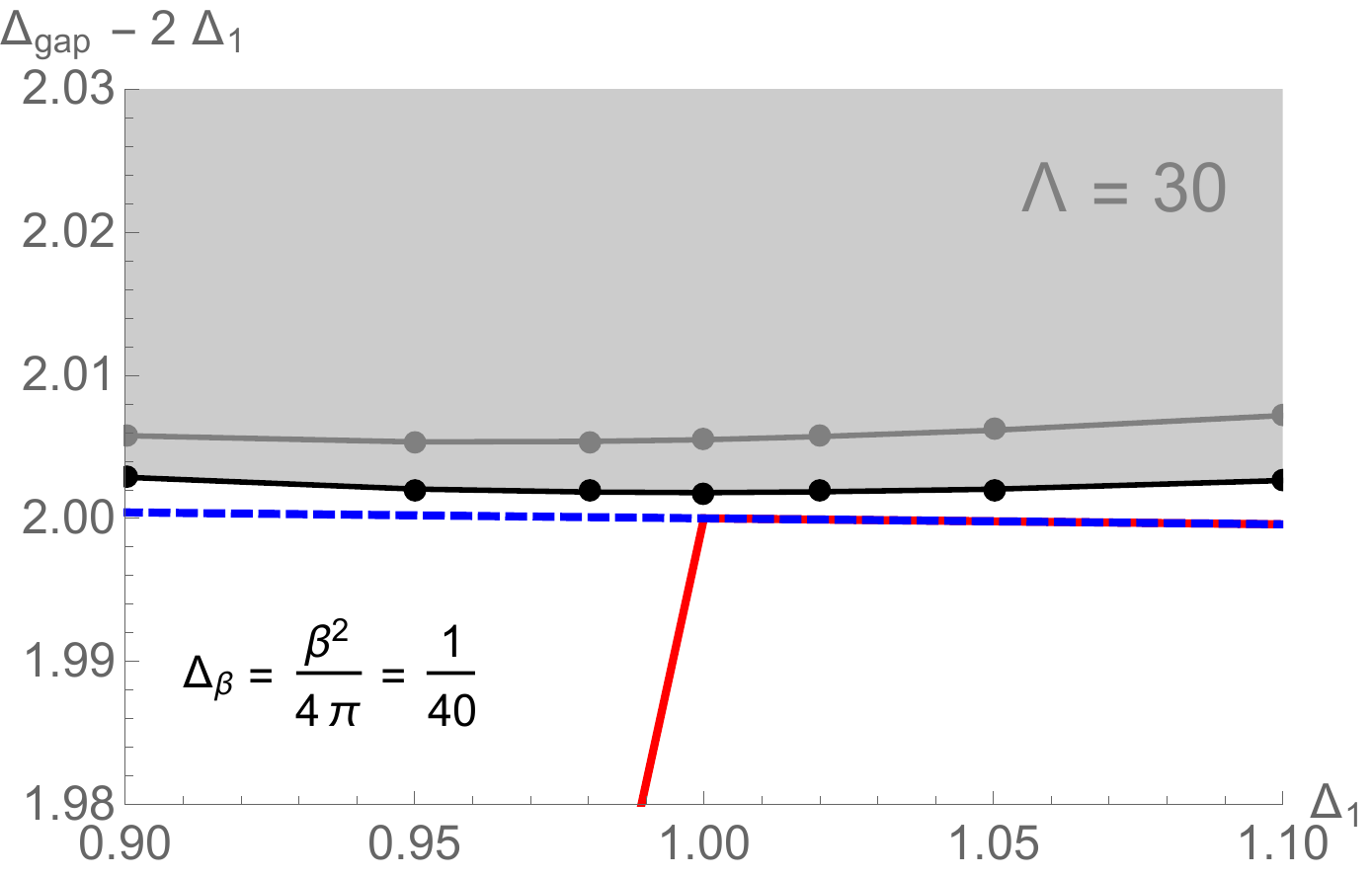}
  \includegraphics[width=7.5cm]{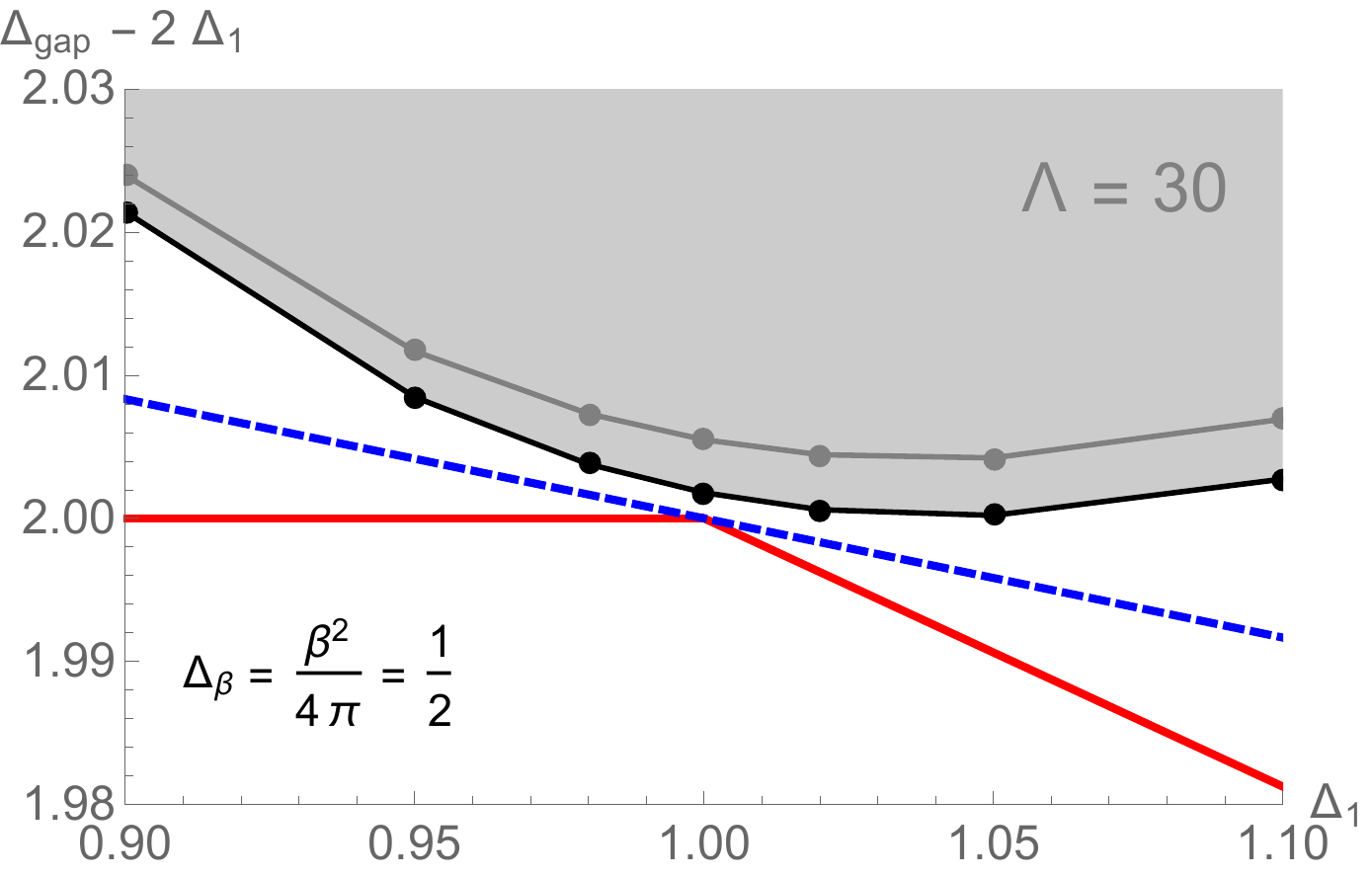}\\
  \includegraphics[width=7.5cm]{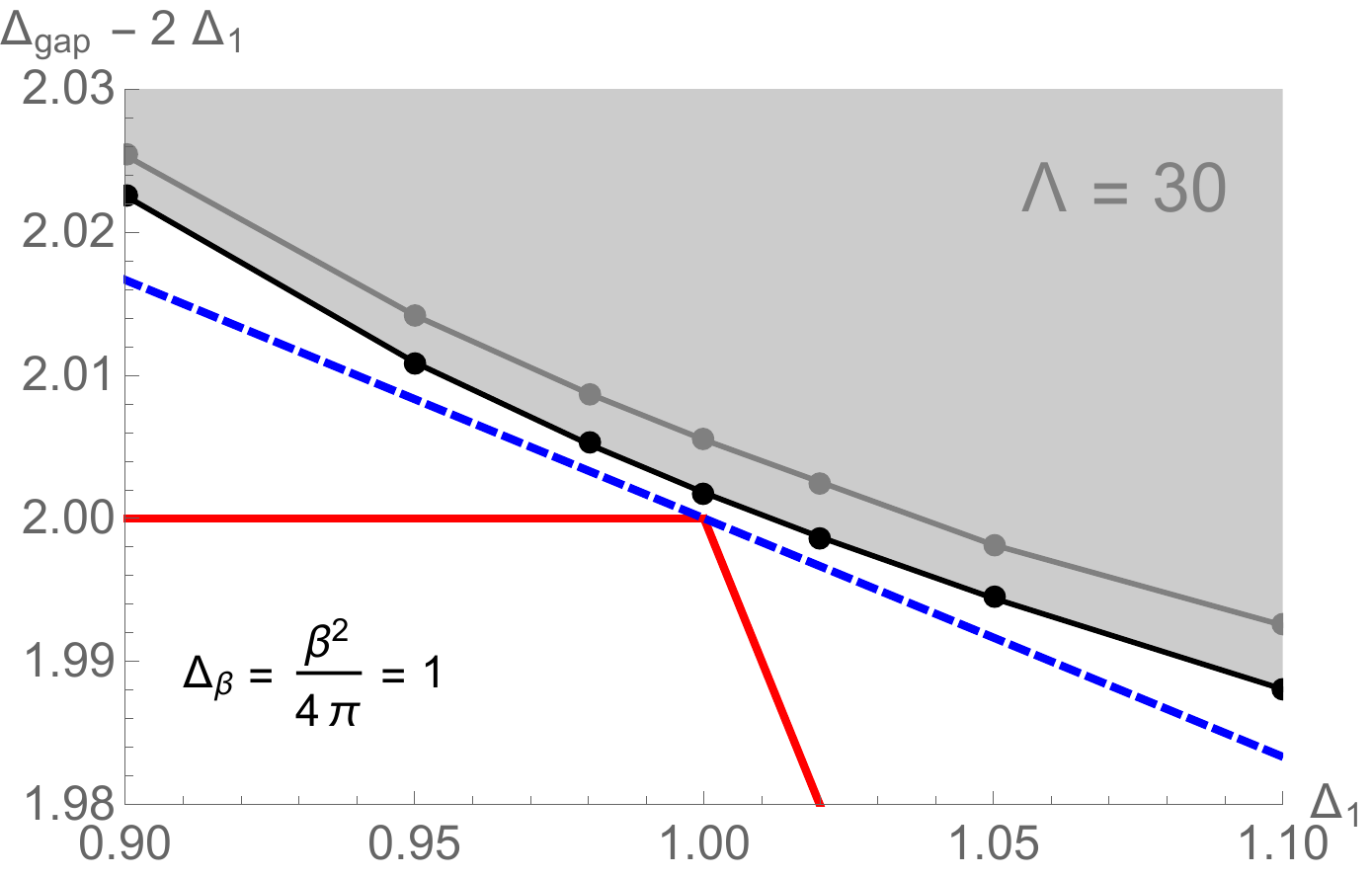}
  \includegraphics[width=7.5cm]{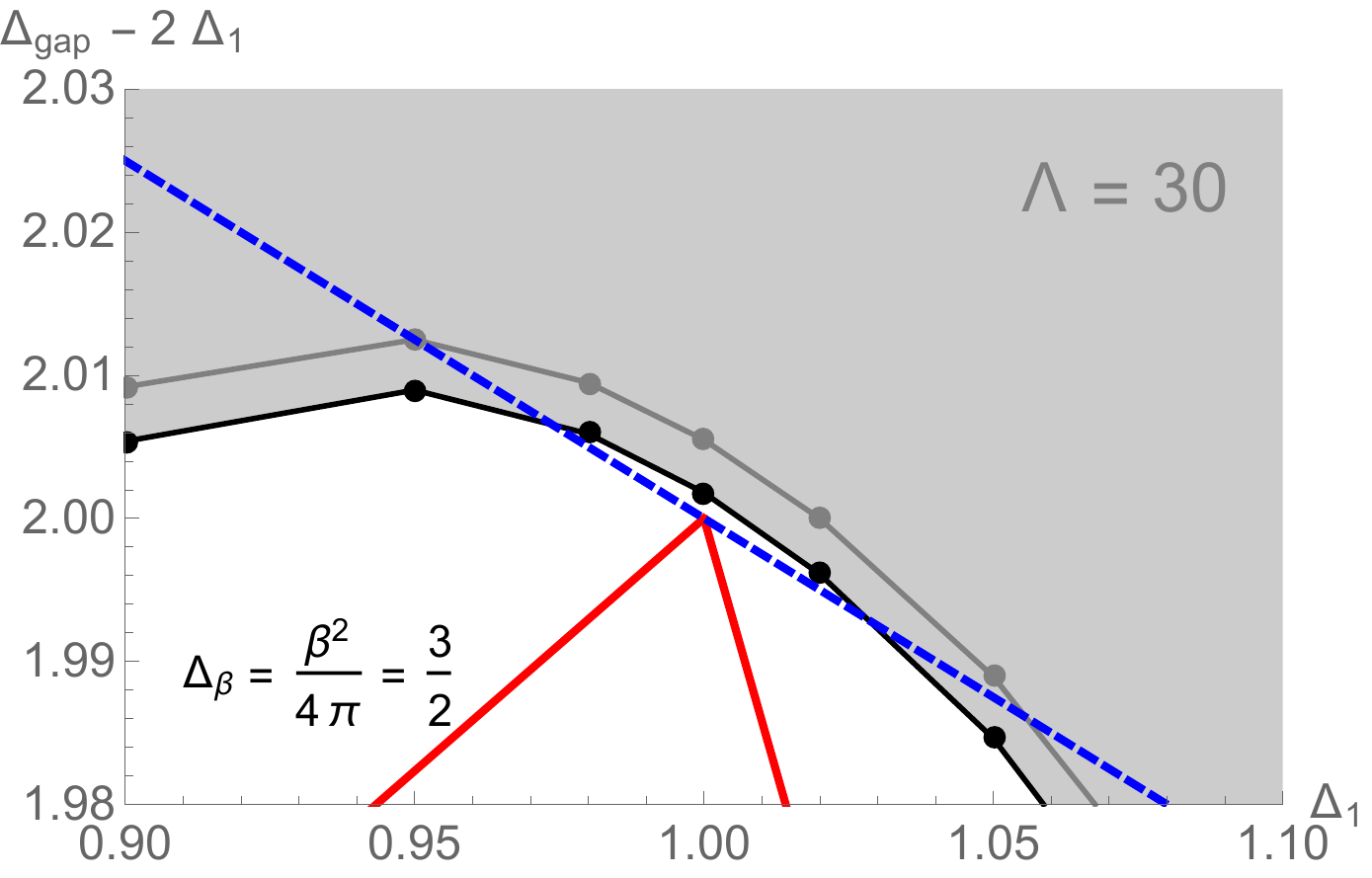}
\end{center}
\caption{\label{fig:sglines}Tracing the maximal gap along the lines given by the sine-Gordon theories with the given values of $\Delta_\beta$. The gray bounds correspond to $\Lambda = 20$ and the black ones to $\Lambda = 30$. The bound is always tangent to the blue lines corresponding to the deformed theory that is obtained by switching on an independent $g_8$. The original sine-Gordon theories, in red, only saturate the numerical bound at the free point. Notice that the vertical axis shows $\Delta_\text{gap} - 2 \Delta_1$ rather than just $\Delta_\text{gap}$ to more clearly show the small deviations from a straight line in the numerical data.}
\end{figure}

Our perturbative analyses can also capture the sine-Gordon theory. We expand
\begin{equation}
  1 - \cos(\beta \phi) = \frac{\beta ^2}{2}\phi ^2  -\frac{\beta ^4}{24}\phi ^4+\frac{\beta ^6}{720}\phi ^6 -\frac{\beta ^8}{40320} \phi ^8 + \ldots\,,
\end{equation}
and then use the fact that, to the first order, the higher-point $\phi^{2n}$ couplings do not contribute to the correlators we are analyzing. Therefore the sine-Gordon lines correspond to
\begin{equation}
  g_2 = -\beta^2, \qquad g_4 = \beta^4, \qquad g_6 = -\beta^6, \qquad g_8 = \beta^8\,.
\end{equation}
For every value of $\beta^2$ this once again traces out a curve in $\cP$. If we trade $\lambda$ for $\Delta_1$ and let $(\Delta_2, c_{112}, c_{222})$ be given by the first-order perturbative result as above, then the gap in the sine-Gordon theories is given by the red lines in figure \ref{fig:sglines}.\footnote{In the physical sine-Gordon theories we should perturb around a minimum of the potential to smoothly connect to the flat-space theory. This means that $\lambda g_2 > 0$, so $\Delta_1 > 1$. Although the part of the red lines for $\Delta_1 < 1$ might not be a sine-Gordon theory, it can still be understood as corresponding to the first-order deformation along the given line in the parameter space.} We see that sine-Gordon does not saturate the multi-correlator bound even to first order, for any of the values of $\beta$ we tested. The tangent lines to the numerical bound instead appear to correspond to the blue dashed lines, which as before correspond to dialing $g_8$ independently to the value that maximizes the gap.\footnote{The blue lines also correspond to the single-correlator perturbative $\phi^4$ result for the maximal gap. It might surprise the reader that the red lines do not automatically saturate this bound even on one side. After all, is one of the two operators $\cO_4$ and $\cO_{4'}$ not the one that appears in the single correlator as well? The resolution to this question is that, with non-zero $g_6$, the operator in the single-correlator bound is actually a linear combination of $\cO_4$ and $\cO_{4'}$. Doing just the single-correlator analysis, one mis-identifies the corresponding block as originating from a single operator with a larger anomalous dimension.}

%!TEX root = draft.tex

\section{Kink scattering}
\label{sec:chargedcorr}
%%%%%%%%%%%%%%%%%%%%%%%%%%%%%%%%%%%%%%%%%%%%

The most elementary excitations of the sine-Gordon model are solitons or kinks that wind once around the compact field space $\phi \sim \phi + 2\pi /\beta$. These transform as vectors under the $O(2)$ (topological) global symmetry of the sine-Gordon theory. In the OPE of a kink and an anti-kink one recovers the breathers of the previous section. These are necessarily $SO(2)$-neutral but can have either sign for the $\mathbb{Z}_2$ center symmetry.

In this section we will look at the numerical bootstrap for $O(2)$ vector operators in one-dimensional CFTs. Our goal is to formulate the analogous problem to the kink anti-kink S-matrix bootstrap of \cite{Cordova:2018uop,Paulos:2018fym}, but for the sine-Gordon theory in AdS. We will again compare the numerical data with the results of a perturbative study around UV theory, which is the compact boson with the relevant sine-Gordon deformation \eqref{SGdef}, but also connect with the flat-space results at very large $\Delta$. 

%%%%%%%%%%%%%%%%%%%%%%%%%%%%%%
\subsection{\texorpdfstring{$O(2)$ covariant correlators in CFT$_1$}{O(2) covariant correlators in CFT1}}
\label{sec:O2CFT1}
%%%%%%%%%%%%%%%%%%%%%%%%%%%%%%
We will consider the crossing equations for the four-point function of $O(2)$ vectors. This has been studied extensively in the literature, specially in the 3d case due to its important applications to condensed matter and statistical physics \cite{Kos:2013tga,Kos:2015mba,Chester:2019ifh,Chester:2020iyt}. We consider external operators $K_i$ of equal dimension $\Delta_v$\footnote{We reserve the symbol $\Delta_K$ for the dimension of the boundary operator in the free compact boson theory.} and write the correlator as
\begin{align}
	x_{12}^{2\Delta_v} x_{34}^{2\Delta_v} \langle K_i(x_1) K_j(x_2) K_k(x_3) K_l(x_4)\rangle %g_{i j}^{k l}(x_i)
	 & =g_{i j k l}(z) \label{4ptKinks}
	\\&=\delta_{i j} \delta_{k l} \, g_{1}(z)+\delta_{i l} \delta_{j k} \, g_{2}(z)+\delta_{i k} \delta_{j l} \, g_{3}(z)\,,
	\nonumber
\end{align}
where  $i,j,k,l\in \{1,2\}$ are $O(2)$ fundamental indices. The crossing equation then becomes
\begin{equation}
	g_{i j k l}(z)=\left(\frac{z}{1-z}\right)^{2 \Delta_v} g_{k j i l}(1-z)\,.
	\label{crossingKinks}
\end{equation}
There are three independent components to this equation, which can be written as
\begin{align}
	\label{eq:O2crossing}
	(1-z)^{2 \Delta_v} g_{2}(z)                        & =z^{2 \Delta_v}g_{2}(1-z)\,,
	\nonumber                           \\
	(1-z)^{2 \Delta_v} \left( g_{1}(z)+g_{3}(z)\right) & =z^{2 \Delta_v}\left(g_{1}(1-z)+g_{3}(1-z) \right) \,,
	\nonumber \\
	(1-z)^{2 \Delta_v} \left( g_{1}(z)-g_{3}(z)\right) & =-z^{2 \Delta_v}\left(g_{1}(1-z)-g_{3}(1-z) \right) .
\end{align}
The correlator \eqref{4ptKinks}  can  be decomposed  into the 3 irreducible representations in the tensor product of $O(2)$ vectors: the symmetric-traceless charge $\mathbf{2}$ representation, the scalar $\mathbf{0}^+$ and the pseudo-scalar/anti-symmetric $\mathbf{0}^-$, where the $\pm$ denotes the transformation properties under $\mathbb{Z}_2\subset O(2)$.
The components of the correlator can be written as
\begin{align}
	g_1(z) & =  \sum_{\mathbf{0}^+} \lambda_{\mathcal{O}}^{2} G_{ \Delta}(z)-\sum_{\mathbf{2}} \lambda_{\mathcal{O}}^{2} G_{ \Delta}(z)\equiv g_{\mathbf{0}^+}(z)-g_{\mathbf{2}}(z) \,,
	\nonumber     \\
	g_2(z) & =\sum_{\mathbf{2}} \lambda_{\mathcal{O}}^{2} G_{ \Delta}(z)-\sum_{\mathbf{0}^-} \lambda_{\mathcal{O}}^{2} G_{ \Delta}(z)\equiv g_{\mathbf{2}}(z)-g_{\mathbf{0}^-}(z)\label{CBdecomp} \,,
	\\
	g_3(z) & =\sum_{\mathbf{2}} \lambda_{\mathcal{O}}^{2} G_{ \Delta}(z)+\sum_{\mathbf{0}^-} \lambda_{\mathcal{O}}^{2} G_{ \Delta}(z)\equiv g_{\mathbf{2}}(z)+g_{\mathbf{0}^-}(z)\,,
	\nonumber
\end{align}
with $G_\Delta(z)$ the 1d conformal block:
\begin{equation}
	G_\Delta(z)= z^\Delta \,_2F_1(\Delta,\Delta;2\Delta,z)\,.
\end{equation}
We will apply  numerical conformal bootstrap methods to this system in section \ref{ssec:numboot} but first let us discuss the perturbative analysis.

%%%%%%%%%%%%%%%%%%%%%%%%%%%%%%
\subsection{Sine-Gordon charged correlators in conformal perturbation theory}
\label{subsec:cptkinks}
%%%%%%%%%%%%%%%%%%%%%%%%%%%%%%
As is customary, we decompose the free boson into its left and right moving components
\begin{equation}
	\phi= \phi_L + \phi_R\,,
\end{equation}
and also define
\begin{equation}
	\tilde{\phi} = \phi_L - \phi_R\,.
\end{equation}
This decomposition makes manifest the two $U(1)$ symmetries: the first is associated to the shift $\phi \to \phi + c$, generated by the Noether current $j_s^{\mu}=\partial^\mu \phi$ whose charge we label by the integer $n$; the second is associated to  the shift $\tilde{\phi}\to \tilde{\phi} +c$ with the current $j_t^\mu= \epsilon^{\mu\nu}\partial_\nu \phi$ whose charge we label by the integer $m$.

With the above decomposition we can write the most general vertex operator as
\begin{equation}
	V_{n,m}=\, :e^{i p_L \phi_L + i p_R \phi_R}:\,,
\end{equation}
with the field space momenta $p_{L,R}$ related to the two $U(1)$ charges through
\begin{equation}
	p_L= \frac{n}{r} + 2 \pi m r\,,\quad p_R= \frac{n}{r} - 2 \pi m r\,.
\end{equation}
The scaling dimension and spin of these operators are given by
\begin{align}
	\Delta_{n,m} &                                                                  
	=\frac{1}{8 \pi}\left(p_L^2+p_R^2\right)= \frac{1}{4 \pi}\left(\frac{n^2}{r^2}+4 \pi^2 m^2 r^2\right) , \nonumber\\
	J_{n,m}      &                                                                  
	=\frac{1}{8 \pi}\left(p_L^2-p_R^2\right)= nm\,.
\end{align}
As an example, in terms of the vertex operators the sine-Gordon potential \eqref{SGdef} $2\cos(\beta \phi) = V_{1,0} + V_{-1, 0}$. Since these are charged under $j_s^\mu$ but not under $j_t^\mu$ we conclude that the sine-Gordon interaction term breaks only the former of the two $U(1)$ symmetries.

In the remainder of this section we will be interested in the correlation functions of the operators
\begin{equation}
	V_{0,\pm1}=\, :e^{\pm\frac{2\pi i}{\beta}\tilde{\phi}}:\,.
\end{equation}
These have the same quantum numbers as the flat space kink and anti-kink and have scaling dimension $\pi/\beta^2$ in the UV.

A major simplification for perturbation theory in AdS$_2$ is that the free boson correlation functions are essentially equivalent to those on the upper half plane $\mathbb{H}$, since the two backgrounds are related by multiplication by a Weyl factor.\footnote{This is obvious in Poincar\'e coordinates:
$ds^2_{AdS_2} = \frac{\LAdS^2}{y^2}(dy^2+dx^2)= \frac{\LAdS^2}{y^2}ds^2_{\mathbb{H}}\,.$}

As before, we will exclusively consider the Dirichlet boundary condition $\phi=0$. This choice also allows us to compute upper half-plane correlators in terms of the full plane correlators, by replacing the right moving modes with left moving modes inserted at the mirror image of the insertion point with respect to the boundary. In particular, for Dirichlet boundary conditions we have
\begin{equation}
	\phi_L(w) \to \phi(x,y)\,, \qquad \phi_R(\bar{w}) \to -\phi(x,-y)\,,
\end{equation}
where $w=x+iy$ is a holomorphic coordinate on the complex plane. We can then treat $\phi$ as a holomorphic field, and compute correlation functions on the plane using standard methods. The boundary correlation functions are then easily obtained as limit of the bulk ones.

%%%%%%%%%%%%%%%%%%%%%%%%%%%%%%%
\subsubsection{Four-point function in  free theory}
%%%%%%%%%%%%%%%%%%%%%%%%%%%%%%%
We start from a four-point function $G(w_i,\bar{w}_i)$ on the upper half plane $\mathbb{H}$, with a particular choice of charges
\begin{equation}
	G_{\mathbb{H}}(w_i,\bar{w}_i)=\left\langle V_{0,+1}(w_1,\bar{w}_1)V_{0,-1}(w_2,\bar{w}_2)V_{0,+1}(w_3,\bar{w}_3)V_{0,-1}(w_4,\bar{w}_4) \right\rangle_{\mathbb{H}}\,.
\end{equation}
By the doubling trick this becomes a holomorphic eight-point function on the plane
\begin{equation}
	G_{\mathbb{H}}=\left\langle e^{i \alpha \phi(w_1)} e^{i \alpha \phi(w_1^*)}e^{-i \alpha \phi(w_2)} e^{-i \alpha \phi(w_2^*)}e^{i \alpha \phi(w_3)} e^{i \alpha \phi(w_3^*)}e^{-i \alpha \phi(w_4)} e^{-i \alpha \phi(w_4^*)}\right\rangle_{\mathbb{R}^2}\,,
\end{equation}
with $\alpha=2 \pi/\beta$. Such holomorphic vertex operator correlation functions can be computed using the formula
\begin{equation}
	\Big\langle\prod_k e^{i \alpha_k \phi(w_k)}\Big\rangle = \prod_{i<j} (w_i-w_j)^{\alpha_i \alpha_j/4\pi}\,,
\end{equation}
which holds when $\sum_i \alpha_i=0$ and vanishes otherwise.
Using this result, and pushing the operators to the boundary, we find
\begin{equation}
	G_{\mathbb{H}}(w_i,\bar{w}_i)|_{y_i \to 0} \approx 2^{\alpha^2/\pi} \prod_{i=1}^4 y_i^{\alpha^2/4\pi} \left( \frac{x_{13}x_{24}}{x_{12}x_{23}x_{14}x_{34}}\right)^{\alpha^2/\pi}\,.
\end{equation}
Crucially, the powers of $y_i$ correspond precisely to the bulk-boundary OPE factor that maps the $V_{0,\pm1}$ operators of dimension $\alpha^2/4\pi=\pi/\beta^2$ from the upper half plane to the boundary. Absorbing an overall power of 2 into the definition of the boundary operators to obtain the canonical normalization, we find our one-dimensional correlator becomes:
\begin{equation}
	G_{+-+-}(x_i)= \frac{1}{(x_{12} x_{34})^{\alpha^2/\pi}}(1-z)^{-\alpha^2/\pi}\,.
\end{equation}
From this, we can read  the dimension of the boundary kink operator $\Delta_K=\alpha^2/2\pi=2\pi/\beta^2$, which is twice the dimension of the corresponding bulk field. Furthermore, the invariant part of the correlator admits a Taylor series at $z=0$, which means that the exchanged operators in the $s$-channel have integer dimension. They are also neutral under the $U(1)$ symmetries, and we recognize them as $\partial_\perp \phi$ and its composites, whose correlation functions we analyzed in the previous section. In particular, we find that the $\mathbb{Z}_2$ odd operator $\partial_\perp \phi$ of dimension 1 is itself exchanged, with an OPE coefficient
\begin{equation}
	c_{K\bar{K}1}^2= 2\Delta_K\,.
\end{equation}
This will be important for comparison with the numerical bootstrap results below. The other OPE channel is equivalent to the $s$-channel of the differently ordered correlator:
\begin{equation}
	G_{++--}(x_i)=\frac{1}{(x_{12} x_{34})^{\alpha^2/\pi}} \left( \frac{z^2}{1-z}\right)^{\alpha^2/\pi} \,.
\end{equation}
The exchanged operators in this channel are vertex operators with winding charge two. In the OPE limit we see the powers $z^{4\Delta_K+n}$; the factor 4 is expected because the dimension of the bulk vertex operators is quadratic in their charge.\footnote{We note in passing that these vertex operators correlation functions are interesting examples of exact CFT correlators which are not of mean field theory type, since the exchanged operators do not have double-particle dimension.}

For later reference, we note that the above correlators are related to the functions $g_{\mathbf{R}}(z)$ introduced previously as:
\begin{align}
	g_{\mathbf{2}}(z)   & = \frac{1}{2}G_{++--}(z)\,, \nonumber           \\
	g_{\mathbf{0^+}}(z) & = \frac{G_{+-+-}(z)+G_{+--+}(z)}{2}\,,
	\label{girrepsfromGpm}
	\\
	g_{\mathbf{0^-}}(z) & = \frac{G_{+-+-}(z)-G_{+--+}(z)}{2}\,.\nonumber
\end{align}

%%%%%%%%%%%%%%%%%%%%%%%%%%%%%%
\subsubsection{First-order corrections}
%%%%%%%%%%%%%%%%%%%%%%%%%%%%%%
It is not hard to extend the previous calculation to first order in $\lambda$. Since our perturbation is $\lambda \int_{AdS_2} d^2x \sqrt{g}\,\cos(\beta \phi)$, all the integrands can still be obtained in terms of correlation functions of vertex operators. However, we must be careful about the fact that our external operators are winding modes, while the perturbation is a sum of two momentum modes $e^{i \beta (\phi_L+\phi_R)} + e^{-i \beta (\phi_L+\phi_R)}$. We can start by computing the first order correction to the kink two-point function, which will allow us to read off its anomalous dimension. We want to compute \begin{equation}
	\langle K(x_1) \bar{K}(x_2)\rangle= x_{12}^{-2\Delta_K} -\lambda \int_{AdS_2}d^2x \sqrt{g} \left\langle K(x_1) \bar{K}(x_2) {\cal O}(x,y)\right\rangle_{AdS_2} + \dots \,,
\end{equation}
where $ {\cal O} = \cos(\beta\phi)-1$ is the relevant deforming operator (with the subtraction of the constant piece necessary to cancel infrared divergences), and the correlator on the right is to be computed in the free theory. To obtain the integrand we use the map to the upper half plane:
\begin{equation}
	\left\langle K(x_1) \bar{K}(x_2) {\cal O}(x,y)\right\rangle_{AdS_2} =
	\left(\frac{\LAdS}{y}\right)^{-\Delta_\beta}
	\left\langle K(x_1) \bar{K}(x_2) {\cal O}(x,y)\right\rangle_\mathbb{H}\,,
\end{equation}
with $\Delta_\beta=\beta^2/(4\pi)$. Then, from the method of images we find
\begin{align}
	 & \left\langle K(x_1) \bar{K}(x_2) {\cal O}(x,y)\right\rangle_\mathbb{H}=
	\lim_{y_1,y_2 \to 0} (2y_1)^{-\half \Delta_K}(2y_2)^{-\half \Delta_K}
	\\
	 & \qquad \frac{1}{2} \left\langle e^{i \alpha \phi(w_1)} e^{i \alpha \phi(w_1^*)}e^{-i \alpha \phi(w_2)} e^{-i \alpha \phi(w_2^*)}\left( e^{i \beta (\phi(w)-\phi(w^*))} +e^{-i \beta (\phi(w)-\phi(w^*))} -2\right) \right\rangle , \nonumber
\end{align}
where we pushed the operators to the boundary and inserted the appropriate bulk-to-boundary power law. Since $\alpha \beta=2\pi$, a remarkable simplification happens, and the first order integrand becomes simply:
\begin{equation}
	\label{eq:kink2pt}
	\left\langle K(x_1) \bar{K}(x_2)\right\rangle= x_{12}^{-2\Delta_K}
	\left(1-\lambda \LAdS^{2 -\Delta_\beta}
	\int_{AdS_2} \frac{dx dy}{y^2} \frac{-2 (x_{12})^2 y^2}{(y^2+(x-x_1)^2)(y^2+(x-x_2)^2)} \right)\,,
\end{equation}
where $\lambda \LAdS^{2 -\Delta_\beta}$ is the dimensionless coupling. From now on, we will set $\LAdS=1$ to avoid cluttering. The integral itself has a logarithmic IR divergence, which, when regularized by stopping the integration a distance $\epsilon$ away from the $AdS$ boundary, allows us to read  the anomalous dimension of the kink operator to be
\begin{equation}
	\Delta_v  =\Delta_K + \gamma \lambda +O(\lambda^2) \,,\qquad
	\qquad \gamma =-2\pi\,. \label{Deltav1storder}
\end{equation}
Importantly, this anomalous dimension is independent of $\beta$.

Our next target is the computation of the four-point functions. This is more involved, but things simplify drastically if we subtract the (one-loop corrected) disconnected parts. For example, in the case of the $+-+-$ correlator we find the clean result
\begin{equation}
	G_{+-+-}(x_i)= \left( \frac{x_{13}x_{24}}{x_{12}x_{23}x_{14}x_{34}}\right)^{2(\Delta_K+ \lambda \gamma)}-\lambda \left( \frac{x_{13}x_{24}}{x_{12}x_{23}x_{14}x_{34}}\right)^{2\Delta_K} G_{+-+-}^{\text{conn},(1)}(z)\,,
\end{equation}
where the connected contribution is simply
\begin{align}
	G_{+-+-}^{\text{conn},(1)}(z) & = -8 x_{12} x_{23} x_{14} x_{34} \\
	 \times&\int_{AdS_2} \frac{dx dy}{y^2} \frac{y^4}{(y^2+(x-x_1)^2)(y^2+(x-x_2)^2)(y^2+(x-x_3)^2)(y^2+(x-x_4)^2)}\,. \nonumber
\end{align}
Remarkably, the quantization of charges once again leads to a rational integrand. In fact, we identify a product of 4 bulk-to-boundary propagators of dimension 1, which leads to the well known D-function $D_{1111}(x_i)$. Carefully collecting all the terms, we obtain
\begin{equation}
	G_{+-+-}(x_i)= \frac{1}{x_{12}^{2(\Delta_K+\gamma \lambda )}x_{34}^{2(\Delta_K+\gamma \lambda )}}(1-z)^{-2\Delta_K}\left(1+ \lambda 4 \pi z  \log\left( \frac{1-z}{z}\right)  \right)\,.
\end{equation}
A similar analysis of the other charge sectors gives
\begin{align}
	G_{+--+}(x_i) & = \frac{1}{x_{12}^{2(\Delta_K+\gamma \lambda )}x_{34}^{2(\Delta_K+\gamma \lambda )}}(1-z)^{2\Delta_K}\left(1+ \lambda 4 \pi    \frac{z}{1-z}\log z  \right)\,,                                                    \\
	G_{++--}(x_i) & = \frac{1}{x_{12}^{2(\Delta_K+\gamma \lambda )}x_{34}^{2(\Delta_K+\gamma \lambda )}}\left(\frac{z^2}{1-z} \right)^{2\Delta_K}\left(1+ \lambda 4 \pi  \left( \frac{\log(1-z)}{z}-\log z\right) \right)\,.\nonumber
\end{align}
From this and equations \eqref{CBdecomp} and \eqref{girrepsfromGpm}, we can extract the value of the correlators at the crossing symmetric point, which will be useful below
	\begin{align}
		& g_2^* \equiv g_2(1/2)=-2^{-2 \Delta_K-1} \left(16^{\Delta_K}-2+8 \pi  \lambda  \log (2)\right)+O(\lambda^2) \,, \nonumber\\  
		& g_1^* \equiv g_1(1/2)=2^{2 \Delta_K-1}+O(\lambda^2) \,.
	\end{align}
Using these equations and \eqref{Deltav1storder}, we can eliminate the Lagrangian parameters $\lambda$ and $\Delta_K$ to obtain the following surface in the 3 dimensional space $(g_1^*,g_2^*,\Delta_v)$,
\begin{equation}
	\label{surfaceCPT}
	\log\left( g_1^* \,2^{1-2\Delta_v} \right) = 1- 2 g_1^* \left( g_1^*+ g_2^* \right) \ll 1\,.
\end{equation}
Notice that the free theories corresponds to setting both sides of this equation to zero, which leads to a line in the space $(g_1^*,g_2^*,\Delta_v)$ parameterised by $\Delta_K$. Switching on the coupling $\lambda$ extends this line to a surface, which is well described by \eqref{surfaceCPT} in the neighbourhood of the entire free theory line.

\subsection{\texorpdfstring{Dirac fermions in AdS$_2$}{Dirac fermions in AdS2}}
\label{subsec:cptdiracfermion}

A Dirac fermion is another example of a bulk QFT that gives rise to boundary correlators with $O(2)$ symmetry. In fact, this theory is at the origin of the well-known duality between the sine-Gordon theory and the Thirring model \cite{Coleman:1974bu}, which corresponds to bosonization in the UV. (We will argue that the duality also holds in AdS$_2$.)

The claim is that sine-Gordon model and a massive fermion with a quartic interaction
$(\bar{\psi} \gamma^\mu \psi)^2$ in AdS$_2$ give rise to the same two-parameter family of QFTs. For example,  we claim that they give rise to the same 
two-dimensional surface in the space $(g_1^*,g_2^*,\Delta_v)$.  However, the weakly coupled description of each theory gives access to a different part of this surface. While sine-Gordon leads to \eqref{surfaceCPT}, the fermionic description leads to
\begin{equation}
	\label{surfaceFermion}
	g_2^*+2^{-2\Delta_v}=2(1-g_1^*)\ll 1\,.
\end{equation}
Notice that both descriptions are weakly coupled around the point
$(g_1^*,g_2^*,\Delta_v) = \left(1,-\half,\half\right)$ corresponding to the free massless fermion. As a consistency check, one can verify that the two surfaces have the same tangent plane at this point.

We outline the calculation of the fermions in AdS$_2$, relegating the details to appendix \ref{sec:fermionsAppendix}. Dirac fermions in AdS$_2$ admit a decomposition into two pieces according their behavior near the boundary
\begin{align}
	\psi(y,x)
	  = \psi_+(y,x) +\psi_-(y,x)                                \,,\qquad \qquad
	\psi_\pm(y,x)
	  \xrightarrow[y\rightarrow 0]{} y^{\Delta_\pm} \psi_{0,\pm}(x)
	\,.
\end{align}
Here, \(\Delta_\pm = \frac{1 }{2 } \pm m\) is the scaling dimension of the fermion, depending on the bulk mass \(m\).  These two pieces individually have a dual interpretation in terms of  vertex operators. We would like to compute the correlators in this theory analogous to the bosonic theory \eqref{girrepsfromGpm}. We need to compute $ G_{++--},G_{+-+-},G_{+--+} $. Zeroth order perturbation theory is done by mere Wick contraction, keeping track of additional minus signs due to the fermionic nature of the fields. However, for the first order perturbation theory, one needs to compute tree level Witten diagrams with fermionic propagators. As reviewed in the appendix \ref{sec:fermionsAppendix}, these diagrams are related to the corresponding scalar Witten diagrams by a shift of one half in the external dimensions. Once the dust settles we obtain the following first-order values for the three observables listed above:
\begin{align}
	&
	g_2^*=-2^{-2 \Delta } \left[1+\frac{4
		\sqrt{\pi } \Gamma \left(2 \Delta
		+\frac{1}{2}\right)
		\bar{D}_{\Delta}^*}{\Gamma \left(\Delta
		+\frac{1}{2}\right)^4}  \lambda
	_f +O(\lambda_f^2) \right], \nonumber\\ 
	& g_1^*=1+\frac
	{\sqrt{\pi } 2^{1-2 \Delta } \Gamma
		\left(2 \Delta +\frac{1}{2}\right)
		\bar{D}_{\Delta}^* }{\Gamma \left(\Delta
		+\frac{1}{2}\right)^4}\lambda_f
		+O(\lambda_f^2)
		\,,
		\\
		&\Delta_v = \Delta + O(\lambda_f^2) \,.
		\nonumber
	\label{eqn:fermionValuesForg}
\end{align}
Here, $\bar{D}_{\Delta}^*= \bar{D}_{\Delta+\frac{1}{2}\,\Delta+\frac{1}{2}\,\Delta+\frac{1}{2}\,\Delta+\frac{1}{2}}(1/2)$ is a special function defined in appendix \ref{sec:fermionsAppendix}, and $\Delta$ is the free fermion dimension. After eliminating $ \lambda_f$ and $\Delta$ this leads to the simpler relation \eqref{surfaceFermion}.

%%%%%%%%%%%%%%%%%%%%%%%%%%%%%%%
\subsection{Numerical bootstrap}
\label{ssec:numboot}
%%%%%%%%%%%%%%%%%%%%%%%%%%%%%%%
Having collected some analytical data on the UV limit of sine-Gordon in AdS$_2$, we can now try to ask whether it is an extremal theory with respect to some bootstrap problem in the one-dimensional boundary theory. Combining equations \eqref{eq:O2crossing} and \eqref{CBdecomp} yields
\begin{equation}
	\label{eq:O2vectoreqt}
	\sum_{\mathbf{0}^+} \lambda_{\mathcal{O}}^{2} V_{\mathbf{0}^+, \Delta}+\sum_{\mathbf{2}} \lambda_{\mathcal{O}}^{2} V_{\mathbf{2}, \Delta}+\sum_{\mathbf{0}^-} \lambda_{\mathcal{O}}^{2} V_{\mathbf{0}^-, \Delta}=0\,,
\end{equation}
with
\begin{equation}
	\label{eq:O2vectors}
	V_{\mathbf{0}^+, \Delta}=\left(\begin{array}{c}
			0              \\
			F_{\Delta}^{-} \\
			F_{\Delta}^{+}
		\end{array}\right), \quad V_{\mathbf{2}, \Delta}=\left(\begin{array}{c}
			F_{\Delta}^{-} \\
			0              \\
			-2 F_{\Delta}^{+}
		\end{array}\right), \quad V_{\mathbf{0}^-, \Delta}=\left(\begin{array}{c}
			-F_{\Delta}^{-} \\
			F_{\Delta}^{-}  \\
			-F_{\Delta}^{+}
		\end{array}\right)\,,
\end{equation}
and
\begin{equation}
	F^{\pm}_\Delta= (1-z)^{2\Delta_v}G_\Delta(z)\pm z^{2\Delta_v}G_\Delta(1-z)\,.
\end{equation}
These can be analyzed with the standard conformal bootstrap methods.

%%%%%%%%%%%%%%%%%%%%%%%%%%%%%%%
\subsubsection*{Bounding the four-point function: single correlator}
%%%%%%%%%%%%%%%%%%%%%%%%%%%%%%%
We are interested in extremizing the values of our correlators at the crossing symmetric point $z = 1/2$. Incorporating this value in the numerical bootstrap was first done in \cite{Lin:2015wcg} and we will essentially follow their approach. To review the method, consider first the analogous problem for a single-correlator setup:\footnote{Analytic bounds on the value of a single correlator were derived in \cite{Paulos:2020zxx}, which state that $g_{\text{GFF}}\leq g(z)\leq g_{\text{GFB}}$ for $\Delta^*\geq 2 \Delta_\phi$. For $z=1/2$, we found that these bounds can be checked, to a high numerical accuracy, using the procedure that we now outline.}
\begin{equation}
	\langle \phi \phi \phi \phi \rangle=\frac{g(z)}{(x_{12} x_{34})^{2\Delta_\phi}}
\end{equation}
and associated crossing symmetry equation:
\begin{equation}
	\sum_\Delta c^2_\Delta \left( (1-z)^{2 \Delta_\phi} G_\Delta(z)-z^{2\Delta_\phi} G_\Delta(1-z)\right) = 0\,.
\end{equation}
Normally one acts with a functional $\alpha(\cdot)$ that is a linear combination of the odd derivatives, so for each block in the above equation we obtain:
\begin{equation}
	\label{eq:oddderivs}
	2\sum_{n=0}^{\Lambda} a_{2n+1} \partial_z^{2n+1}\left( (1-z)^{2 \Delta_\phi} G_\Delta(z)\right) |_{z=1/2}\,,
\end{equation}
with $\alpha_{2n+1}$ the components of the functional. Suppose that now we want to formulate impose that the correlator takes the value $g(1/2)=g^*$ at the crossing symmetric point. This implies that
\begin{equation}
	\sum_{\Delta} c^2_\Delta 2^{-2\Delta_\phi} G_\Delta(1/2) = 2^{-2\Delta_\phi} g^*\,,
\end{equation}
or, more suggestively
\begin{equation}
	\sum_{\Delta} c^2_\Delta \partial^0_z\left( (1-z)^{2 \Delta_\phi} G_\Delta(z)- \delta_{\Delta,0} \,2^{-2\Delta_\phi} g^*\right)|_{z=1/2} =0\,,
\end{equation}
where the choice to assign $g^*$ to the identity block is arbitrary but convenient. Upon comparison with the original problem, we conclude that we should (a) add the zero derivative component to the basis of odd derivatives \eqref{eq:oddderivs}, and (b) work with shifted blocks such that
\begin{equation}
	(1-z)^{2 \Delta_\phi} G_\Delta(z) \to (1-z)^{2 \Delta_\phi} G_\Delta(z) -\delta_{\Delta,0}(1/2)^{2\Delta_\phi} g^* \equiv F^*_\Delta(z)\,.
\end{equation}
Note that the shift does not alter any of the equations corresponding to odd derivatives. The complete functional must then obey:
\begin{equation}
	\alpha\big(F^*_\Delta(z)\big) = \sum_{n=0,1,3,5,\ldots} a_n \partial_z^n\big(F^*_\Delta(z)\big)|_{z=1/2} >0
\end{equation}
for all $\Delta$ in the assumed spectrum, including the identity operator. We can then perform a binary search in $g_*$ to find its extremal allowed values for a given spectrum.

\subsubsection*{Bounding the four-point function: correlator of $O(2)$ vectors}
As discussed in section \ref{sec:O2CFT1}, in the $O(2)$ case the correlator has three components $g_{1,2,3}(z)$. At the crossing symmetric point $z=1/2$, equation (\ref{eq:O2crossing}) implies that $g_3(1/2)=g_1(1/2)$. This is automatically imposed in the zero-derivative part of the third component of equation \eqref{eq:O2vectoreqt}, since  $F^+_\Delta(z)$ contains the information about even derivatives. This leaves us with two independent values which we can take to be $g_1(1/2)$ and $g_2(1/2)$. Using the block decomposition and the third crossing equation, we have that
\begin{align}
	g_2(1/2)  & =\sum_{\mathbf{2}} \lambda_{\mathcal{O}}^{2} G_{ \Delta}(1/2)-\sum_{\mathbf{0}^-} \lambda_{\mathcal{O}}^{2} G_{ \Delta}(1/2)\,,\nonumber \\
	2g_1(1/2) & =\sum_{\mathbf{0}^+} \lambda_{\mathcal{O}}^{2} G_{ \Delta}(1/2)+\sum_{\mathbf{0}^-} \lambda_{\mathcal{O}}^{2} G_{ \Delta}(1/2)\,,
\end{align}
where the right hand sides are exactly in the form of the first and second components of the crossing equation \eqref{eq:O2vectoreqt}. Now we can just extend the above single-correlator procedure to the first and second components of \eqref{eq:O2vectoreqt}; we allow the functional to include the zero-derivative component of these equations and add constant shifts to the blocks. For the second component (corresponding to $g_1(1/2)$) the replacement reads:
\begin{equation}
	(1-z)^{2 \Delta_v} G_\Delta(z) \to (1-z)^{2 \Delta_v} G_\Delta(z) -\delta_{\Delta,0}(1/2)^{2\Delta_v} 2g_1^* \equiv F^*_{1,\Delta}(z)\,,
\end{equation}
which once again does not alter the odd-derivative components. For the first component, whose zero derivative term corresponds to $g_2(1/2)$, we must be more careful because the identity operator is not exchanged in this equation. The resolution is to shift the blocks as
\begin{equation}
	(1-z)^{2 \Delta_v} G_\Delta(z) \to (1-z)^{2 \Delta_v} G_\Delta(z) -\delta_{\Delta,0}(1/2)^{2\Delta_v} (g_2^*+1) \equiv F^*_{2,\Delta}(z)\,,
\end{equation}
and to \emph{also} add an identity operator in this channel. The extra `1' then cancels this identity block, and the zero-derivative component of the first equation does end up imposing the correct values of $g_2^*$. The higher-derivative components of course do normally see this extra identity operator, but this is easily fixed by setting them to zero by hand in the vector corresponding to the action of the linear functional on the identity operator. Altogether this shows that the problem for a fixed $g_1^*$ and $g_2^*$ can be formulated entirely analogously to the single-correlator case. 

We will explore the allowed values of $g_1^*$ and $g_2^*$ for a given gap in the spectrum. It is convenient to first maximize the gap in a grid of $g_1^*$ and $g_2^*$, and then find a central value of $g_1^*$ and $g_2^*$ where the problem is primal feasible for the desired gap. Then one can parametrize the $(g_1^*, g_2^*)$ plane in polar coordinates centered at that point, and do a radial bisection for several angles to find the boundary of the allowed space in this plane.\footnote{Note that the allowed region in the $(g_1^*, g_2^*)$ plane is convex. Proof: pick two points $p_1$ and $p_2$ in the plane that are allowed, so at each point there is a good solution to crossing symmetry. Now take a linear combination of these two solutions with positive weights and total weight one. These are still good solutions (crossing symmetric, positive OPE coefficients, unit operator appears with coefficient 1), but by varying the relative weight we cover the entire line connecting $p_1$ and $p_2$. That line is therefore also in the allowed region.}

\subsubsection{Numerical maximization results: the O(2) menhir}
We will impose a gap of $2\Delta_v$ in all sectors. Physically we have in mind that there are no bound states (in the flat-space limit), and in practice this makes the number of free parameters more manageable. In the UV theory this condition is obeyed in the interval $1/4\leq \Delta_v=\Delta_K\leq 1/2$, or equivalently $4\pi \leq\beta^2\leq 8\pi$.

\begin{figure}
	\centering
	\includegraphics[width=0.8\linewidth]{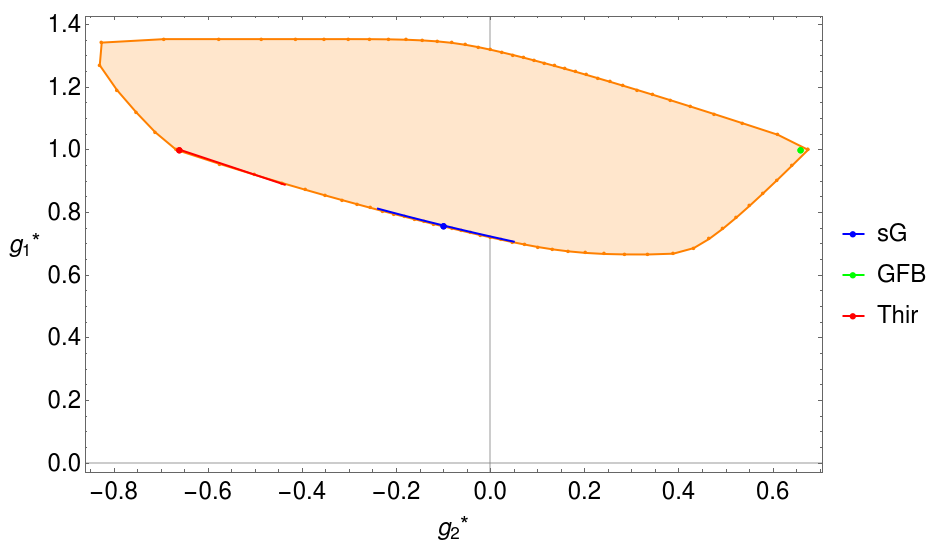}
	\caption{Allowed region in the space of correlation function values for $\Delta_v=0.3$with a gap of $2\Delta_v$ in all sectors. 
	The plot is computed with $\Lambda=25$ but it would not change significantly for higher $\Lambda$.
	The plot contains several interesting kinks. Two of them can be identified with the generalized free fermion in red and the generalized free boson in green. In blue, we find the correlator of boundary vertex operators with winding number 1 in the compact boson CFT with Dirichlet bondary conditions.
	The small segments in red and blue correspond to the first order deformations discussed above.
	\label{fig:slateat03}}
\end{figure}

As a first result, we show in figure \ref{fig:slateat03} the allowed region in the $(g_1^*, g_2^*)$ plane for a representative value $\Delta_v= 0.3$.\footnote{Related bounds were obtained in \cite{Ghosh:2021ruh}, and our slate nicely fits in the leftmost region of the convex hull shown in figure 6 of \cite{Ghosh:2021ruh}. However, our bounds are far stricter since we only allow for the identity exchange in the singlet channel and we always impose a gap of $2\Delta_v$ in all sectors.}

The slate contains several interesting features, including a few kinks. Two of them are easily identified with the generalized free boson and fermion solutions. Remarkably, the vertex operator correlation function also sits right at the boundary of the allowed region. We also plot the first-order perturbative results around the free boson as given in equation \eqref{surfaceCPT}, and around the free fermion as given in equation \eqref{surfaceFermion}. They are nicely tangent to the bound, but for the free fermion we see that the Thirring coupling has to be positive to stay within the allowed region. The other sign is forbidden since it leads to a negative anomalous dimension for the two fermion operator of dimension $2\Delta_v$, violating our gap assumption.

\begin{figure}
	\centering   	
	\includegraphics[trim=0cm 5cm 0cm 5cm , width=0.99\linewidth]{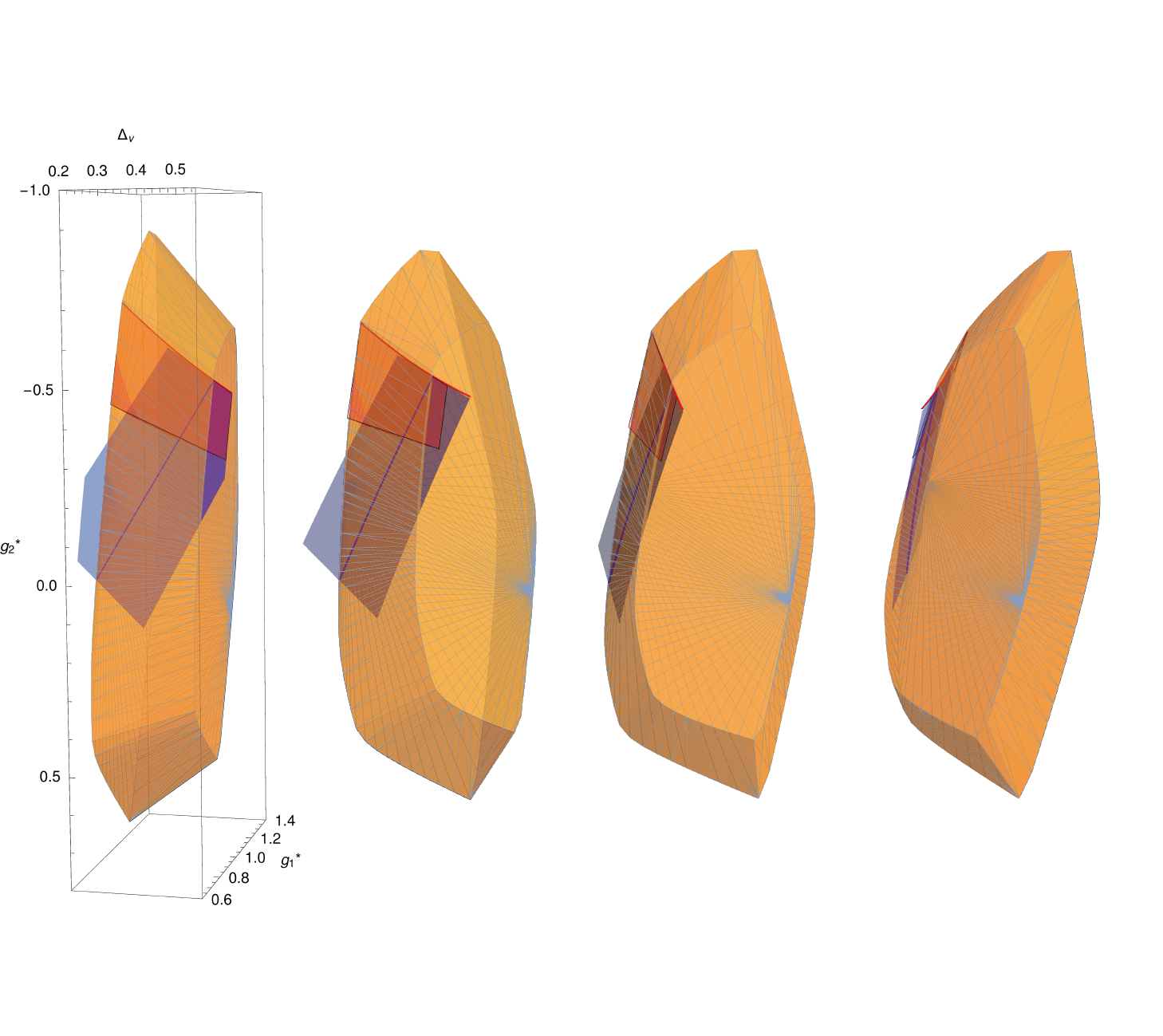}
	\caption{Allowed region in the $(g_1^*\,,\,g_2^*)$ space of O(2) symmetric correlators. The three-dimensional shape is a tower of allowed space for external dimensions $1/4\leq\Delta_v\leq1/2$. The blue line and attached surface correspond to the free vertex operator correlator and its first order correction \eqref{surfaceCPT}, both of which are tangent to the bound. In red, we have the massive fermion line and the surface corresponding to the first-order Thirring perturbation \eqref{surfaceFermion}. Again, these are tangent to the bound.
	}
	\label{fig:menhirs}
\end{figure}

We also studied how the slate changes as we vary the dimension of the external operator $\Delta_v$. The resulting three-dimensional figure is shaped like a menhir and is shown in figure \ref{fig:menhirs}. The kinks that were visible in the $\Delta_v = 0.3$ plot remain present in the full interval. An interesting fact is that when $\Delta_v=1/2$, the vertex operator correlator is equal to the generalized free fermion correlator. This is the boundary version of the elementary bosonization relation between a free boson and a free fermion.

As shown by the blue surface in figure \ref{fig:menhirs}, the first-order sine-Gordon perturbative surface \eqref{surfaceCPT} is tangent to the bound in a remarkably extended region. The same is true for the first-order Thirring perturbative surface \eqref{surfaceFermion}, which is shown in red in figure \ref{fig:menhirs}. We also see that at $\Delta_v=1/2$ the $\lambda \cos(\beta \phi)$ perturbation is related to the mass deformation of the free fermion as expected from the bosonization map from sine-Gordon to the fermionic Thirring model. This can be checked by comparing the tangent vectors associated to the two deformations.

\begin{figure}[h]
	\centering   	\includegraphics[width=0.49\linewidth]{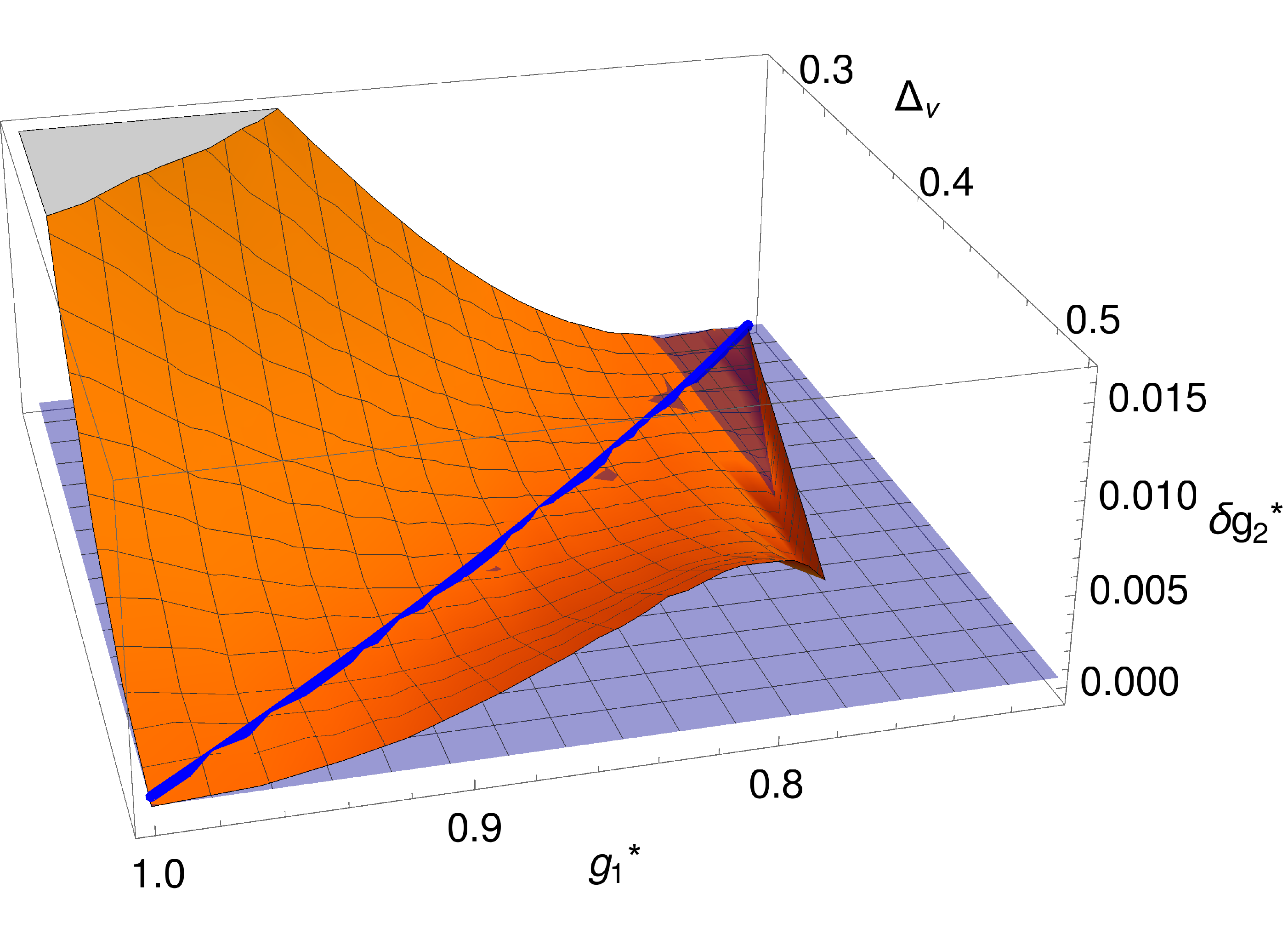}
	\centering
	\includegraphics[width=0.49\linewidth]{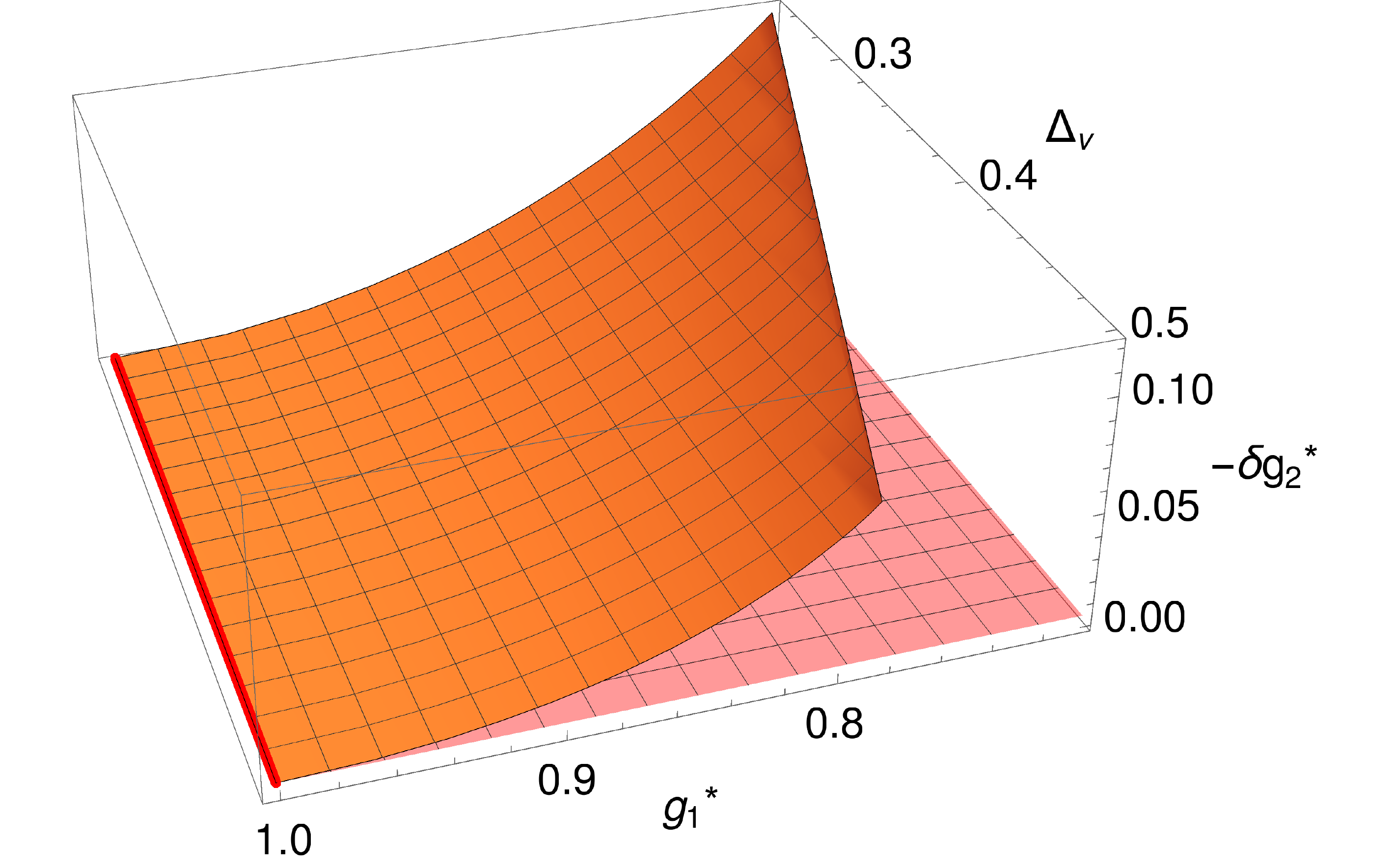}
	\caption{Difference between the perturbative and extremal numerical value for $g_2^*$ as a function of $g_1^*$ and $\Delta_v$. The left plot coresponds to the vertex operator formulation of sine-Gordon, and the right to the fermionic Thirring model description. The error is small near each description's perturbative region. Both descriptions work well near the massless free fermion point $g_1^*=1\,,\, \Delta_v=1/2 $.
	}
	\label{fig:errors}
\end{figure}

To more carefully quantify the saturation of the bounds by the bosonic and fermionic formulations of the sine-Gordon theory, we present in figure \ref{fig:errors} the difference between the values of $g_2^*$ for the perturbative results and the numerical bound ($\delta g_2^*$) for each fixed value of $g_1^*$ and $\Delta_v$, which specify the two free parameters in the perturbative theories. We find a remarkable match in the respective regions of validity of the perturbative description which are rather complementary. However, we find first-order perturbation theory in the bosonic theory to be more effective in a larger region of observable space.

\subsubsection*{Comments on the flat-space limit}
It is also interesting to ask what happens as we increase the external dimension $\Delta_v$, where we expect to connect to the flat space limit and to the sine-Gordon kink S-matrix. For this, we need to be able to relate the CFT correlator to the flat space S-matrix. Let us consider first the four-point function of identical operators of dimension $\Delta_\phi$. According to the work of \cite{Komatsu:2020sag} there is an elementary relation between  the connected correlation function and the scattering amplitude in flat space. In our $O(2)$ case this relation becomes:
\begin{align}
	\sigma_{1}(s) & =\left.\lim _{\LAdS \rightarrow \infty} z^{-2 \Delta_v}\left(g_{1}(z)-1\right)\right|_{z=1-s / (4 m^2)} \,, \nonumber \\
	\sigma_{2}(s) & =\left.\lim _{\LAdS \rightarrow \infty} z^{-2 \Delta_v} g_{2}(z)\right|_{z=1-s / (4 m^2)}\,.
\end{align}
Here the $\sigma_{i}$ are the components of the $O(2)$ S-matrix in the same conventions as our CFT correlators (same as in \cite{Kos:2013tga}). The extra prefactors are simply due to the one-dimensional contact Witten diagram at large $\Delta_v$, which should be divided out according to the prescription in \cite{Komatsu:2020sag}. We also observe that the value of the correlator at the conformal crossing symmetric point $z=1/2$ maps to the massive crossing symmetric point $s=2m^2\equiv2$.

\begin{figure}
	\centering   	\includegraphics[width=0.65\linewidth]{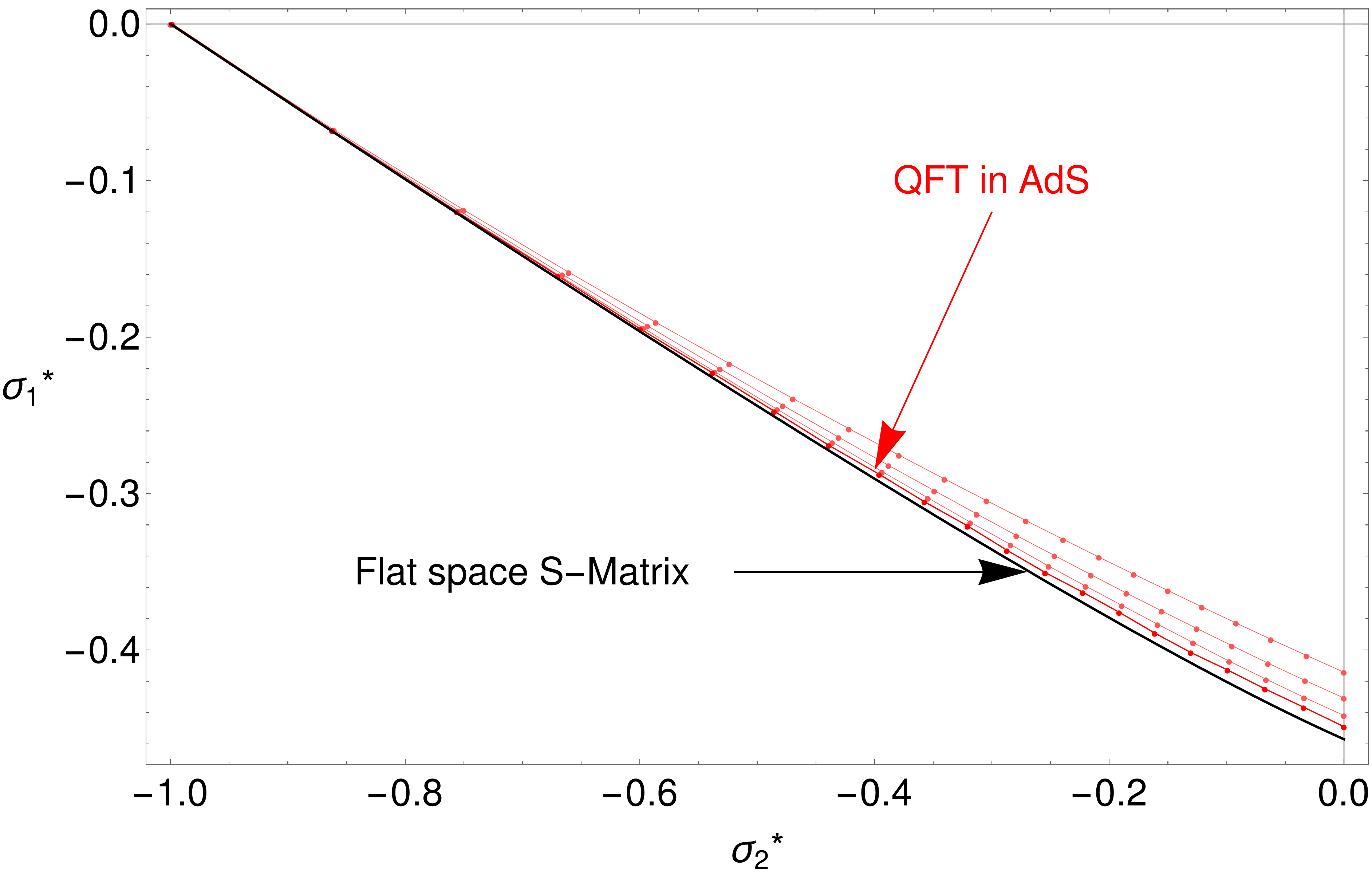}
	\caption{Bounds on the rescaled variables $\sigma_2^*,\sigma_1^*$ for $\Delta_v=1/4,1/2,1,2$, from the interior to the exterior. The black line, corresponds to the flat space values of the sine gordon kink S-matrix, in the parameter range $1/4\leq\Delta_K = 2\pi/\beta^2 \leq1/2$, which is the no-bound state range. }
	\label{fig:sigmas}
\end{figure}

\begin{figure}
	\centering
	\includegraphics[width=0.65\linewidth]{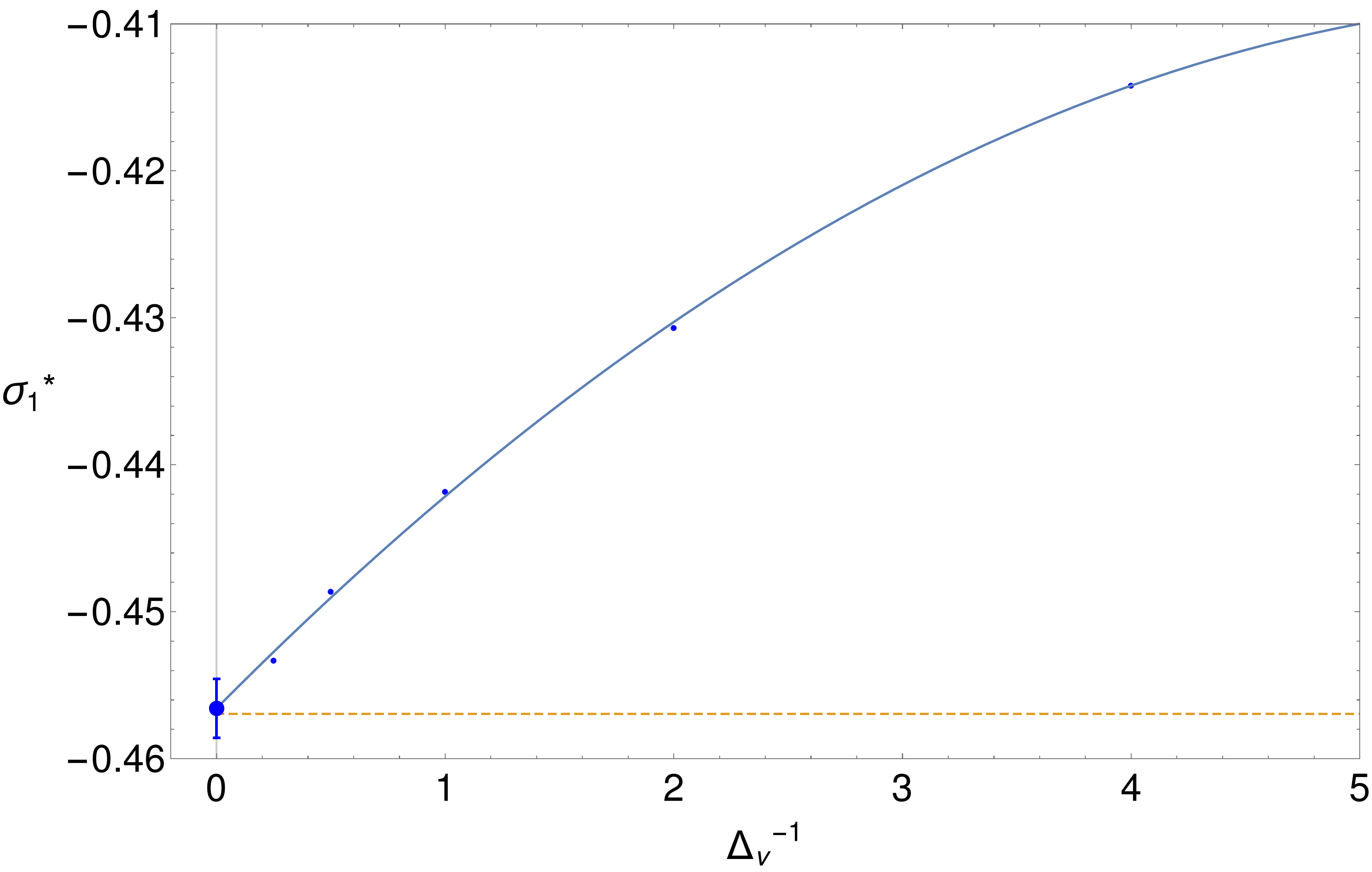}
	\caption{Lower bounds on the rescaled variable $\sigma_1^*$ as a function of $\Delta_v^{-1}$, for $\sigma_2^*=0$. The blue line is a quadratic interpolation in $\Delta_v^{-1}$. The extrapolation to the flat space limit is presented as a larger blue point with a non-rigorous error-bar, which we estimated by performing extrapolations of different degree in $\Delta_v^{-1}$. We observe an excellent match with the flat-space value, represented by the yellow line.}
	\label{fig:sigmaextrap}
\end{figure}

Although the flat-space limit is really only valid in the large $\LAdS$ and therefore large $\Delta_v$ limit, it is still interesting to plot the quantities $\sigma_i(1/2)\equiv \sigma_i^*$ at finite $\Delta_v$. We do so in figure \ref{fig:sigmas}.
Remarkably, in these variables, the UV and IR regions become extremely close! In particular, the free fermion line collapses into a single point. We can also extrapolate these results to $\Delta_v \to \infty$. Upon doing so we find a reasonably good match with the expected flat space sine-Gordon values, which can be obtained by numerically evaluating the Zamolodchikov-Zamolodchikov S-matrix \cite{Zamolodchikov:1978xm} and which saturates the S-matrix bounds of \cite{Cordova:2019lot}. Some numerical data and the associated extrapolation for the case of $\sigma_2^*=0$ is presented in figure
\ref{fig:sigmaextrap}.

Our proposal is that sine-Gordon in AdS$_2$ provides a two parameter family of correlators which approximately saturate the bounds in the $(\sigma_1^*,\sigma_2^*)$ plane (or equivalently the $(g_1^*, g_2^*)$ plane) for all values of the AdS radius. The saturation is sharp in the UV, where it corresponds to the winding vertex operator correlators, but also in the IR where it describes the flat space sine-Gordon kink S-matrix. In addition, the bounds are also saturated along the free fermion line. At intermediate values we expect the sine-Gordon correlators to be close to the bounds but perhaps not exactly saturating them because extremal solutions typically have a sparser spectrum of exchanged operators than any physical theory (see discussion in \ref{sec:sparse}). It would be interesting to understand this in more detail, and in particular study the effect of including the constraints of multiple correlators which should bring the bootstrap bound closer to the real QFT in AdS.
 
%!TEX root = draft.tex

%%%%%%%%%%%%%%%%%%%%%%%%%%%%%%%%%%%%%%%%%%%%
\section{Conclusions}
\label{sec:conclusions}
%%%%%%%%%%%%%%%%%%%%%%%%%%%%%%%%%%%%%%%%%%%%

Studying quantum field theory in Anti-de Sitter space is a worthwhile endeavour. Its conformally covariant boundary observables allow us to leverage the conformal bootstrap axioms for non-conformal theories. This work is the first step towards the goal of bootstrapping an RG flow using conformal techniques.

We started by studying the simplest possible setup: $\mathbb{Z}_2$ symmetric deformations of a massless free boson in AdS$_2$. In flat space, the canonical example of an RG flow between this boson and a gapped phase is the sine-Gordon theory. The integrable S-matrix of the lightest breathers in this theory maximizes the coupling to their bound state. This led us to analyze the AdS version of this problem, which amounts to the maximization of the OPE coefficient $c_{112}^2$ between the two lightest $\Z_2$ odd operators in the boundary theory and their $\Z_2$ even ``bound state''. We found that this OPE coefficient is extremized both in the free UV limit and to first order in perturbation theory. However, at second order in the lambda expansion, the sine-Gordon theory moves to the interior of the bound and stops being extremal. Instead, we find that the extremal theory is associated to Witten diagrams with only quartic vertices. 

However, the extremality of these physical theories cannot last forever. The extremal solutions to the crossing equations are observed to have a sparse spectrum with ``one operator per bin'' (of width 2 in $\Delta$ space), much like a generalized free theory. In physical theories perturbation theory does not allow for this possibility, since three loop diagrams allow for unitarity cuts which are known to contain four-particle operators \cite{Ponomarev:2019ofr,Meltzer:2019nbs}. This means that while we are able to track sine-Gordon theory in the endpoints of the RG flow, we cannot control it in between, as the extremal spectrum cannot coincide with the physical one.

Our next step was to include multiple correlators in the numerical bootstrap study. While this analysis did lead to the discovery of interesting features in the space of CFT data, we did not improve on the single-correlator bounds in the region where we are able to make contact with the perturbative RG flows.

To find sine-Gordon, there was fortunately another path to take. In the flat space theory, the breathers are in fact a composite state of two more elementary excitations: kinks and anti-kinks. These form a doublet under a topological $O(2)$ symmetry, and are therefore sensitive to the radius of the UV compact boson theory. This clearly singles out sine-Gordon in the zoo of all the $\Z_2$ symmetric deformations. In the UV the kinks overlap with winding mode operators, and their correlators therefore provided a new target for a perturbative and numerical analysis. In this case we decided to numerically bound the values of these correlators at the crossing symmetric point, with the allowed region taking a menhir-like shape shown in figure \ref{fig:menhirs}. Once again, it is known that these bounds are saturated by the sine-Gordon theory in the deep IR and we found that they are also saturated to the first order in perturbation theory. It would be nice if we could show that the sine-Gordon theories remain near the boundary of the space also for intermediate points along the flow, but to do so we need more perturbative and numerical data.

Amusingly, we could also perturbatively saturate the bounds on the correlator by studying quartic deformations of a Dirac fermion. This is related to the duality between sine-Gordon theory and the Thirring model, which we explored further in AdS$_2$. In the future it would be interesting to explore other aspects of this duality in hyperbolic space, for example how the boundary conditions are mapped to each other.

A recurring theme in this paper was the difference between the spectrum of a physical theory and the spectrum of extremal solutions to crossing. For the single-correlator bounds we appear to obtain a rather sparse extremal spectra with one operator per bin, which we showed to be unphysical because the local quantum field theories we analyzed have a denser spectrum. The multi-correlator analysis is less obvious. The optimistic expectation is that the inclusion of more external operators is bound to reveal the presence of more exchanged operators in the spectrum. Unfortunately this expectation is sometimes plagued by the existence of spurious solutions to crossing, an example of which we described in appendix \ref{app:multicorrbounds}. It would be interesting to avoid having to deal with these solutions and to explictly extract an extremal spectrum with more than one operator per bin. This would be the first step in a hierarchy of multi-correlator problems, which would hopefully approach a realistic, dense, CFT spectrum.

Finally it would be nice to see how this all connects to the integrability of flat-space S-matrices. S-matrix integrability is defined as the absence of particle production along with factorization of higher-point processes determined by the Yang-Baxter equations. Is there a form of integrability that can survive in AdS? If so, then what would be the precise signature of integrability\footnote{One possibly useful example was studied in \cite{Cavaglia:2021bnz}, where the spectrum of a one-dimensional conformal theory can be computed using integrability methods imported from $\mathcal{N}=4$ SYM. The spectrum shown in their figure 2 is much richer than one operator per bin once the coupling is large enough for the lifting of degeneracies to be visible and includes many level crossings.} in its one-dimensional boundary CFT data? And is there some connection to the solutions that extremize the bootstrap bounds? It would be interesting to address these questions in the future.

\section*{Acknowledgments}
 We would like to thank Connor Behan, Barak Gabai, Tobias Hansen, Shailesh Lal, Edoardo Lauria, Zhijin Li, Andrea Manenti, Marco Meineri, David Meltzer, Marten Reehorst, Sourav Sarkar, Jo\~{a}o Silva and Pedro Vieira for useful discussions.
 This research received funding from the Simons Foundation grants \#488637 (MC, AS), \#488649 (JP) and \#488659 (BvR) (Simons collaboration on the non-perturbative bootstrap). Centro de F\'\i sica do Porto is partially funded by 
 Funda\c c\~ao para a Ci\^encia  e a Tecnologia (FCT) under the grant UID-04650-FCUP. AA is funded by FCT under the IDPASC doctoral program with the fellowship PD/BD/135436/2017.
JP is   supported  by the Swiss National Science Foundation through the project 200020$\_$197160
and through the National Centre of Competence in Research SwissMAP.
\appendix

%!TEX root = draft.tex

%%%%%%%%%%%%%%%%%%%%%%%%%%%%%%%%%%%%%%%%%%%%
\section{Conformal perturbation theory for sine-Gordon breathers in $AdS_2$}
\label{sec:appsG}
%%%%%%%%%%%%%%%%%%%%%%%%%%%%%%%%%%%%%%%%%%%%

In this appendix we recover the results of section \ref{sec:1stOrder} in the language of conformal perturbation theory instead of using the Feynman-Witten rules. This is of course somewhat of an overkill, since only the mass shift and the $\phi^4$ vertex contribute at this order, but it will greatly simplify the analysis of the second order calculation, where all $\phi^{2n}$ vertices contribute simultaneously.
We start from the following action
\begin{equation}
	S =\int_{AdS_2} d^2x \sqrt{g}\left[  \frac{1}{2}(\partial \phi)^2 + \lambda \cos (\beta \phi)\right].
\end{equation}
Recall that demanding that the boson is $2 \pi r $ periodic, requires $\beta = n/r$, with $n$ as an integer. We take $n=1$, which means deforming by the most relevant operator. We will use the notation $\cos(\beta \phi) =  (V_\beta+ V_{-\beta})/2$,  with both the chiral and anti-chiral components, 
where $V$ denotes  the full vertex operators $V_{\beta}= :e^{i \beta \phi}:$. The  space of relevant scalar vertex operator deformations is determined by $\beta$. We find that there are $\lfloor\sqrt{8\pi}/\beta \rfloor$ pairs of momentum modes and $\lfloor \sqrt{2/\pi}\,\beta\rfloor$ pairs of winding modes. In particular, there is exactly one deformation preserving the symmetries of the RG flow in the range of $\beta$ discussed in section \ref{sec:chargedcorr}: the sine-Gordon potential $\cos(\beta \phi)$. The parameter $\beta$ also determines the flat space spectrum of particles. In particular, the number of bound states is given by $\lfloor 8 \pi/\beta^{2} \rfloor -1$. Note that for $\Delta_\beta= \beta^2/(4\pi)<2/3$ there are at least two bound states as mentioned in the introduction. Additionally there are no bound states in the range $4\pi<\beta^2<8\pi$, a fact that will be important in section \ref{sec:chargedcorr}.

At short distances, the curvature of AdS plays no role, and the UV theory is just a free boson in $AdS_2$. In Euclidean signature, and in Poincar\'{e} coordinates, the geometry is related by a Weyl transformation to that of a half-plane, leading to the statement that we can do perturbative calculations around the free-boson BCFT. This will lead to perturbation theory calculations more similar to conformal perturbation theory rather than Feynman-Witten rules. The relation between the two is obtained by expanding the cosine potential in its argument and using Wick contractions, as done in the main text.

In addition, we required a choice of boundary condition which we took to be Dirichlet. As discussed in the main text, the boundary operator of lowest dimension is the restriction of $\partial_{\perp} \phi$ to the boundary, with dimension 1. This boundary condition also implies that a bulk insertion of $V_\beta(z,\bar{z})$ is mapped to the two insertions $V_\beta(z), V_{-\beta}(z^*)$ by the Cardy doubling trick/method of images.
We will be interested in the CFT data of these boundary operators which we will extract from their correlation functions.
We focus on the following observable:
\begin{equation}
	\langle \partial \phi(x_1) \partial \phi(x_2) \partial \phi(x_3) \partial \phi (x_4)\rangle_{\mathbb{R}}\,.
\end{equation}
The answer will be given in perturbation theory by a power series in $\lambda$. The conformal perturbation theory prescription instructs us to compute terms that organize as 
\begin{align}
	 & \langle \partial \phi(x_1) \partial \phi(x_2) \partial \phi(x_3) \partial \phi (x_4)\rangle \nonumber                                                                                                      \\
	 & =\sum_n \frac{(-1)^n}{n!}\lambda^n \int_{AdS} d^2 z_1 \dots \int_{AdS} d^2 z_n \langle \partial \phi \,\partial \phi\,\partial \phi\,\partial \phi \,V_{\pm \beta}(z_1,\bar{z}_1) \dots  V_{\pm \beta}(z_n,\bar{z}_n)\rangle_{AdS}~.
\end{align}
From the Weyl-rescaling we have that $\langle \mathcal{O}_1\dots \mathcal{O}_n \rangle_{AdS} = \prod_i \Omega(z_i)^{-\Delta_i}\langle \mathcal{O}_1\dots \mathcal{O}_n \rangle_{BCFT}$, where $\Omega(z_i) = L_{AdS}/y_i$.
Therefore, the fundamental objects for this procedure are correlation functions of the boundary $\partial \phi$ operator with bulk operators  $V_{\pm\beta}$ in the free boson Dirichlet BCFT. This can be done with Wick contractions, which we systematize by using the following trick
\begin{equation}
	\label{eq:trick}
	\partial (e^{i \alpha \phi}) = i \alpha (\partial \phi) e^{i \alpha \phi} \implies \partial \phi= \frac{\partial(e^{i \alpha \phi})}{i \alpha}|_{\alpha \rightarrow 0}~.
\end{equation}
The idea is to use this convenient formula along with the formula for correlators of chiral vertex operators in free theory with chiral dimension $2h_i=\alpha_i^2/4\pi$,
\begin{equation} \label{eq-vertexcorr}
	\langle V_{\alpha_1} \dots V_{\alpha_n} \rangle_{\mathbb{R}^2} = \prod_{i<j} |z_i-z_j|^{ \alpha_{i} \alpha_j/4\pi} \,.
\end{equation}
We replace the $\partial \phi$ by a single derivative of a chiral vertex operator since chiral fields don't need the insertion of the mirror image. After the replacement of a bulk vertex operator by the two mirror replicas with opposite charge, we have a simple prescription to compute the required correlators
\begin{align}
	 & \langle \partial_1 \phi\, \partial_2 \phi\,\partial_3 \phi\,\partial_4 \phi\, V_{\pm \beta}(z_1, \bar{z}_1) \dots V_{\pm \beta}(z_n, \bar{z}_n) \rangle_{BCFT}  \nonumber                                                                                                        \\
	 & =\lim_{\alpha \rightarrow 0} \alpha^{-4} \partial_1 \partial_2\partial_3\partial_4 
	 \left\langle V_{\alpha}(x_1)V_{\alpha}(x_2)V_{\alpha}(x_3)V_{\alpha}(x_4) V_{\pm \beta}(z_1) V_{\mp \beta}(z^*_1) \dots  V_{\pm \beta}(z_n) V_{\mp \beta}(z^*_n)  \right\rangle_{\mathbb{R}^2} \,.
\end{align}
Here $\partial_i = \partial_{y_i}$, $x_i$ are boundary points and $z_i$ are bulk points. To take these derivatives, we put the auxiliary vertex operators at $(x_i,y_i)$, then differentiate with respect to $y_i$ and only then set $y_i=0$. After this, one can take the limit of $\alpha$ going to zero.

%%%%%%%%%%%%%%%%%%%%%%%%%%%%%%%%%%%%%%%%%%%%
\subsection{First-order perturbation theory}
\label{sec:appfirstord}
%%%%%%%%%%%%%%%%%%%%%%%%%%%%%%%%%%%%%%%%%%%%

Typically, one requires charge conservation with the insertion of vertex operators. But in Dirichlet boundary conditions, this is automatically satisfied as the mirror operator has opposite charge. In particular, we will have a non-vanishing first order correction to the four-point function. Note that $\cos(\beta \phi) = (V_\beta + V_{-\beta})/2$ is a sum of two contributions. The two vertex operators turn out to give identical results, so the factor of half in the cosine means we just need to compute the following term
\begin{align}
	 & -\langle \partial_1 \phi \,\partial_2 \phi\,\partial_3 \phi\,\partial_4 \phi \,V_{\beta}(z, \bar{z}) \rangle_{BCFT} = -\langle \partial_1 \phi \,\partial_2 \phi\,\partial_3 \phi\,\partial_4 \phi \,V_{\beta}(z)V_{-\beta}(z^*) \rangle_{\mathbb{R}^2} = \nonumber \\
	 & =-\lim_{\alpha \rightarrow 0} \alpha^{-4} \partial_1 \partial_2\partial_3\partial_4 \langle V_{\alpha}(x_1)V_{\alpha}(x_2)V_{\alpha}(x_3)V_{\alpha}(x_4) V_{ \beta}(z) V_{-\beta}(z^*)  \rangle_{\mathbb{R}^2}  \label{eqn:partialPhiCorr}	=        \\
	 & =\frac{-\lambda}{(2y)^{\frac{\beta^2}{4\pi}}}  \Biggl[ \left( \frac{1}{x_{12}^2 x_{34}^2}+ \text{2 perms} \right)  - \frac{\beta^2}{\pi}\left(\frac{1}{x_{12}^2}\Pi_3 \Pi_4 + \text{5 perms}\right)
		+ \frac{\beta^4}{\pi^2} \Pi_1 \Pi_2 \Pi_3 \Pi_4	\Biggr] \,,
	\nonumber
\end{align}
where we  identified $\Pi_i$ as the bulk to boundary propagator for $\Delta=1$ as given in (\ref{eq:BBpropagator}). 
To study the correlator in AdS, we must multiply by the Weyl factors of the bulk insertion points, that is:
\begin{equation}
	\langle \partial_1 \phi \,\partial_2 \phi\,\partial_3 \phi\,\partial_4 \phi \,V_{\beta}(z, \bar{z}) \rangle_{AdS} =
	\left( \frac{y}{L_{AdS}}\right)^{\frac{\beta^2}{4\pi}}   \langle \partial_1 \phi\, \partial_2 \phi\,\partial_3 \phi\,\partial_4 \phi \,V_{\beta}(z, \bar{z}) \rangle_{BCFT}~.
	\label{eqn:weyl}
\end{equation}
It is important to note that one vertex operator corresponds to two chiral insertions, such that we get the right power of $y$ to kill the prefactor in \ref{eqn:partialPhiCorr}. After this, the expression is covariant in AdS, depending only on objects that can be written as scalar products in the embedding space. 

Recall that now we have to integrate over the Poincare patch, with the appropriate measure: $dxdy (L_{AdS}^2/y^2)$. The integral of the first term in (\ref{eqn:partialPhiCorr}) is just the free answer times the volume of AdS which diverges like $\text{Vol}(\mathbb{R})/\epsilon$, in holographic regularization, where we stop the $y$ integral at a distance $\epsilon$ from the boundary. We can of course ignore this term by subtracting the constant part of the potential in the bulk.
The integral of the second term corresponds to a mass shift-diagram. In fact, writing only the position dependence, the answer is
\begin{equation}
\int_{AdS}dxdy\frac{L_{AdS}^2}{y^2}\left(\frac{1}{x_{12}^2}\Pi_3 \Pi_4 + \text{5 perms}\right) =\left(\pi\frac{  \log(\frac{x_{12}^2}{4\epsilon^2})+ \log(\frac{x_{34}^2}{4\epsilon^2})}{x_{12}^2 x_{34}^2} + \text{2\ perms} \right) .
\end{equation}
We have omitted terms that go to zero as $\epsilon$ goes to zero.
Now we have divergences which are logarithmic in $\epsilon$, along with $\log(x_{ij}^2)$ dependence which gives rise to the first order anomalous dimension of the external operator $\partial \phi$. Because this is linear in $\lambda$ we see that this is dual to the small mass of the bulk field.
Finally, the last term in (\ref{eqn:partialPhiCorr}) is just a D-function, or a contact Witten diagram. These integrals are finite and are given by
\begin{equation}
	\int_{AdS}dxdy\frac{1}{y^2}\left(\Pi_1 \Pi_2 \Pi_3 \Pi_4\right) = D_{1111}(x_i)|_{d=1} = \frac{\pi}{4}  \frac{1}{x_{12}^2 x_{34}^2} z^2 \bar{D}_{1111}(z)\,,
\end{equation}
where
\be
\bar{D}_{1111}(z) &= \frac{1}{z-1} \log(z^2) - \frac{1}{z} \log \Big((1-z)^2\Big) \,,
\ee
where $z$ is the 1d cross-ratio.
This term will lead to a change in the conformal block expansion, generating anomalous dimensions and OPE coefficients for all the exchanged operators. In this case they are just two-particle operators with perturbative corrections.
A neat way to pick the anomalous dimensions is to use the following orthogonality relation 
\begin{equation}
	\oint \frac{dz}{2 \pi i} \frac{1}{z^2} z^{\Delta + n} F_{\Delta+n}(z) z^{1- \Delta-n'} F_{1- \Delta - n'}(z) = \delta_{n,n'} \,,
\end{equation}
where we use the notation $F_{h}(z) \equiv \,_2 F_1(h,h;2h;z)$.
This allows one to pick  anomalous dimensions from the log terms in the Witten diagram
\begin{equation}
	\gamma_{2n}^{(1)} = \frac{1}{\left(c_{\partial \phi \partial\phi,2n}^{(0)}\right)^2} \oint \frac{dz}{2\pi i} z^{-3-2n}F_{-1-2n}(z) G(z)|_{\log z} \,.
\end{equation}
Here $2n$ labels the number of derivatives in the two-particle operator, $c_{\partial \phi \partial\phi,2n}^{(0)}$ is the OPE coefficient in the free theory, and the $G(z)|_{\log z}$ is the piece of the correlator that multiplies $\log z$, after extracting the usual $x_{12}^{-2}x_{34}^{-2}$ prefactor. In fact, from expanding the free four-point function
\begin{equation}
	\langle (\partial_{\perp}\phi)(\partial_{\perp}\phi)(\partial_{\perp}\phi)(\partial_{\perp}\phi)\rangle= \frac{1}{x_{12}^2 x_{34}^2}\left(1 + z^2 + \frac{z^2}{(1-z)^2} \right) ,
\end{equation}
in conformal blocks, one gets
\begin{equation}
	\left(c_{\partial \phi \partial\phi,2n}^{(0)}\right)^2 = \frac{2 \Gamma(2+2n)^2 \Gamma(2n+3)}{\Gamma(2n+1)\Gamma(4n+3)} \,.
\end{equation}
This matches the usual GFF answer with $d=1,\Delta=1$. Next, the contribution from the contact Witten diagram is
\begin{equation}
	-(\lambda L^{2-\Delta_{\beta}})2^{-\Delta_{\beta}} \frac{\beta^4}{4\pi} \frac{1}{x_{12}^2 x_{34}^2} \,z^2 \bar{D}_{1111}(z)\,.
\end{equation}
The $2^{-\Delta_{\beta}}$ factor appears as an overall factor in the perturbative calculation, so it can be absorbed in the definition of lambda.
Removing the $x_i$ dependent prefactor and looking at the coefficient of $\log(z)$ gives:
\begin{equation}
	-(\lambda L^{2-\Delta_{\beta}})2^{-\Delta_{\beta}} \frac{\beta^4}{4\pi}   \frac{2z^2 }{z-1}\,.
\end{equation}
We need to compare this term to the $\log(z)$ piece of the perturbed conformal block expansion
\begin{equation}
	\sum_{n=0}^{\infty} \left(c_{n}^{(0)}\right)^2 \gamma_n^{(1)}z^{2n}  F_{2+2n}(z)  = G^{(1)}(z)|_{\log z}\,.
\end{equation}
Therefore, to compute the anomalous dimension of the first double-trace operator ($n=0$), since the power series of the contribution starts at order $z^2$ and $F_{-1}(z)$ is analytic around $z=0$, with $F_{-1}(0)=1$, we get
\begin{equation}
	\gamma_{n=0}^{(1)} = -\frac{1}{2}	(\lambda L^{2-\Delta_{\beta}})2^{-\Delta_{\beta}}\frac{\beta^4}{4\pi} \frac{2}{(-1)}
	=	(\lambda L^{2-\Delta_{\beta}})2^{-\Delta_{\beta}}\frac{\beta^4}{4\pi}\,.
\end{equation}
Here, we have used $	c_{\partial \phi \partial\phi,2n=0}^{(0)}=2$. Generally, for higher dimensional double-particle operators there is a similar prefactor, but the $n$ dependence would be $\gamma^{(1)}_n \sim \frac{1}{(2n+1)(n+1)}$.
Given this anomalous dimension it is also easy to compute the associated OPE coefficient, by noticing the following
\begin{equation}
	G^{(1)}(z)|_{no-\log(z)}
	 =\sum_{n=0}^{\infty} \left(c_{n}^{(1)}\right)^2  z^{2+2n}  F_{2+2n}(z) 
	 + \left(c_n^{(0)}\right)^2 z^{2+2n} \gamma_{n}^{(1)} \frac{1}{2} \partial_{n}[ F_{2+2n}(z)] \,.
\end{equation}
Note that $\partial_{n}F_{2+2n}(z)$ starts its Taylor series at order $z^1$, so looking at the $z^2$ coefficient of this equation we get
\begin{align}
	\left[ G^{(1)}(z)|_{no-\log(z)}\right] |_{z^2}           & =
	\left(c^{(1)}_{n=0}\right)^2 \cdot (1) +\left(c^{(0)}_{n=0}\right)^2 \gamma^{(1)}_{n=0}\cdot(0)           \\
	\implies  \left[ G^{(1)}(z)|_{no-\log(z)}\right] |_{z^2} & =\left(c^{(1)}_{n=0}\right)^2\,.
\end{align}
We have
\begin{equation}
	\left[ G^{(1)}(z)|_{no-\log(z)}\right] = 	-(\lambda L^{2-\Delta_{\beta}})2^{-\Delta_{\beta}} \frac{\beta^4}{4\pi} z^2 (- \frac{2}{z} \log((1-z))) \,.
\end{equation}
Therefore, expanding the logarithm we get
\begin{equation}
	\left(c^{(1)}_{n=0}\right)^2 = -2(\lambda L^{2-\Delta_{\beta}})2^{-\Delta_{\beta}} \frac{\beta^4}{4\pi}= -2 \gamma^{(1)}_{n=0} \,.
\end{equation}

Finally, we need to extract the anomalous dimension of the external operator, as discussed when we renormalized it. The corrected 2-pt function, which is read from the disconnected piece of the 2-pt function is
\begin{equation}
	(\lambda L^{2-\Delta_{\beta}})2^{-\Delta_{\beta}} \frac{\beta^2}{\pi} \frac{\pi \log(x_{12}^2)}{x_{12}^2} \,.
\end{equation}
Now recall from before that the order $\lambda$ term from $ \frac{1}{x_{12}^{2(1+ \gamma)}}$ is $-  \gamma \frac{\log(x_{12}^2)}{x_{12}^2}$. This implies
that
\begin{equation}
	\gamma =-(\lambda L^{2-\Delta_{\beta}})2^{-\Delta_{\beta}} \beta^2 \,.
\end{equation}
In these conventions the anomalous dimension is negative, but this is  not surprising since the cosine perturbation has a negative mass.

%%%%%%%%%%%%%%%%%%%%%%%%%%%%%%%%%%%%%%%%%%%%
\subsection{Second-order perturbation theory }
\label{sec:appsecondorder}
%%%%%%%%%%%%%%%%%%%%%%%%%%%%%%%%%%%%%%%%%%%%

Now we will be interested in contributions of the form
\begin{equation}
	\lim_{\alpha \rightarrow 0} \alpha^{-4} \partial_1 \partial_2\partial_3\partial_4
	\left\langle V_{\alpha}(x_1)V_{\alpha}(x_2)V_{\alpha}(x_3)V_{\alpha}(x_4) V_{\pm \beta}(z) V_{\mp \beta}(z^*)  V_{\pm \beta}(z') V_{\mp \beta}(z'^*)  \right\rangle_{\mathbb{R}^2} \,,
\end{equation}
where we recall that $\partial_j V_\alpha (x_j)$ really means $(\partial_{y_j}V_\alpha(z_j=x_j+i y_j))|_{y_j \rightarrow 0}$. After calculating this object we must multiply by the Weyl factors and perform two integrals, over the AdS points $z_1$ and $z_2$ respectively.
As a warmup, let us consider the two-point function
\begin{equation}
	\lim_{\alpha \rightarrow 0} \alpha^{-2} \partial_1 \partial_2
	\left\langle V_{\alpha}(x_1)V_{\alpha}(x_2) V_{+ \beta}(z) V_{- \beta}(z^*)  V_{+ \beta}(z') V_{- \beta}(z'^*)  \right\rangle_{\mathbb{R}^2}~.
\end{equation}
Using our faithful companion, equation(\ref{eq-vertexcorr}), we obtain
\begin{align}
	2^{-2 \Delta_\beta} y^{-\Delta_\beta}y'^{-\Delta_\beta}\left( \eta^{\Delta_\beta} x_{12}^2 -4\Delta_{\beta}(\Pi_{1}\Pi_{2}\eta^{\Delta_\beta}+\Pi_{2}\Pi'_{1}\eta^{\Delta_\beta}+\Pi_{1}\Pi'_{2}\eta^{\Delta_\beta}+\Pi'_{1}\Pi'_{2}\eta^{\Delta_\beta})\right) .
\end{align}
Here $\Pi_i$ and $\Pi'_{i}$ are  the bulk-to-boundary propagators, but now with an index that labels the boundary point and a prime (or not) that labels the bulk point, for example: $\Pi_{1}= \frac{y}{y^2+(x-x_1)^2}$ and $\Pi'_{2}=\frac{y'}{y'^2+(x'-x_2)^2}$. Also, $\eta^{\Delta_\beta}$ plays the role of an effective bulk to bulk propagator, because $\eta= \frac{\zeta}{\zeta+4}$ is a function only of the chordal distance $\zeta= \frac{(x-x')^2+(y-y')^2}{y y'}$.
For explicitness let us also write
\begin{equation}
	\eta^{\Delta_\beta} =\left( \frac{\left(x-x'\right){}^2+\left(y-y'\right){}^2}{\left(x-x'\right){}^2+\left(y+y'\right){}^2}\right)^{\Delta_\beta}~.
\end{equation}

\begin{figure}
	\centering
	\includegraphics[width=0.7\linewidth]{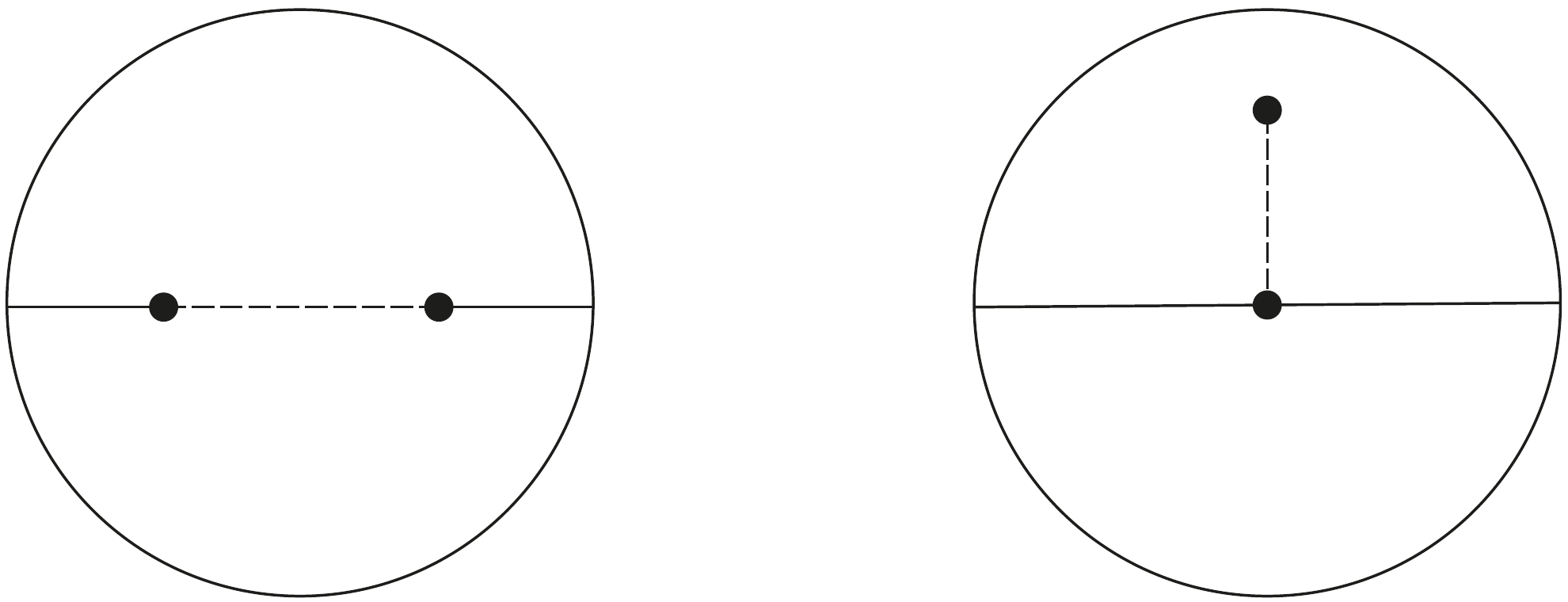}
	\caption{Connected diagrams contributing to the two-point function. The combinatorics and the $\pm$ signs of the bulk-to-bulk propagator are not explicit.}
	\label{fig:2pt}
\end{figure}

\begin{figure}
	\centering
	\includegraphics[width=0.95\linewidth]{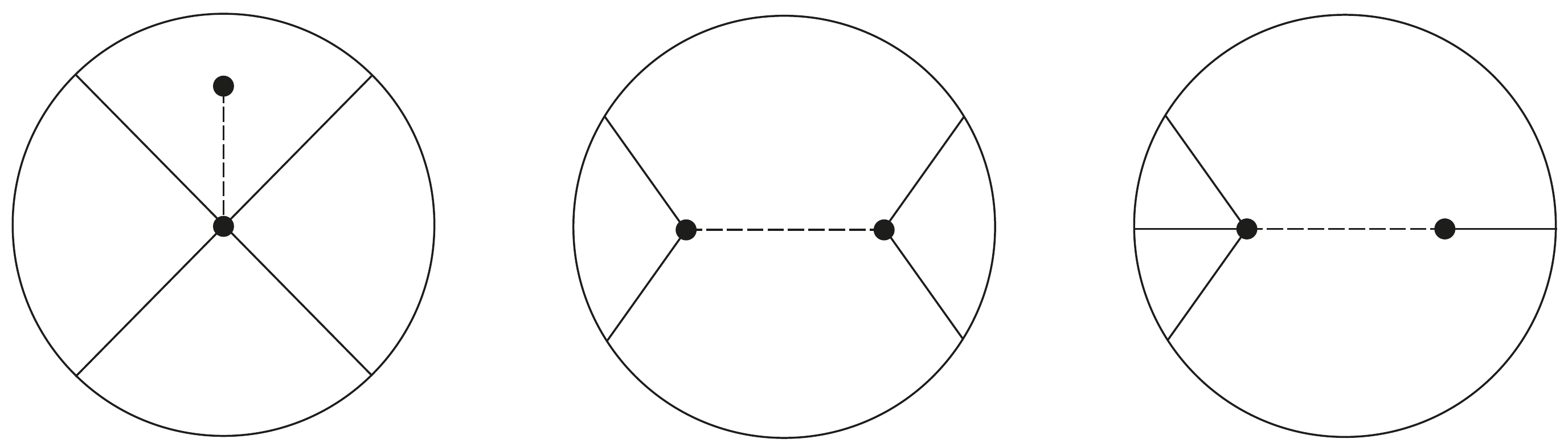}
	\caption{Connected diagrams contributing to the four-point function. The combinatorics and the $\pm$ signs of the bulk-to-bulk propagator  are not explicit.}
	\label{fig:4pt}
\end{figure}

Note that at this order there are four possible orderings for the $V_{\beta}$, which are grouped into two pairs that give the same result. The other inequivalent choice is
\begin{align}
	 & \lim_{\alpha \rightarrow 0} \alpha^{-2} \partial_1 \partial_2
	 \left\langle V_{\alpha}(x_1)V_{\alpha}(x_2) V_{+ \beta}(z) V_{- \beta}(z^*)  V_{- \beta}(z') V_{+ \beta}(z'^*)  \right\rangle_{\mathbb{R}^2} \\
	 & =(4yy')^{- \Delta_\beta} \left[\eta^{-\Delta_\beta} x_{12}^2 -4\Delta_{\beta}(\Pi_{1}\Pi_{2}\eta^{-\Delta_\beta}-\Pi_{2}\Pi'_{1}\eta^{-\Delta_\beta}-\Pi_{1}\Pi'_{2}\eta^{-\Delta_\beta}+\Pi'_{1}\Pi'_{2}\eta^{-\Delta_\beta})\right] .
	  \nonumber
\end{align}
Comparing to the first term, $\eta \rightarrow 1/ \eta$ and there is an extra minus sign on the terms where the two bulk-to-boundary propagators end in different bulk points. This structure of terms calls for a diagrammatic representation in terms of Witten Diagrams with a full line for the bulk-to-boundary propagator and a dashed line for the {\em effective bulk-to-bulk propagator} $\eta^{\Delta_\beta} \pm \eta^{-\Delta_\beta}$ (the $+$ is for an even number of bulk to boundary propagator ending in each integration point and the $-$ when there is an odd number of bulk-boundary propagators in each point), with a dot denoting the integration point and a power of $\lambda$.
In fact, the two point contributions can be written diagrammatically as in figure \ref{fig:2pt}, and the four-point as in figure
\ref{fig:4pt}.

\begin{figure}
	\centering
	\includegraphics[width=0.75\linewidth]{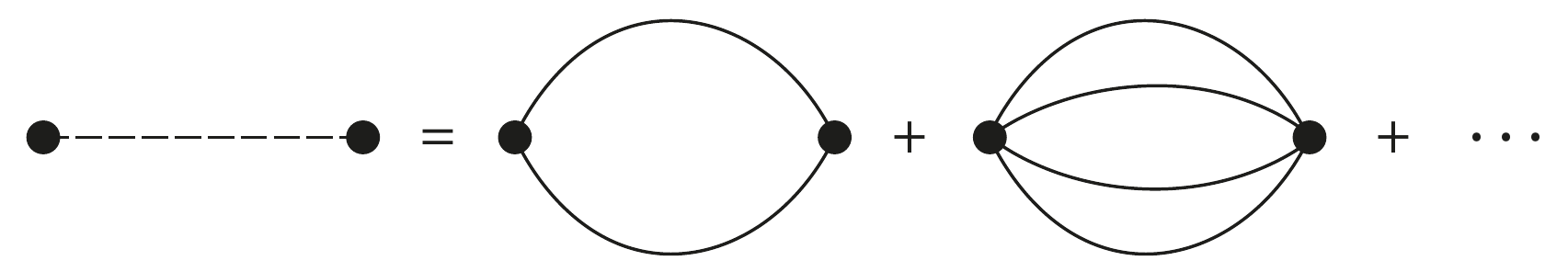}
	\caption{Graphical representation of the effective bulk-to-bulk propagator as an infinite sum of sets of $2n$ propagators of dimension 1.}
	\label{fig:bubblechain}
\end{figure}

In both cases, one must count all possible arrangements of the external points in the given diagrams and write the bulk-to-boundary propagators accordingly.
This $\eta^{\Delta_\beta} \pm \eta^{-\Delta_\beta}$ object is related to the usual bulk-to-bulk propagator, which as a function of the chordal distance given by
\begin{equation}
	G_{\Delta}= \mathcal{C}_\Delta \zeta^{-\Delta}~ _2F_1\left(\Delta,\Delta,2\Delta,\frac{-4}{\zeta}\right) ,
\end{equation} 
where we already used the fact that $d+1=2$.
The effective bulk-to-bulk propagator should somehow ressum the effects of all powers in the expansion of the cosine potential.
First, we introduce the following notation:
\begin{equation}
	g_{\beta,\pm}(\zeta)= \left( \frac{\zeta}{\zeta+4}\right) ^{\Delta_\beta}\pm~\left( \frac{\zeta}{\zeta+4}\right) ^{-\Delta_\beta}~.
\end{equation}
In fact, one can check that the effective propagator $g_{\beta,+}(\zeta)$ is an exponentiation of the single particle propagator:
\begin{align}
	g_{\beta,+}(\zeta) & =  \left( \frac{\zeta}{\zeta+4}\right) ^{\Delta_\beta}+~\left( \frac{\zeta}{\zeta+4}\right) ^{-\Delta_\beta} \nonumber       \\
	                   & = 2 \cosh \left(\frac{\beta ^2 \log \left(\frac{4}{\zeta }+1\right)}{4 \pi }\right) = 2 \cosh\Big(\beta^2 G_{\Delta=1}(\zeta)\Big) \,.
\end{align}
This provides a graphical interpretation for the effective bulk-to-bulk propagator that we represent in figure \ref{fig:bubblechain}.
Similarly, $g_{\beta,-}$ is proportional to the $\sinh$ of the single particle propagator. 

We can now proceed with the calculation. By using the isometries of AdS, most of the diagrams reduce to objects that have already appeared in the first order calculation. First, we note that the second diagram of figure \ref{fig:2pt}, can be written as
\begin{equation}
	\int_{AdS_2}d^2X\left[\int_{AdS_2}d^2X'g_{\beta,+}(X \cdot X') \right] \frac{1}{(P_1\cdot X)(P_2 \cdot X)}\,.
\end{equation}
Here, using the standard embedding formalism notation,  the $P_i$ denote boundary points and $X,X'$ the bulk integration points. Thus $P_i$ and 
$X$ are 2+1 dimensional vectors satisfying $(P_i)^2=0$ and $X^2=-L_{AdS}^2$. 
Therefore, the $X'$ integral which is an invariant function of $X$ alone must be a constant, let's say $C_0$,
\begin{equation}
	\int_{AdS_2}d^2X'g_{\beta,+}(X \cdot X') = C_0\,.
\end{equation}
As expected, this constant is infinite and must be properly regulated, but we will deal with that later. Proceeding we obtain
\begin{equation}
	C_0	\int_{AdS_2}d^2X\frac{1}{(P_1\cdot X)(P_2 \cdot X)}\,,
\end{equation}
which is proportional to the mass-shift diagram of the first order calculation.

The other diagram that contributes to the two-point function (left of figure \ref{fig:2pt}) can be written as
\begin{equation}
	\int_{AdS_2}\frac{d^2 X}{(P_1\cdot X)}\left[ \int_{AdS_2}d^2X' g_{\beta,-}(X\cdot X') \frac{1}{(P_2\cdot X')}\right] .
\end{equation}
The $X'$ integral must be an invariant function of $X$ and $P_2$ and therefore must be a function only of the scalar product $(P_2\cdot X)$, and since the function must be homogeneous of degree $-1$ with respect to $P_2$, this fixes the answer to be
\begin{equation}
	\int_{AdS_2}d^2X' g_{\beta,-}(X\cdot X') \frac{1}{(P_2\cdot X')}= \frac{C_1}{(P_2\cdot X)}\,,
\end{equation}
where $C_1$ is another (infinite) constant. The final form of the contribution is then
\begin{equation}
	C_1	\int_{AdS_2}d^2X\frac{1}{(P_1\cdot X)(P_2 \cdot X)} \,,
\end{equation}
which again was already calculated at first order.

Using these results, it is straightforward to compute the left and right diagrams of figure \ref{fig:4pt}, which contribute to the four-point function. For the left diagram, we integrate over the top point, to get
\begin{equation}
	C_0 \int_{AdS_2} d^2X \frac{1}{(P_1 \cdot X)(P_2 \cdot X)(P_3 \cdot X)(P_4 \cdot X)}\,,
\end{equation}
which is proportional to a contact Witten diagram which has already appeared. Similarly, on the right hand side diagram, by performing the integral over the right-most point, we will be left with
\begin{equation}
	C_1 \int_{AdS_2} d^2X \frac{1}{(P_1 \cdot X)(P_2 \cdot X)(P_3 \cdot X)(P_4 \cdot X)}\,,
\end{equation}
which again has been calculated. This leaves the middle diagram. By using the spectral representation
\begin{equation}
	g_{\beta,\pm}(X \cdot X') = \int_{-\infty}^{\infty} d \nu \tilde{g}_{\beta,\pm}(\nu) \Omega_{i \nu}(-\cosh(\rho)) \,,
\end{equation}
where we have used the isometries of AdS to set one of the points at the {\em center} in global coordinates, such that $X\cdot X'=- \cosh \rho$.
We are left with a standard calculation familiar from exchange Witten diagrams:
\begin{equation}
	\int_{-\infty}^{\infty} d \nu \tilde{g}_{\beta,\pm}(\nu)\int_{AdS_2}d^2X d^2X'\frac{1}{(P_1\cdot X)(P_2\cdot X)} \,\Omega_{i \nu}\big(-\cosh(\rho)\big) \,\frac{1}{(P_3\cdot X')(P_4\cdot X')}~.
\end{equation}
Using the split representation for the harmonic function, with \linebreak $\Pi_{\frac{d}{2}+ i \nu}(P_0,X)= (P_0\cdot X)^{-\frac{d}{2}-i \nu}$,
\begin{equation}
	\Omega_{i\nu}\big(X\cdot X'\big)=\frac{\nu^{2} \sqrt{\mathcal{C}_{\frac{d}{2}+i \nu} \mathcal{C}_{\frac{d}{2}-i \nu}}}{\pi} \int d P_{0} \Pi_{\frac{d}{2}+i \nu}\left(P_{0}, X\right) \Pi_{\frac{d}{2}-i \nu}\left(P_{0}, X'\right) .
\end{equation}
We can perform the integral over the AdS points which are proportional to 3-pt functions in the CFT. One is left with the spectral integral, and the integral over the boundary, introduced by the split representation:
\begin{equation}
	\int_{-\infty}^{\infty} d \nu\, \frac{ \tilde{g}_{\beta,\pm}(\nu) \alpha(\nu)}{\left(P_{12}\right)^{\Delta-\frac{1}{4}-\frac{i \nu}{2}}\left(P_{34}\right)^{\Delta-\frac{1}{4}+\frac{i \nu}{2}}} \int \frac{ dP_{0}}{\left(P_{10}\right)^{\frac{1}{4}+\frac{i \nu}{2}}\left(P_{20}\right)^{\frac{1}{4}+\frac{i \nu}{2}}\left(P_{30}\right)^{\frac{1}{4}-\frac{i \nu}{2}}\left(P_{40}\right)^{\frac{1}{4}-\frac{i \nu}{2}}}\,.
\end{equation}
Here $\alpha(\nu)$ is a completely kinematical object, which has, however, poles in $\nu$ (they will be related to the double trace contribution to this diagram), and $\Delta=1$ is the free dimension of the external operator, kept general for clarity.
The $P_0$ integral is the shadow representation of the conformal partial wave, so the result becomes
\begin{equation}
	\label{eq:allpoles}
	\int_{-\infty}^{\infty} d \nu \,\frac{\tilde{g}_{\beta,\pm}(\nu)}{\left(P_{12}\right)^{\Delta}\left(P_{34}\right)^{\Delta}}  \frac{\Gamma_{\Delta-\frac{d}{4}-\frac{i \nu}{2}}^{2} \Gamma_{\Delta-\frac{d}{4}+\frac{i \nu}{2}}^{2} }{64 \pi^{\frac{d}{2}+1} \Gamma_{\Delta}^{2} \Gamma_{1-\frac{d}{2}+\Delta}^{2}}\left[\frac{\Gamma_{\frac{d}{4}+\frac{i \nu}{2}}^{4} \mathcal{G}_{\frac{d}{2}+i \nu}(z, \overline{z})}{\Gamma_{\frac{d}{2}+i \nu} \Gamma_{i \nu}}+\frac{\Gamma_{\frac{d}{4}-\frac{i \nu}{2}}^{4} \mathcal{G}_{\frac{d}{2}-i \nu}(z, \overline{z})}{\Gamma_{\frac{d}{2}-i \nu} \Gamma_{-i \nu}}\right].
\end{equation}
We have used $\mathcal{G}$ to denote the usual conformal block, which is really only a function of one cross-ratio in 1d. We also used $\Gamma_a \equiv \Gamma(a)$ to save space and everywhere we should set $d=1$. It is important to note the existence of double trace poles in the overall Gamma functions. The only thing left to determine is $\tilde{g}_{\beta,\pm}(\nu)$.

%%%%%%%%%%%%%%%%%%%%%%%%%%%%%%%%%%%%%%%%%%%%
\subsubsection{Evaluating the AdS diagrams}
%%%%%%%%%%%%%%%%%%%%%%%%%%%%%%%%%%%%%%%%%%%%

Let us know study the integrals in detail. First we consider
\begin{equation}
	\int_{AdS_2}d^2Xg_{\beta,+}(X \cdot X') \,.
\end{equation}
Since this is a constant, we can choose the location of $X'$ at our convenience. In particular, in global coordinates, with $X'$ at the center, we have $X\cdot X'= - \cosh \rho$ and, using $\cosh \rho= 1 + \frac{\zeta}{2}$, we can write
\begin{equation}
	\label{eq:firstint}
	\int_{0}^{\infty}\int_{0}^{2 \pi} d\theta d\rho \sinh \rho \left[ \left( \frac{\cosh \rho -1}{\cosh \rho +1}\right) ^{\Delta_{\beta}} + (\Delta_\beta \rightarrow -\Delta_\beta)\right]  .
\end{equation}
Let us focus on the first term. The integral is manifestly rotationally invariant, so we have
\begin{equation}
	2 \pi \int_{0}^{\infty} d \rho \sinh\rho\left( 1-\frac{2}{1+\cosh \rho}\right)^{\Delta_\beta} \,.
\end{equation}
The expression is now amenable to generalized binomial expansion, which is convenient, because it makes the integral easy to compute, but mostly because it provides a natural way to study the IR divergences, and to renormalize UV divergences by a suitable analytic continuation in $\Delta_\beta$. To see why, let us  note that in \eqref{eq:firstint}, as $\rho \rightarrow 0$ the integrand goes to 0, since $\Delta_\beta\geq0$, so there is no UV divergence for this term. When $\Delta_\beta \rightarrow -\Delta_\beta$, we have a UV Divergence for $\Delta_{\beta}>1$, but we can just analytically continue the result for positive $\Delta$, which essentially amounts to performing the binomial expansion with power $-\Delta_{\beta}$.

Next, for the IR  there is an obvious problem. When $\rho \rightarrow \infty$, the propagator goes to 1 and the measure makes the integral blow up exponentially at large $\rho$, this is easily dealt with by subtracting the constant, but, in fact, it is easy to just introduce a hard cutoff $L$  and use the binomial expansion. This isolates the constant, and also shows that there is another, weaker divergence, which is linear in $L$. This should be thought of as an anomalous dimension log-like divergence, since the leading divergence is exponential in $L$, corresponding to the second term in the expansion. After that all the integrals converge and we can resum back the binomial expansion. We obtain, not writing the overall factor of $2 \pi$,
\begin{equation}
	\label{eq:int}
	\left(\frac{e^L}{2}-1\right)+ \Big(4 \Delta_{\beta}\log(2)-2\Delta_\beta L\Big)+2\Delta_{\beta}\Big( H(\Delta_\beta)-1\Big)+ O\Big(e^{-L}\Big)\,.
\end{equation}
Equivalently, the integral can be done directly, and it is of hypergeometric type. After expanding at large values of the cutoff, one also recovers (\ref{eq:int}). 
The terms in  (\ref{eq:int}) are grouped by their order in the binomial expansion, with the last one ressuming from the third term to infinity. $H(\Delta) = \gamma + \Psi(\Delta+1)$ is the analytic continuation of the Harmonic numbers, with $\gamma$ the Euler-Mascheroni constant and $\Psi(a)= \Gamma'(a)/\Gamma(a)$, the DiGamma function. We can now analytically continue to negative $\Delta_\beta$ and add the contribution of the second term, yielding, finally
\begin{equation}
	C_0=2\pi \left( 2(\frac{e^L}{2}-1) + 2\Delta_\beta(\frac{1}{\Delta_\beta }-\pi  \cot (\pi  \Delta_\beta ))\right) .
\end{equation}
Subtraction of the constant value at infinity gets rid of the first term in the sum inside the bracket.

For the next integral we have
\begin{equation}
	\int_{AdS_2}d^2X g_{\beta,-}(X\cdot X') \frac{(P_2\cdot X')}{(P_2\cdot X)}= C_1\,.
\end{equation}
Making the same choice as before, $\rho'=0$, gives
\begin{equation}
	\int_{0}^{\infty}\int_{0}^{2 \pi} d\theta d\rho \sinh \rho \left[ \left( \frac{\cosh \rho -1}{\cosh \rho +1}\right) ^{\Delta_{\beta}} - (\Delta_\beta \rightarrow -\Delta_\beta)\right] 
	\frac{1}{\cosh\rho - \sinh \rho \cos(\theta- \theta_2)} \,.
\end{equation}
Since the function is periodic in $\theta$, we can shift $\theta\rightarrow \theta + \theta_2$, without changing the integration region. (Note that our parametrization is $X=(-\cosh \rho, \sinh \rho \cos(\theta),\sinh \rho \sin \theta)$ and $P_2=(-1,\cos(\theta_2),\sin(\theta_2)$). The $\theta$ integral just gives $2 \pi$, as the $\rho$ dependence cancels out, and we are left with exactly the same result as in the previous integral, but with a relative minus sign between the $+\Delta_\beta$ and the $-\Delta_\beta$ terms. Namely
\begin{equation}
	C_1=2 \Big(4 \Delta_{\beta}\log(2)-2\Delta_\beta L\Big) + 2\Delta_\beta\Big(2(\gamma-1)+\Psi(1+\Delta_\beta)+\Psi(1-\Delta_\beta)\Big)\,.
\end{equation}
In this case there is no volume term, as the constant terms cancel at infinity, but  one would still need to account the first non-zero term in the binomial expansion, which corresponds to the $log^2$ singularity in second order perturbation theory for the anomalous dimension.

Now we just need to compute the spectral representation of $g_{\beta,+}(-\cosh\rho)$.
In $H_{d+1}$ we have
\begin{equation}
	\tilde{g}(\nu) = \frac{2 \pi^{\frac{d+1}{2}}}{\Gamma(\frac{d+1}{2})} \int_{0}^{\infty} d \rho \sinh(\rho)^{d}~_2F_1\left( \frac{d}{2}-i\nu,\frac{d}{2}+i\nu;\frac{d+1}{2};-\sinh(\rho/2)^2\right)  g(\rho)\,.
\end{equation}
It is convenient to notice the following identity
\begin{equation}
	\label{eq:3f2asint2f1}
	_{3} F_{2} \left[ \begin{array}{c}{a_{1},a_{2}, c} \\ {b_{1}, d}\end{array};z\right]=\frac{\Gamma(d)}{\Gamma(c) \Gamma(d-c)} \int_{0}^{1} t^{c-1}(1-t)^{d-c-1}~_{2}F_{1} \left[ \begin{array}{c}{a_{1}, a_{2}} \\ {b_{1}}\end{array};tz\right] d t \,,
\end{equation}
and to change to the   variable $x=4/(4+\xi)$. Details of the transform for a power of the chordal distance were given in appendix B of \cite{Carmi:2018qzm}.
Following a similar calculation, the spectral  transform for our effective propagator is given by
\begin{equation}
	\frac{2 \pi^{\frac{d+1}{2}}}{\Gamma(\frac{d+1}{2})}2^{d} \int_{0}^{1} d x x^{-\frac{d+3}{2}+\Delta_\beta}\left(\frac{1}{x}-1\right)^{\frac{d-1}{2}+\Delta_\beta}~ _{2}F_{1}\left(\frac{d}{2}+i \nu, \frac{d}{2}-i \nu, \frac{d+1}{2}, \frac{x-1}{x}\right) .
\end{equation}
Now it is convenient to use a Pfaff identity for the $_2F_1$
\begin{equation}
	_{2} F_{1}(a, b ; c ; z)=(1-z)^{-b}~ _{2}F_{1}\left(c-a, b ; c ; \frac{z}{z-1}\right) ,
\end{equation}
Using in the identity above $z= (x-1)/x$, we get an extra power of $x$ and a Hypergeometric of argument $1-x$. Finally we can change the integration variable to $x'=1-x$ and we get an integral exactly of the form of (\ref{eq:3f2asint2f1}). With this technique we can easily reproduce the results of \cite{Carmi:2018qzm}. Furthermore, the result for our effective propagator is (note that this avoided any singularities as $d\rightarrow1$) 
\begin{equation}
	\label{eq:niceformula}
	4 \pi  \Gamma (\Delta_\beta +1) \Gamma \left(-i \nu -\frac{1}{2}\right) \,
	_3\tilde{F}_2\left(\frac{1}{2}-i \nu ,\Delta_\beta +1,\frac{1}{2}-i \nu ;\Delta_\beta -i \nu
	+\frac{1}{2},1;1\right) .
\end{equation}
Note that the hypergeometric is only balanced for $Im(\nu)<-1/2$. We will find a hypergeometric transformation which provides a suitable analytic continuation and furthermore restores manifest $\nu \leftrightarrow -\nu$ symmetry.
The one that gets the job done is
\begin{equation}
	_{3} \tilde{F}_{2}\left[a,b,c;e,f;1\right] ~=~ _{3}\tilde{F}_{2}\left[e-c,f-c,r;r+a,r+b;1\right] \frac{\Gamma(r)}{\Gamma(c)}\,,
\end{equation}
with $r=e+f-a-b-c$.
The balance of this $_3F_2$ is $1+\Delta_\beta$, which means it converges for all positive values of $\Delta_\beta$ (in the full $\nu$ plane).  Furthermore, for $\Delta_\beta<1$ 
we can add the negative power piece since it will still converge. This means that we have the final answer
\begin{equation}
	\label{eq:spectransf}
	\tilde{g}(\nu)=-4 \pi ^2 \Delta_\beta  \text{sech}(\pi  \nu ) \, _3F_2\left(1-\Delta \beta ,\frac{1}{2}-i \nu ,i \nu +\frac{1}{2};1,2;1\right) + (\Delta_\beta \leftrightarrow -\Delta_\beta) \,,
\end{equation}
where we added the piece with $\Delta_\beta \leftrightarrow - \Delta_\beta$. We have checked that this expression has simple poles at $\frac{1}{2}+i \nu=2+2n$, and only there. It looks like it also has poles at $1+n$ in general, but the negative $\Delta_\beta$ term cancels the poles at odd exchanged dimension, which we know cannot exist. The simple poles will multiply the double pole already present from (\ref{eq:allpoles}), and generate triple poles, which will give second derivatives of the conformal block with respect to dimension, that are 
associated to both $\log^2$ and $\log$ terms which are important for anomalous dimensions.
In particular, if we pick  the pole at $i \nu= 3/2$, which corresponds to the first double-particle operator ($\Delta=2$), we get the expected $\log^2,\log$ and regular term. The $\log^2$ piece is
\begin{equation}
	\frac{3 i \pi  \Delta_\beta ^2 ((z-2) \log (1-z)-2 z) \log ^2(z)}{z} \,,
\end{equation}
whose small $z$ expansion starts with a  $z^2 \log^2(z) $ term, as expected from the computation below. (We are everywhere failing to write a prefactor of $\Delta_\beta^2 2^{4-2 \Delta_\beta}$ that comes from the vertex operator calculation). In fact, comparing to (\ref{eq:check}) below, the result has the right $\beta$ dependence. This is consistent with the conformal block expansion, which relates the coefficient to the first order anomalous dimension squared.

For convenience, we write here the second order expansion of the conformal block decomposition, which determines the second order CFT data. The  conformal block expansion is
\begin{equation}
	\label{eq:CBdecomposition}
	\sum_{\Delta' \in S}c_{\phi \phi \Delta'}^2 \,  z^{\Delta'}F_{\Delta'}(z) = G(z)\,,
\end{equation}
where we use again the short hand $_2F_1(\Delta,\Delta,2\Delta,z)\equiv F_\Delta(z)$.
Our spectrum is
\begin{equation}
	\Delta' = \Delta_n= 2+2n + \lambda \gamma_n^{(1)}+\lambda^2 \gamma_n^{(2)} \,,
\end{equation}
We have of course already computed $\gamma_0^{(1)}$. The OPE coefficients squared are written as
\begin{equation}
	c_{\phi\phi\Delta'}^2=c_n^2=  \left(c_{n}^{(0)}\right)^2 +\lambda \left(c_{n}^{(1)}\right)^2 + \lambda^2 \left(c_n^{(2)}\right)^2 \,,
\end{equation}
and the correlation function computed in perturbation theory as 
\begin{equation}
	G(z)= G^{(0)}(z) + \lambda G^{(1)}(z) + \lambda^2 G^{(2)}(z) \,.
\end{equation}
Expanding the $z^{\Delta'}$ term in (\ref{eq:CBdecomposition}) will generate $\log(z)$ and $\log(z)^2$ terms, which satisfy a separate equation. The $\log^2(z)$ terms give
\begin{equation}
	\label{eq:log2}
	\sum_n \left(c_n^{(0)}\right)^2 z^{2+2n}\frac{1}{2!} \left(\gamma_n^{(1)}\right)^2  F_{2+2n}(z) = G^{(2)}(z)|_{\log^2z} \,.
\end{equation}
In particular, the power of $z^2$ fixes a relation with the first order data of the $(\partial_\perp \phi)^2$ operator
\begin{equation}
	\label{eq:check}
	\left(c_{0}^{(0)}\right)^2 \frac{1}{2} \left(\gamma_0^{(1)}\right)^2 =G^{(2)}(z)|_{\log^2z}|_{z^2}\,.
\end{equation}
This is a non-trivial consistency check. The $\log(z)$ equation already fixes the second order anomalous dimension
\begin{equation}
	\sum_n z^{2+2n} \left[ \left( \left(c_n^{(1)}\right)^2\! \gamma_n^{(1)} + \left(c_n^{(0)}\right)^2  \!\gamma_n^{(2)} \right) F_{2+2n}(z) +\left(c_n^{(0)}\gamma_n^{(1)}\right)^2\frac{1}{2}\partial_n F_{2+2n}(z)\right] = G^{(2)}(z)|_{\log z}\,.
\end{equation}
Again the power of $z^2$ is enough to determine the first operator
\begin{equation}
	\left(c_0^{(1)}\right)^2 \gamma_0^{(1)} + \left(c_0^{(0)}\right)^2\gamma_0^{(2)}=G^{(2)}(z)|_{\log z}|_{z^2}\,.
\end{equation}
Finally, the equation for the regular term gives
\begin{align}
	 & \sum_n z^{2+2n}\left[ \left(c_n^{(2)}\right)^2 F_{2+2n}(z) + \left(c_n^{(1)}\right)^2 \frac{\gamma_n^{(1)}}{2} \partial_n F_{2+2n}(z) \right.                                                 \\
	 & \left.   \phantom{ \frac{\gamma_n^{(1)}}{2}}
	 + \left(c_n^{(0)}\right)^2\left(\gamma_n^{(2)}\frac{1}{2}\partial_n F_{2+2n}(z) + \frac{1}{8}\left(\gamma_n^{(1)}\right)^2 \partial_n^2 F_{2+2n}(z)\right)\right]=G^{(2)}|_{reg}\,.
\end{align}
It then follows that the $z^2$ piece fixes the OPE coefficient
\begin{equation}
	\left(c_0^{(2)}\right)^2 = G^{(2)}(z)|_{reg}|_{z^2}~.
\end{equation}

The previous expansion encapsulates the $\lambda$ dependence, but we still have a parameter $\beta$. Thus
 it is also convenient to expand  in $\beta$ to cross-check the calculation with $\phi^n$ theories.
We have that, order by order in a small $\beta$ expansion, the effective propagator generates products of the single particle propagator, as expected from expansion of the potential
$\cos(\beta \phi)= 1- \frac{\beta^2}{2}\phi^2 + \frac{\beta^4}{4!}\phi^4- \frac{\beta^6}{6!}\phi^6+ \dots $. We might wonder if this property holds after the spectral transform, and indeed it does. By taking the first piece (this corresponds to the exponential instead of the $\cosh$ of the single propagator) of (\ref{eq:spectransf}), and expanding in small $\beta$, the first term is proportional to
\begin{equation}
	\tilde{G}(\nu) = \frac{1}{\nu^2+(1-\frac{1}{2})^2} \,,
\end{equation}
which is the spectral representation of the propagator of a  scalar field dual to an operator of dimension 1 in $CFT_1$.
The next term is
\begin{equation}
	\frac{2 \left(H\big(-\frac{i \nu }{2}-\frac{1}{4}\big)+H\big(\frac{i \nu }{2}-\frac{1}{4}\big)+\log (4)\right)}{4 \pi  \nu ^2+\pi } \,,
\end{equation}
which matches with the spectral function for the product of two propagators (as in a bubble diagram), which was computed in \cite{Carmi:2018qzm}. This seems like a non-trivial check, and makes it reasonable to propose that the formula (\ref{eq:spectransf}) is a generating function (by expansion in $\beta$) of the spectral representation of any number of propagators. Although it always has poles in the double-particle locations, and a higher number of propagators should correspond to multi-particle poles, this is compatible, because double/multi-particle operators are degenerate for external dimension 1, as in our case.

Furthermore, with this spectral function one can pick the poles in the spectral integral of (\ref{eq:allpoles}) and get the conformal block decomposition. We can look, for simplicity, to the coefficient of $\log^2(z)~ z^{2+2n}~ _2F_1(2+2n,2+2n,4+4n,z)$ in this expansion and read off
\begin{equation}
	 \frac{\left(c^{(0)}_n\right)^2 \! \!i \pi  \Delta \beta  \Big(\!\, _3F_2(-2 n-1,2 n+2,1-\Delta \beta ;1,2;1)-\!\,_3F_2(-2 n-1,2 n+2,\Delta \beta +1;1,2;1)\Big)}{8 (n+1) (2 n+1)},
\end{equation}
where we factorized the free theory OPE coefficients, to make the comparison to (\ref{eq:log2}) easier. In particular, the remaining terms should be first order anomalous dimensions squared. Indeed, in the small $\beta$ expansion, to first non-trivial order, one recovers the result from $\phi^4$ theory $\gamma_n \propto 1/((n+1)(2n+1))$. However, there are interesting corrections from higher orders in $\beta$, which should correspond to first order anomalous dimensions of multi-particle operators, which are generated by the $\phi^{2n}$ $2n$-point functions ($2n$-point contact diagrams). This is not visible in the four-point function at first order. More rigorously, we have mixing among multi-particle operators and the results should be interpreted as averages over degenerate operators. We can also try computing these anomalous dimension averages at finite $\beta$, for special values of $\Delta_\beta$ where the equations simplify. For example, for $\Delta_\beta=1/2$ we get
\begin{equation}
	\langle(\gamma_n^{(1)})^2\rangle = \frac{i \pi  \left(\left(\frac{1}{2}\right)_n\right){}^2}{16 \left((2)_n\right){}^2}\,,
\end{equation}
whose large $n$ behavior is $\langle\gamma_n\rangle \sim 1/n^{3/2}$. One can study the general large $n$ behavior of these dimension for general $\Delta_\beta$ and obtains
\begin{equation}
	\langle\gamma_n\rangle \sim \frac{1}{n^{2-\Delta_\beta}} \,.
\end{equation}
This follows the general expectations of \cite{Heemskerk:2009pn, Fitzpatrick:2010zm}, which essentially states that the large $n$ behaviour of the anomalous dimensions is controlled by the mass dimension of the bulk coupling. It appears that this is not visible in the 4-point function at first order (where only the $\phi^4$ term contributes), because effectively the beta expansion truncates at $\beta^4$, which corresponds to $\Delta_\beta \rightarrow 0$ and gives $\gamma_n \sim \frac{1}{n^{2-0}}$. More carefully, this means that the solution to the mixing problem is not fixed by the first order single-particle correlator, which is compatible with a pure $\phi^4$ interaction and an only two-particle spectrum. When we go to second order in $\lambda$, the $n$-particle interactions kick in, and the mixing problem becomes apparent, bringing all multi-particle operators to the limelight.

Note that to analyze the $\log^2 z$ behavior it was enough to study the s-channel block expansion of the s-channel generalized bubble diagram. This is because the t- and u- channel blocks can analogously be expanded in their respective channel's conformal blocks, which have only single-log singularities in the $s-$channel OPE limit. Equivalently, we can take the s-channel block expansion and consider the behavior of the blocks around the t- and u- OPE limits.
To simplify this procedure, it is important to notice that the s-channel bubble diagram is invariant under permutations of the external points $x_1$ and $x_2$. This means that the u-channel contribution is directly related to the t-channel, so it is enough to consider the t-channel OPE limit and include a factor of 2. In fact, by using invariance of the s-channel diagram under the permutation $x_1 \leftrightarrow x_2$ one can derive
\begin{equation}
	G_{(s)}(z)= G_{(s)}\left(\frac{z}{z-1}\right) ,
\end{equation}
where $G_{(s)}$ denotes the s-channel generalized bubble diagram. In fact, by further using permutations to get to the other channels, one obtains
\begin{equation}
	G_{(t)}\left(\frac{z}{z-1}\right)= G_{(u)}(z)\,,
\end{equation}
From which it is clear that the behaviour as $z\rightarrow0$ of the two channels is the same.

Unlike the case for the $\log^2 z$ terms,  the t-channel contributes to both the $\log z$ and regular terms, which means it will contribute to the second order anomalous dimension and the second order OPE coefficient. Furthermore, we have that the t-channel OPE limit of the s-channel blocks is given by
\begin{equation}
	z^\Delta~ _2F_1(\Delta,\Delta,2\Delta,z) \sim	-\frac{\Gamma (2 \Delta ) \left(2 \psi ^{(0)}(\Delta )+\log (1-z)+2 \gamma \right)}{\Gamma
		(\Delta )^2} + O(1-z)~,
\end{equation}
which means that all operators of all dimensions contribute at the same order in the small $z$ expansion, so one needs to perform an infinite sum in the t-channel to get the contribution for one operator in the s-channel. Given the form of the spectral function (\ref{eq:spectransf}), for general $\Delta_\beta$ these sums are hard to perform explicitly (we computed the sum over residues numerically for several values of $\Delta_\beta$), but in the small beta expansion, where the leading contribution comes from $\phi^4$ bubble diagrams, we were able to reproduce the known loop data
\begin{equation}
	\gamma_0^{(2)}= -\frac{1+4 \zeta(3)}{2} \,,  \qquad c_0^{(2)}= \frac{\pi^4}{15} + \frac{7}{2}\,.
\end{equation}
In our conventions, the normalization is actually proportional to $\beta^8$, as expected from expanding the cosine potential and counting powers of $\beta$ in the $\phi^4$ bubble diagram, but in our normalization this gets divided by the square  of $\gamma_0^{(1)}$. 

\section{Multiple correlators and numerical bounds}
\label{app:multicorrbounds}
In \cite{Homrich:2019cbt} the correlation functions of two operators were analyzed, which we will call $\phi$ and $\chi$. It was assumed that there existed a $\Z_2$ symmetry under which $\phi$ is odd and $\chi$ is even. With an eye towards the flat-space limit, the assumed OPEs were
\begin{equation}
\begin{split}
\phi \times \phi &= \mathbf 1 + \lambda_{\phi \phi \chi} \chi + (\ldots \text{operators with }\De > 2 \De_{\phi} \ldots)  \,,\\
\phi \times \chi &= \phantom{\mathbf 1 + \vphantom{1}}\lambda_{\phi \phi \chi} \phi + (\ldots \text{operators with }\De > \De_{\phi} + \De_{\chi} \ldots)\,,\\
\chi \times \chi &= \mathbf 1 + \lambda_{\chi \chi \chi} \phi + (\ldots \text{operators with }\De > 2 \De_{\phi} \ldots)\,.\\
\end{split}
\end{equation}
Also, both $\phi$ and $\chi$ were assumed to be Lorentz scalars, which in one dimension simply means that they are parity even.

Section 4 in \cite{Homrich:2019cbt} was concerned with obtaining upper bounds on the couplings $\lambda_{\phi \phi \chi}$ and $\lambda_{\chi \chi \chi}$ from the conformal bootstrap, extrapolating these to the flat-space limit, and comparing them with multi-amplitude S-matrix bootstrap bounds that were also obtained in that paper. Since operator ordering matters in one Euclidean dimension, the correlation functions that were analyzed were:
\begin{equation}
\<\phi \phi \phi \phi\> \,,  \quad\<\phi \phi \chi \chi\>\,, \quad\< \phi \chi \phi \chi\>\, ,\quad\<\chi \chi \chi \chi\>\,,
\end{equation}
and the authors of \cite{Homrich:2019cbt} also analyzed the corresponding flat-space amplitudes
\begin{equation}
S_{\phi \phi \to \phi \phi}\,, \quad  S_{\phi \phi \to \chi \chi} \qquad \text{ and } \qquad S_{\phi \chi \to \chi \phi}\, , \quad S_{\phi \chi \to \phi \chi}\, ,\quad  S_{\chi \chi \to \chi \chi}\,,
\end{equation}
with analytic S-matrix bootstrap methods.

Although in many cases a good match between the two bootstrap approaches was found, this was no longer true when the mass ratio $m_2/m_1$ was slightly larger than about $\sqrt{2}$. (In fact, tested points were 1.5 and 1.6, and stability requires $m_2 / m_1  < 2$.) For these mass ratios the multi-correlator analysis resulted in exactly the same bound as that obtained from $\< \phi \phi \phi \phi \>$ alone. On the other hand, the S-matrix bootstrap method applied to just the $S_{\phi \chi \to \chi \phi}$ scattering amplitude already resulted in a bound that was significantly better, up to about a factor of three. (This problem was quite general, but for the particular case where $\lambda_{\chi \chi \chi}$ is assumed to equal $- \lambda_{\phi \phi \chi}$ it is clearly illustrated on the right-hand side of figure 12 of \cite{Homrich:2019cbt}.)

This difference leads to a natural puzzle: if correlators become scattering amplitudes in the flat-space limit, then why do bounds obtained from correlators not always reduce to bounds obtained from amplitudes? In the next few paragraphs we explain the resolution to this puzzle. It will also help us to understand why many of the multi-correlator bounds in the main text do not improve on the single-correlator bounds.

If $\lambda_{\phi \phi \chi}$ saturates the single-correlator bound then the solution to the $\<\phi \phi \phi \phi\>$ crossing equation must be the solution that converges to the sine-Gordon amplitude in flat space. Our aim is now to show that the other crossing equations can also be solved if $\De_\chi / \De_\phi$ is large enough, and therefore yield no further constraints on $\lambda_{\phi \phi \chi}$.

We begin with the $\< \chi \chi \chi \chi\>$ crossing equation. This equation in itself is decoupled from the $\< \phi \phi \phi \phi\>$ equation. For the present discussion we only need to assume that this bound is weak, in the sense that if we fix
\begin{equation}
\alpha = \frac{\lambda_{\chi \chi \chi}}{\lambda_{\phi \phi \chi}}
\end{equation}
and use it to trade $\lambda_{\chi \chi \chi}$ for $\lambda_{\phi \phi \chi}$, then the bound obtained from the $\< \chi \chi \chi \chi\>$ correlator is weaker than that obtained from the $\< \phi \phi \phi \phi \>$ correlator.\footnote{In fact, we can observe that the maximization of $\lambda_{\chi \chi \chi}$ is precisely the same as that of scenario II of \cite{Paulos:2016fap}. In that paper it was shown that there was no upper bound (in the flat-space limit) as soon as the gap, which in our case is $2 m_\phi$, was smaller than $\sqrt{3} m_\chi$. Therefore, for 	$\De_\chi/\De_\phi > 2/\sqrt{3} \approx 1.15$ and sufficiently close to the flat-space limit this correlator in itself does not give us a useful bound at all. The assumption stated in the main text is therefore certainly satisfied.}

Now consider the $\<\phi \phi \chi \chi\>$ correlator. Since its $s$-channel conformal block decomposition features coefficients of the form $\lambda_{\phi \phi k}\lambda_{\chi \chi k}$, it can only feature operators that appear \emph{both} in the $\<\phi \phi \phi \phi \>$ four-point function and in the $\< \chi \chi \chi \chi\>$ four-point function. This provides a non-trivial link between the correlation functions under normal circumstances, but we will not outline a loophole that can avoid this connection.

The main idea is that there might exist solutions to the crossing equations that exist {\em purely in the continuum part of the spectrum}. For example, consider the crossing symmetry equation for $\< \chi \chi \chi \chi \>$,
\begin{equation}
(1-z)^{2 \De_{\chi}} \left( 1 + \lambda^2_{\chi \chi \chi}  \,g(\De_\chi,z) + \sum_{k,\,\De_k \geq 2 \De_\phi} \lambda^2_{\chi \chi k} \,g(\De_{k},z) \right) = (z \leftrightarrow 1 - z)\,,
\end{equation}
and suppose there exists a function $f_{\chi}(z)$ that obeys
\begin{equation}
\label{fchiconditions}
\begin{split}
f_{\chi}(z) &= \sum_{k,\,\De_k \geq 2 \De_\phi} \mu_{k}^2 \, g(\De_{k},z)\,,\\
(1-z)^{2\De_\chi} f_\chi(z) &= (z \leftrightarrow 1 - z)\,,
\end{split}
\end{equation}
thus this  function has  a conformal block decomposition obeying crossing and unitarity but without the fixed part consisting of the identity and, in this case, the block corresponding to $\chi$ itself. Then we can add this function with an \emph{arbitrarily large (positive) coefficient} to the $\< \chi \chi \chi \chi \>$ equation without violating the bootstrap axioms. For the system of correlators at hand, doing so buys us the freedom to add any operators in $f_\chi(z)$ to the $\< \phi \phi \chi \chi \>$ correlation function as well. Indeed, even if the operators in $f_\chi(z)$ do not strictly speaking appear in the $\< \phi \phi \phi \phi \>$ four-point function, we can imagine adding them there with a very small coefficient, and if we simultaneously add $f_\chi(z)$ with a very large coefficient to $\<\chi \chi \chi \chi\> $ then we can get these operators with an arbitrary coefficient in the $s$-channel of $\< \phi \phi \chi \chi\>$.

\begin{figure}
	\centering
	\includegraphics[width=0.42\linewidth]{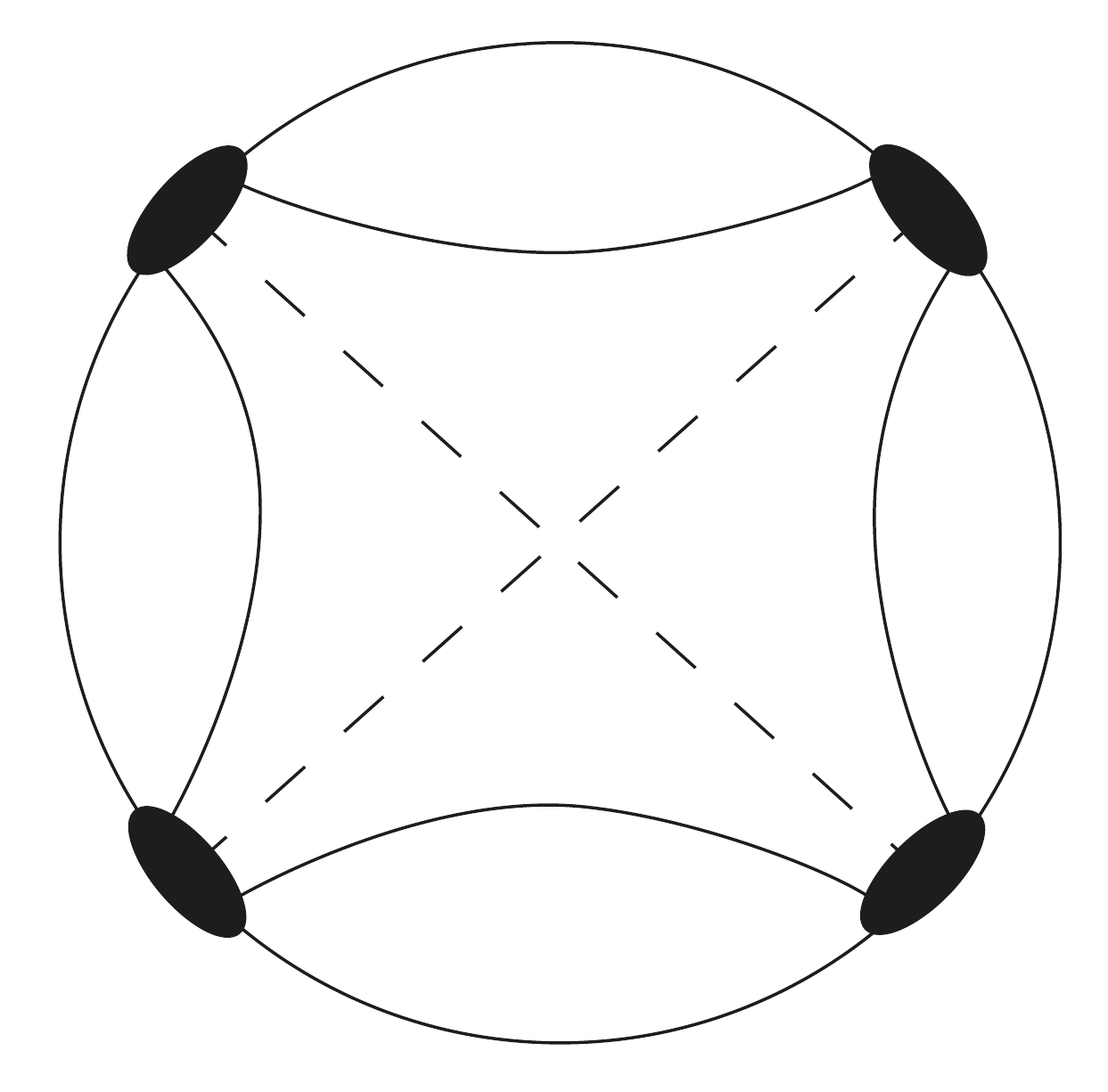}
	\caption{Witten diagram representation of the $f_\chi$ correlator, where $\chi$ is interpreted as a "triple trace" of the form $\chi=[\phi_1 \phi_2 \phi_1]$. The solid lines denote the $\phi_1$ propagators and the dashed lines denote the $\phi_2$ propagators. The diagram is manifestly $s\leftrightarrow t$ crossing symmetric.}
	\label{fig:fchidiagram}
\end{figure}

Instead of a single function $f_\chi(z)$, we propose the following family of functions
\begin{equation}
\label{fchiconcrete}
f_\chi(z) = \frac{z^{\De_\chi + \a}}{(1-z)^{\De_\chi - \a}}\,,
\end{equation}
which has a conformal block decomposition with positive coefficients if the parameter $0 \leq \alpha \leq \De_\chi / 3$.\footnote{A closed form for the conformal block coefficients appears in \cite{Hogervorst:2017sfd}. We have checked that the first 40 coefficients are positive.} This function has a Witten diagram interpretation: it is the four-point function obtained from a completely connected Witten diagram (see figure \ref{fig:fchidiagram}) where $\chi$ is interpreted as a ``triple-trace'' operator of the form $\chi=[\phi_1 \phi_2 \phi_1]$, with dimension $\Delta_\chi= 2\Delta_1+ \Delta_2$ and $\alpha=\Delta_2$. Its conformal block decomposition begins with an operator with dimension $\De_\chi + \a$ so consistency with \eqref{fchiconditions} requires $\De_\chi + \a \geq 2 \De_\phi$, leading to
\begin{equation}
\De_\chi  \geq \frac{3}{2} \De_\phi\,,
\end{equation}
as a necessary condition for $f_\chi(z)$ to exist. This precisely agrees with the observation mentioned above that the QFT in AdS bound differs from the S-matrix bound  only for $\De_\chi/ \De_\phi$ equal to 1.5 and 1.6.

With $\alpha$ a free parameter we now have the freedom to add arbitrary conformal blocks of dimensions at least $2 \Delta_\phi$ in the $s$-channel of the $\< \phi \phi \chi \chi\>$ correlator using the procedure outlined above: we add $f_\chi$ for a suitable $\a$ with a large coefficient and select the relevant block by switching on a non-zero small coefficient in $\<\phi \phi \phi \phi\>$. But then all we are left with are the two crossing equations from $\< \phi \phi \chi \chi\>$ and $\< \phi \chi \phi \chi\>$ where there is not sufficient positivity to obtain any meaningful bound. Altogether then, we must conclude that it is impossible to improve on the single-correlator bound for the parameter ranges stated above.

Finally, it is interesting to make contact with the flat-space limit. The main culprit is clearly $f_\chi(z)$ in equation \eqref{fchiconcrete}. In the flat-space limit, according to the dictionary of \cite{Komatsu:2020sag}, the corresponding contribution to the scattering amplitude would become
\begin{equation}
T_{\chi \chi \to \chi \chi} = \lim_{R \to \infty}  z^{-2\De_\chi}f_\chi(z) = \lim_{\De, \a \to \infty} \frac{1}{\big(z(1-z)\big)^{\De_\chi - \a}}\,.
\end{equation}
Since $\De_\chi - \a > 0$, we find that the limit is zero if $|z(1-z)| > 1$ but becomes infinite otherwise. As explained in \cite{Komatsu:2020sag}, this is a familar complication: in the flat-space limit not every possible correlator becomes a good scattering amplitude, and we now see how that can also limit the bounds obtained from the QFT in AdS construction.
It would be interesting to understand more systematically when do the conformal bootstrap bounds for QFT in AdS converge to the corresponding S-matrix bootstrap bounds.

%%%%%%%%%%%%%%%%%%%%%%%%%%%%%%%%%%%%%%%%%%%%%
\section{Fermions in AdS}
\label{sec:fermionsAppendix}
%%%%%%%%%%%%%%%%%%%%%%%%%%%%%%%%%%%%%%%%%%%%%

In this appendix we describe the details of the calculation involving  fermions in \(AdS_2\) outlined in the main text.
\subsection{Bosonization in \(AdS_2\)}
The bosonization duality in  flat space relates the observables in the fermionic theory to the bosonic theory as
\begin{align}
	\psi_\mp                                                                         & \leftrightarrow  e^{\pm i\phi_\mp} ,                                                 \\
	\left( \bar{\psi } \gamma^\mu \psi, \bar{\psi } \gamma^\mu \gamma^3 \psi \right) & \leftrightarrow  \left( \epsilon^{\mu\nu}\partial_\nu \phi,\partial^\mu \phi \right)
	.
\end{align}
In order to test its natural generalization to \(AdS\), we would like to perform perturbation theory in \(AdS\) around the free fermion.
We consider the massive Thirring interaction in \(AdS_2 \). In flat space, it is dual to the sine-Gordon interaction \(\cos(\beta \phi)\). The Thirring interaction is a specific interaction of four fermions in flat space given by
\begin{align}
	\mathcal{L} = \lambda_f \left( \bar{\psi} \gamma^\mu_{flat} \psi \right)  \left(  \bar{\psi} \gamma_{\mu,flat} \psi\right)  
	.
\end{align}

To generalize the fermion interactions and propagators to \(AdS_2 \), we use the shorthand notation \(Z=(y,x)\) to denote a generic bulk point as well as the vielbein \(e^a_\mu\)  
\cite{Beccaria:2019dju}. We can write the gamma matrices in \(AdS_2\), \(\gamma^\mu_{AdS} = e^\mu_a \Gamma^a\).
Let \(\psi\) denote the Dirac fermion in \(AdS_2\). When one takes the limit of this field to the boundary, one of the components dominates \cite{Faller:2017hyt,Carmi:2018qzm}
\begin{equation}
	\psi(y,x)
	  = \psi_+(y,x) +\psi_-(y,x) \,,
\end{equation}	 
with 
\begin{equation}	
	\psi_\pm
	  \rightarrow_{y\rightarrow 0} y^{d/2\pm m} \psi_{0,\pm}(x)
	\,.
\end{equation}
Note that these components are individually dual to vertex operators in the bosonic theory.
The bulk to boundary propagators for the fermions in \(AdS_{2}\) are \cite{Kawano:1999au,Faller:2017hyt}
\begin{align}
	\label{eqn: fermionProduct}
	\Sigma_{\Delta}  \left( y,x;x_i \right)
	 & = \frac{\gamma_0 y + \gamma_1 (x-x_i) }{\sqrt{y}} \Pi_{\Delta+\half}\left( y,x;x_i\right) \mathcal{P}^-  \,, \\
	\bar{\Sigma}_{\Delta} \left( y,x;x_i \right)
	 & =\mathcal{P}^+  \frac{\gamma_0 y + \gamma_1 (x-x_i) }{\sqrt{y}} \Pi_{\Delta+\half}\left( y,x;x_i  \right)
	\,.
	\nonumber
\end{align}
Here \(x,x_i\) are one-dimensional positions on the boundary. We have used the chiral projector \(\mathcal{P}^{\pm} = \left( 1\pm \gamma_0  \right)\)/2, while \(K\) denotes the corresponding propagator of the scalar operator in \(AdS\). For the purposes of perturbation theory, we note the following identity  for the product of propagators \cite{Kawano:1999au,Faller:2017hyt,Carmi:2021dsn}
\begin{align}
	\bar{\Sigma}_{\Delta} \left(  y,x;x_1  \right) \Sigma_{\Delta} \left(  y,x;x_2  \right)
	=\left( \bar{x}_{12}^\mu\gamma_\mu  \mathcal{P}^-\right)  \Pi_{\Delta + \half } \left(   y,x;x_1 \right) \Pi_{\Delta + \half} \left(   y,x;x_2 \right)
	.
\end{align}
The tensor structure $ \bar{x}_{12}^\mu\Gamma_\mu  \mathcal{P}^- = \bar{x}_{12}^\alpha \bar{\gamma}_\alpha$, where \(\bar{\gamma }\) are the boundary gamma matrices. In one-dimensional CFTs, this corresponds simply to \(x_{12}\).

%%%%%%%%%%%%%%%%%%%%%%%%%%%%%
\subsection{Perturbation theory}
%%%%%%%%%%%%%%%%%%%%%%%%%%%%%

We would like to compute the contribution in the free theory of fermions using the standard fermionic mean field theory formula.
This corresponds to simple wick contractions in $AdS_2$. We define the cross ratio $z$ in the 1d CFT as in the main text (\ref{eq:cross_ratio}).
Performing the Wick contractions using the correct negative signs for massive fermions leads to
\begin{align}
	G_{ +--+}
	 & = \frac{1}{x_{12}x_{34}} \left[ 1-z^{2 \Delta} \right],    \nonumber                                    \\
	G_{ ++-- }
	 & = \frac{-1}{x_{12}x_{34}} \left[z^{2\Delta}-\left( \frac{z}{1-z}\right) ^{2\Delta}   \right], \nonumber \\
	G_{ +-+- }
	 & =\frac{1}{x_{12}x_{34}} \left[1+\left( \frac{z}{1-z} \right)^{2\Delta}  \right]
	.
	\nonumber
\end{align}
For the first order perturbation theory, it is useful to define the $D$ function as
\begin{equation}
	D_{1111}
	  =\frac{\pi}{4} \frac{z^2}{x_{12}^2 x_{34}^2} \bar{D}\left( z \right)                                                        
	  = \frac{\pi}{4} \frac{z^2}{x_{12}^2 x_{34}^2}\left[ \frac{\log \left( z^2 \right)}{z-1}  - \frac{\log\left( 1-z \right)^2}{z}\right]
	.
\end{equation}
This function  is used in scalar contact Witten diagrams.
The first order corrections to the correlation functions can be computed using Witten diagrams. Schematically, the contact Witten diagram is written as
\begin{align}
	W_{\text{fermion}} = \lambda_f
	\int_{AdS}   \bar{\Sigma}_{\Delta} \left( Z, x_{15} \right)  \Sigma_{\Delta} \left( Z, x_{25} \right)  \bar{\Sigma}_{\Delta} \left( Z, x_{35} \right)  \Sigma_{\Delta} \left( Z, x_{45} \right)
	.
\end{align}
Consider first the case of massless free fermion, \(\Delta=\half\). Using (\ref{eqn: fermionProduct}), the product of the fermion propagators can be converted into the product of scalar propagators. They will be multiplied by the appropriate tensor structure. Thus, the fermionic contact Witten diagram can be written in terms of scalar contact Witten diagram  \( W_{\text{fermion}} \propto D_{1111}\)
\cite{Carmi:2021dsn,Carmi:2019ocp}. We compute the correlation functions using appropriate Witten diagram to arrive at the following correlation functions
\begin{align}
	G_{ +--+} 
	 & = \lambda_f\frac{\pi}{4} \frac{z(1-z)}{x_{12} x_{34}} \bar{D}_{1111}\left( z \right) , \\
	G_{ ++-- }
	 & =\lambda_f\frac{\pi}{4} \frac{z^2}{x_{12} x_{34}} \bar{D}_{1111}\left( z \right),      \\
	G_{ +-+- }
	 & =\lambda_f\frac{\pi}{4} \frac{-z}{x_{12} x_{34}} \bar{D}_{1111}\left( z \right)
	.
\end{align}
It is possible to compute the first order correction also for massive fermions, using the identity
\begin{align}
	D_{\Delta \Delta \Delta \Delta}
	 & =\frac{\pi^\half \Gamma\left( 2 \Delta-\half \right)}{2 \Gamma^4\left( \Delta \right)}
	\frac{z^{2 \Delta } }{x_{12}^{2 \Delta }x_{34}^{2 \Delta }}
	\bar{D}_{\Delta \Delta \Delta \Delta}\left( z \right)
	.
\end{align}
The corresponding correlators are as follows
\begin{align}
	G_{ +--+}
	 & =\lambda_f \left( x_{12} x_{34} - x_{13}x_{24} \right) D_{
	\Delta+\half \,\Delta+\half \,\Delta+\half \,\Delta+\half  }
	\\                                                   & =\frac{\pi^\half \Gamma\left( 2 \Delta + \half \right)}{2 \Gamma^4\left( \Delta+\half \right)}
	\frac{z^{2 \Delta} \left( 1-z \right)}{x_{12}^{2 \Delta }x_{34}^{2 \Delta }}
	\bar{D}_{	\Delta+\half \,\Delta+\half \,\Delta+\half \,\Delta+\half  }\left( z \right)
	,                                                                                                                                                      \nonumber \\
	G_{ ++-- }
	 & = \lambda_f\left( -x_{14} x_{23} + x_{13}x_{24} \right) D_{
	\Delta+\half \,\Delta+\half \,\Delta+\half \,\Delta+\half  }
	\\                                                   & =\frac{\pi^\half \Gamma\left( 2 \Delta + \half \right)}{2 \Gamma^4\left( \Delta+\half \right)}
	\frac{z^{2 \Delta+1  }  }{x_{12}^{2 \Delta }x_{34}^{2 \Delta }}
	\bar{D}_{			\Delta+\half \,\Delta+\half \,\Delta+\half \,\Delta+\half  }\left( z \right)
	,             \nonumber                                                                                                                                          \\
	G_{ +-+-} 
	 & =\lambda_f\left( x_{12} x_{34} + x_{14}x_{23} \right) D_{
	\Delta+\half \,\Delta+\half \,\Delta+\half \,\Delta+\half  }
	\\                                                   & =\frac{-\pi^\half \Gamma\left( 2 \Delta + \half \right)}{2 \Gamma^4\left( \Delta+\half \right)}
	\frac{ z^{2  \Delta -1 } }{x_{12}^{2 \Delta }x_{34}^{2 \Delta }}
	\bar{D}_{			\Delta+\half \,\Delta+\half \,\Delta+\half \,\Delta+\half  }\left( z \right)
	.\nonumber
\end{align}

%%%%%%%%%%%%%%%%%%%%%%%%%%%%%%%%%%%%%%%%%%%%%
\section{OPE coefficient maximization for O(2) correlators}
\label{sec:appopemax}
%%%%%%%%%%%%%%%%%%%%%%%%%%%%%%%%%%%%%%%%%%%%%

In the main text we probed the sine-Gordon kink S-Matrix by extremizing the correlator at the crossing symmetric point. This is the natural observable in the scenario where there are no bound states, which can be achieved by tuning the sine-Gordon parameter $\beta$. Working with bound states in $AdS$ is complicated at finite radius, since we have no control over the dimensions of the dual operators, except in perturbation theory. However, the existence of bound states provides another natural quantity to maximize: the coupling. This was done in the $\mathbb{Z}_2$ symmetric S-Matrix context in \cite{Paulos:2016fap}, leading to the S-matrix of the lightest breather in the sine-Gordon model. In the $O(2)$ symmetric case, the authors of \cite{Cordova:2018uop,Paulos:2018fym} were able to pinpoint the sine-Gordon kink S-matrix by maximizing the coupling between kink anti-kink and the lightest breather, which is $U(1)$ neutral and $\mathbb{Z}_2$ odd.\footnote{ In the parameter region where this is the only stable bound state, maximizing the coupling is not enough to obtain this S-Matrix and one needs to input additional information about resonances in the physical sheet \cite{Cordova:2018uop} to get saturation of the bounds. On the other hand, in the parameter region where there are two, or more bound states, by inputting the exact values of their masses, one directly recovers the sine-Gordon S-matrix upon maximizing the coupling \cite{Paulos:2018fym}.} In this appendix, we will study the natural generalization of this problem: maximize the OPE coefficient between the external operators and the lightest exchanged operator with the right quantum numbers.

The charged external operators have dimension $\Delta_K=2\pi/\beta^2$, which we can tune by changing the boson radius $r=1/\beta$. We consider $\Delta_K>1/4$ where the deformation is relevant. For any value of $\Delta_K$, the charge zero sectors of the free boson correlators contain only operators of integer dimension, with the first few $\mathbb{Z}_2$ odd operators having odd dimension, and $\mathbb{Z}_2$ even operators having even dimension. \footnote{This can be checked using $SL_2(\mathbb{R})$ characters.}

\begin{figure}
	\centering
	\includegraphics[width=0.65\linewidth]{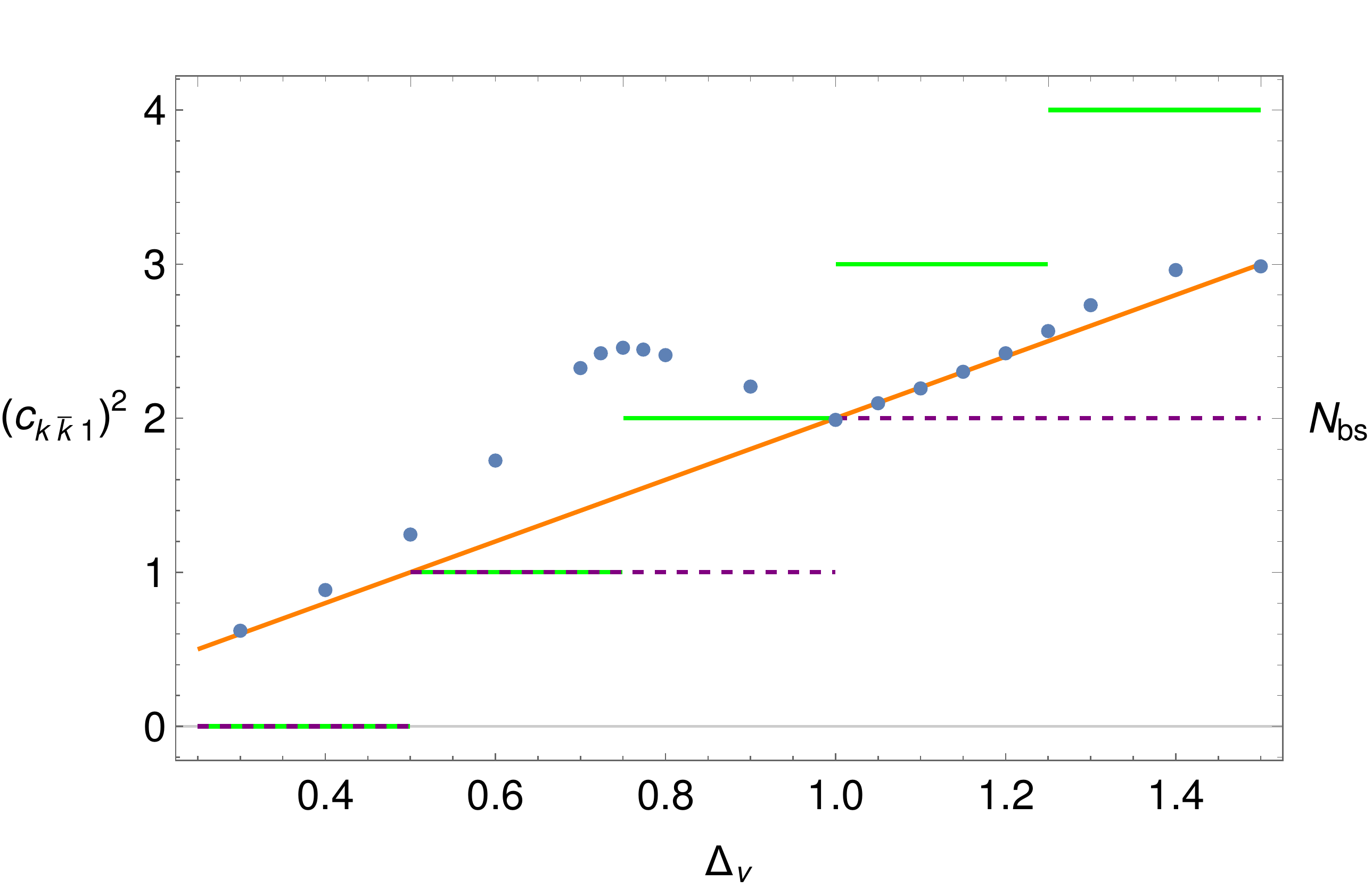}
	\caption{The blue points represent the upper bound on the kink anti-kink $\mathbb{Z}_2$ odd breather OPE coefficient as a function of the kink dimension $\Delta_v$, assuming a $\mathbb{Z}_2$ odd bound state of dimension 1, a $\mathbb{Z}_2$ even bound state of dimension 2, and all gaps to be $2\Delta_v$. In orange we plot the analytic result for the winding mode correlator. Finally in green and purple, we plot the number of bound states in the IR and UV sine-Gordon theories, respectively.}
	\label{fig:noextragaps}
\end{figure}

We can then impose the dimension of the bound states, i.e, of the $U(1)$ neutral operators with dimension smaller than $2\Delta_v$, and maximize, for each value of $\Delta_v$, the OPE coefficient $c^2_{K\bar{K}1}$.
We think of our freedom to vary $\Delta_v$ as the analogue of the choice of the sine-Gordon parameter $\beta$.
We begin by imposing a $\mathbb{Z}_2$ odd operator of dimension 1, a $\mathbb{Z}_2$ even operator of dimension 2, and take the gaps in all 3 sectors to be $2\Delta_v$, the {\em two-particle threshold}. Note that for $\Delta_v<1$ the $\mathbb{Z}_2$ even bound state gets absorbed into the kink--anti-kink {\em continuum}, and the same happens for the $\mathbb{Z}_2$ odd bound state at $\Delta_v<1/2$. We present the bounds on the OPE coefficient in figure \ref{fig:noextragaps}.
As a consistency check, we see that our one parameter family of free correlators has an OPE coefficient which is always below the bound, and in fact saturates it for $\Delta_v$ slightly above 1.
The bound has a maximum at $\Delta_v=3/4$, which curiously corresponds to the value of $\beta$ at which the flat space theory gets a second bound state. There is also a kink at $\Delta_v=1$ associated to the fact that $\Delta_2$ becomes a true {\em bound state} of the UV theory. We indicate the number of bound states in the UV and IR by purple and green step functions to clarify these facts. 

\begin{figure}
	\centering
	\includegraphics[width=0.65\linewidth]{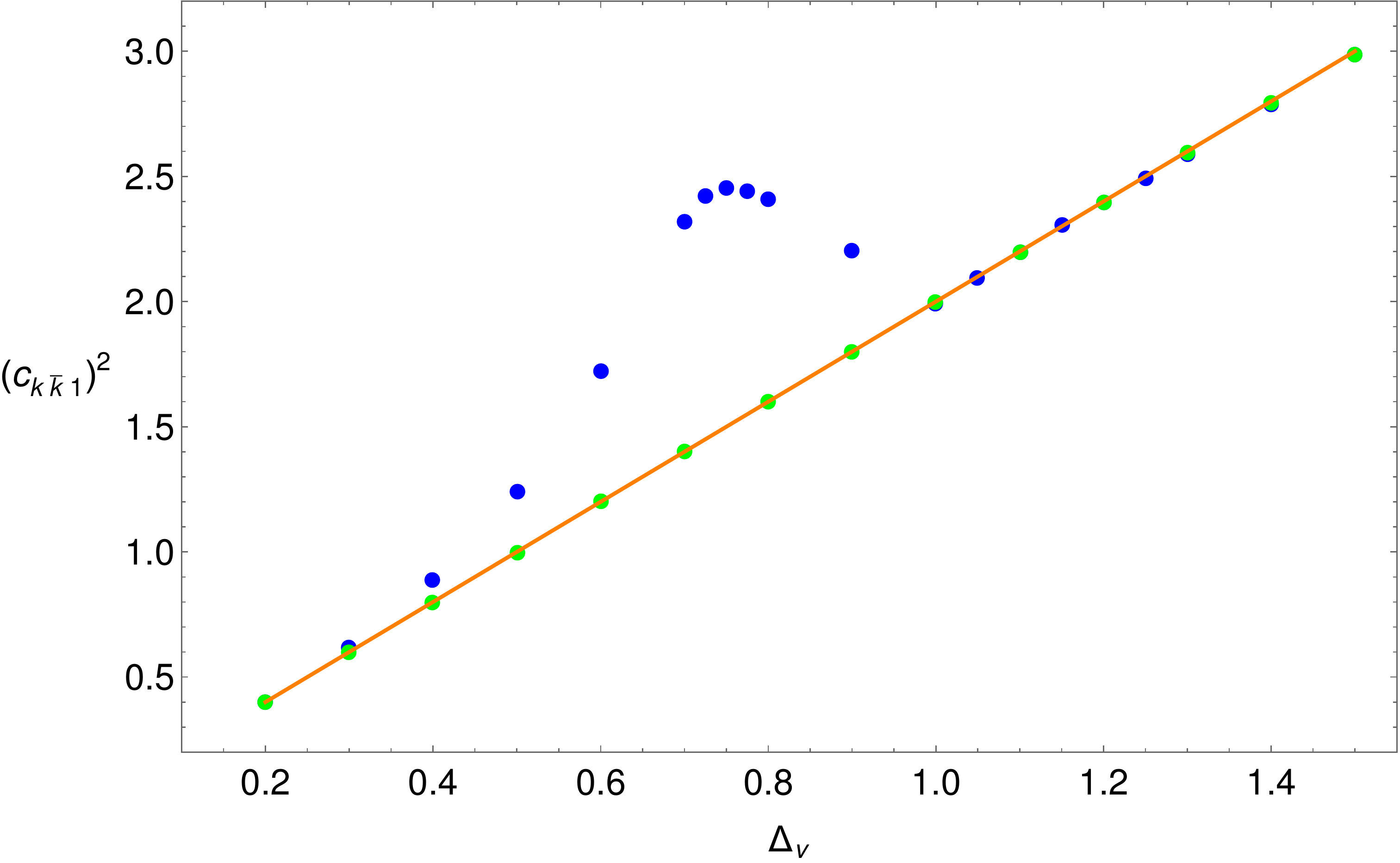}
	\caption{Same plot as before, with the stronger assumption that the $\mathbb{Z}_2$ odd gap is 3 for the blue points, and additionally that the $\mathbb{Z}_2$ even gap is 4 for the green points. The blue points coincide with the ones of the previous plot for $\Delta_v<1$ and match the analytic winding mode correlator for $\Delta_v\geq1$. The green points match the analytic correlator in the full range of $\Delta_v$.}
	\label{fig:gapodd3}
\end{figure}

For our correlator to saturate the bounds, we need to introduce more information about the spectrum. We know that the next $\mathbb{Z}_2$ odd operator after the lightest one has dimension 3. We can impose this gap in the $\mathbb{Z}_2$ odd sector to obtain the blue dots in figure \ref{fig:gapodd3}.

\begin{figure}
	\centering
	\includegraphics[width=0.65\linewidth]{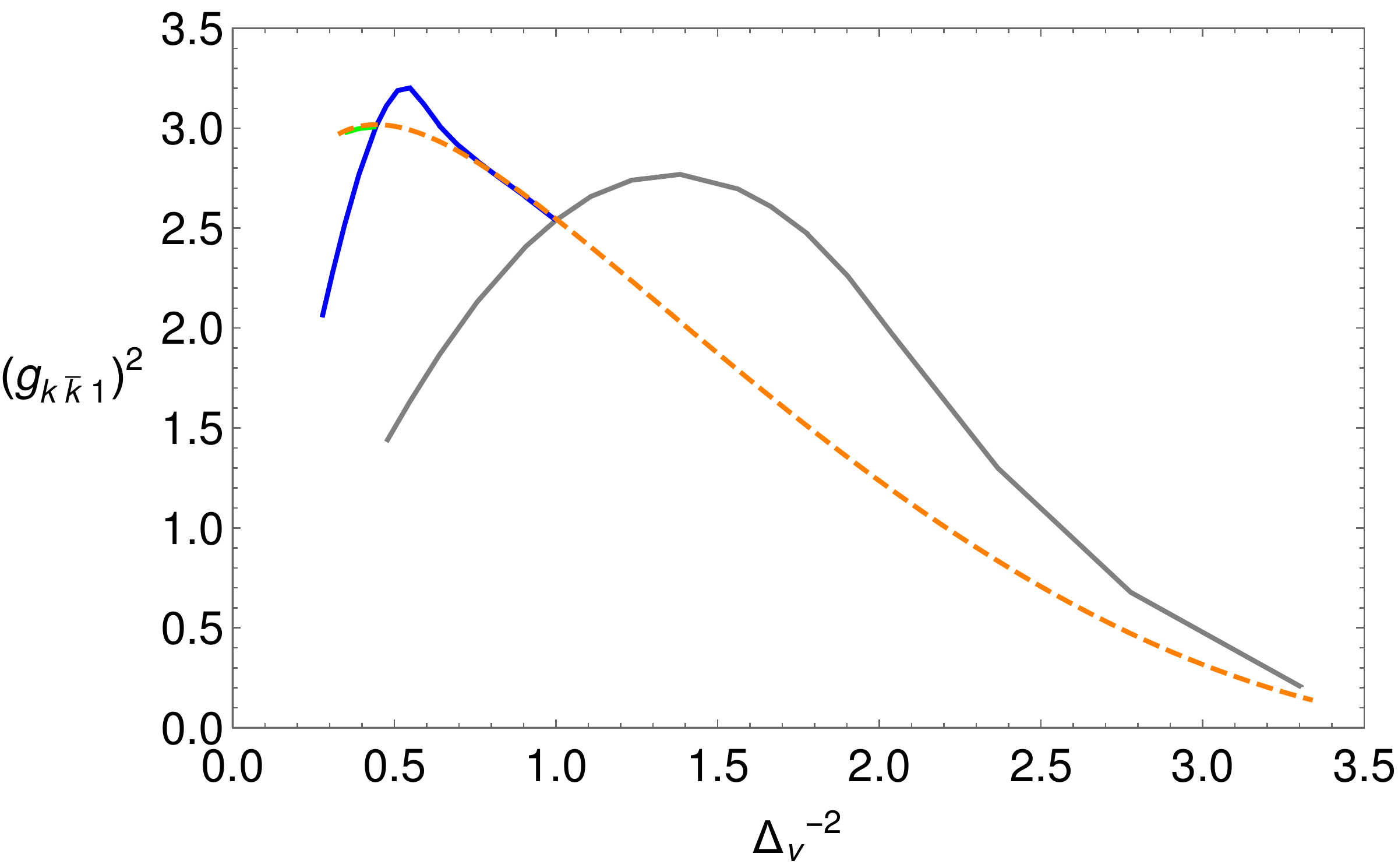}
	\caption{Bounds in terms on the AdS coupling $g_{K \bar{K}1}^2$ as a function of the AdS {\em mass} $\Delta_v^{-2}$. In grey are the bounds assuming only the bound state of dimension, in blue the bounds when we add the bound state of dimension 2 and in green when we further include the bound state of dimension 3. The orange dashed curve is the sine-Gordon correlator for zero $AdS$ radius ($\lambda=0, \Delta_1=1$), which saturates the bound in parts of the two and three bound state regions.}
	\label{fig:AdScouplingbounds}
\end{figure}

This has the effect of lowering the bound on the region $1<\Delta_v<3/2$ to the extent that the vertex operator correlator now saturates it, but gives the same result as the previous plot for $\Delta_v<1$. Finally, we increase the gap in the $\mathbb{Z}_2$ even sector to 4 which ensures that our correlator is now extremal for any value of $1/4
	\leq\Delta_v\leq3/2$. This is presented in the green dots of figure \ref{fig:gapodd3}.
We see that just like in the flat space S-Matrix analysis, one needs to introduce specific data about the  resonance spectrum, namely the gaps in the $\mathbf{0}^+$ and $\mathbf{0}^-$ sectors for the correlation function to saturate the bounds on the OPE coefficient/coupling.
Therefore, this is a less optimal question than correlator maximization, where no extra gaps were needed. This is related to the fact that at $z=1/2$ the correlator is not just dominated by the leading operator in the OPE, and therefore maximizing its OPE coefficient is not necessarily equivalent to maximizing the value of the full correlator.
We can also perform a qualitative comparison between the results at zero radius and the flat space limit. For this it is convenient to rescale the OPE coefficients into AdS couplings and to plot the mass squared ratio instead of the external dimension\footnote{In fact, since the bound state is dual to a massless particle in the free limit, we actually plot $\Delta_v^{-2}$}. We now compare the flat space results of \cite{Paulos:2018fym} (their figure 2) to our small AdS radius results (figure \ref{fig:AdScouplingbounds}). The plots are qualitatively similar, with sine-Gordon failing to be extremal in the one bound state region but matching the maximum allowed value, at least in some part of the parameter range where more bound states are taken into account.
It would be interesting to take a scaling limit where we increase the bound state dimension and try to quantitatively match to the flat space results.

\bibliographystyle{JHEP}
\bibliography{z2Inv.bib}
\end{document}